\documentclass[twocolumn,tighten]{aastex61}

\shorttitle{Neutrino-cooled
  accretion disks from neutron star mergers}
\shortauthors{Siegel \& Metzger}

\usepackage{amsfonts,amsmath,amssymb,mathrsfs}
\usepackage{graphicx}
\usepackage{dcolumn}
\usepackage{bm}
\usepackage{hyperref}
\hypersetup{
  colorlinks=true,        
  linkcolor=blue,         
  citecolor=blue,         
}

\usepackage{natbib}

\newcommand{\nb}{n_\mathrm{b}}
\newcommand{\nel}{n_\mathrm{e}}
\newcommand{\el}{\mathrm{e}}
\newcommand{\nue}{{\nu_\mathrm{e}}}
\newcommand{\nua}{{\bar{\nu}_\mathrm{e}}}
\newcommand{\nux}{{\nu_x}}
\newcommand{\nuax}{{\bar{\nu}_x}}
\newcommand{\nui}{{\nu_i}}
\newcommand{\nuia}{{\bar{\nu}_i}}

\begin{document}

\title{Three-dimensional GRMHD simulations of neutrino-cooled
  accretion disks from neutron star mergers}

\author{Daniel M. Siegel}
\altaffiliation{NASA Einstein Fellow}
\affiliation{
 Department of Physics and Columbia Astrophysics
  Laboratory, Columbia University, New York, NY 10027, USA
} 
\author{Brian D. Metzger}%
\affiliation{
 Department of Physics and Columbia Astrophysics
  Laboratory, Columbia University, New York, NY 10027, USA
}




\date{\today}

\begin{abstract}
Merging binaries consisting of two neutron stars (NSs) or an NS and a stellar-mass black hole typically form a massive accretion torus around the remnant black hole or long-lived NS. Outflows from these neutrino-cooled accretion disks represent an important site for $r$-process nucleosynthesis and the generation of kilonovae.  We present the first three-dimensional, general-relativistic magnetohydrodynamic (GRMHD) simulations including weak interactions and a realistic equation of state of such accretion disks over viscous timescales ($380\,\mathrm{ms}$). We witness the emergence of steady-state MHD turbulence, a magnetic dynamo with an $\sim\!20\,\mathrm{ms}$ cycle, and the generation of a `hot' disk corona that launches powerful thermal outflows aided by the energy released as free nucleons recombine into $\alpha$-particles.  We identify a self-regulation mechanism that keeps the midplane electron fraction low ($Y_\el\sim0.1$) over viscous timescales.  This neutron-rich reservoir, in turn, feeds outflows that retain a sufficiently low value of $Y_\el\approx 0.2$ to robustly synthesize third-peak $r$-process elements. The quasi-spherical outflows are projected to unbind $40\%$ of the initial disk mass with typical asymptotic escape velocities of $0.1c$, and may thus represent the dominant mass ejection mechanism in NS--NS mergers.  Including neutrino absorption, our findings agree with previous hydrodynamical $\alpha-$disk simulations that the entire range of $r$-process nuclei from the first to the third $r$-process peak can be synthesized in the outflows, in good agreement with observed solar system abundances.  The asymptotic escape velocities and the quantity of ejecta, when extrapolated to moderately higher disk masses, are consistent with those needed to explain the red kilonova emission following the NS merger GW170817.
\end{abstract}


\section{Introduction}
\label{sec:introduction}

When a binary system consisting of two neutron stars (NSs) or an NS and a rapidly spinning stellar-mass black hole (BH) merges into a single compact object following a prolonged inspiral driven by gravitational-wave (GW) radiation, the outcome is a violent interaction that releases mass and energy into the surrounding environment \citep{Lee&RamirezRuiz07,Lehner&Pretorius14,Baiotti&Rezzolla17}.  Neutron-rich matter ejected into space during this process subsequently synthesizes elements much heavier than iron via the rapid capture of neutrons onto nuclei ($r$-process; \citealt{Lattimer&Schramm74,Symbalisty&Schramm82,Freiburghaus1999,Goriely+11}).  The highest-mass nuclei reached by the $r$-process depends on the neutron abundance in the ejecta, as quantified by its electron fraction $Y_\el = n_{\rm p}/n_{\rm b}$, where $n_{\rm p}$ and $n_{\rm b}$ are the proton and total baryon densities, respectively.  Exclusively light $r$-process nuclei with atomic mass $A \lesssim 140$ are created for $0.25 \lesssim Y_\el \lesssim 0.40$, while heavier isotopes with $A \gtrsim 140$ are also produced if the ejecta is sufficiently neutron-rich, $Y_\el \lesssim 0.25$ \citep{Lippuner2015}.

The first detection of GWs from an NS--NS merger \citep{LIGO+17DISCOVERY} and the subsequent localization of this event---dubbed GW170817---to a galaxy at a distance of only $\approx$ 40 Mpc (e.g.~\citealt{LIGO+17CAPSTONE} and references therein) provides a golden opportunity to test theoretical predictions for the electromagnetic and nucleosynthetic signatures of these events.  Eleven hours after the merger, an optical counterpart was discovered \citep{Coulter2017,Soares-Santos2017,Valenti+17,Arcavi2017,Lipunov2017,Evans2017} with a luminosity, thermal spectrum, and rapid temporal decay consistent with ``kilonova" (KN) emission powered by the radioactive decay of $r$-process nuclei synthesized in the merger ejecta \citep{Li1998,Metzger+10,Metzger17}.  Visual (``blue") KN emission \citep{Metzger+10} was detected at early times, which then faded and was supplanted after a few days by a second distinct emission component at near-infrared (``red") wavelengths \citep{Barnes&Kasen13,Tanaka&Hotokezaka13,Wollaeger+17}, thus implicating the presence of at least two separate ejecta components.  The blue KN is well-modeled as being powered by $\approx 1.5\times 10^{-2}M_{\odot}$ of light $r$-process nuclei (ejecta with an initial electron fraction $Y_\el \gtrsim 0.25$) moving at high velocities $\approx 0.2-0.3\,c$, while the red KN requires a greater quantity $\approx 4\times 10^{-2}M_{\odot}$ of ejecta that also contains heavy $r$-process nuclei ($Y_\el \lesssim 0.25$) expanding at a lower velocity $v \approx 0.1\,c$ (e.g.~\citealt{Kasen2017,Cowperthwaite2017,Tanvir2017,Shappee2017,Kilpatrick2017,Kasliwal2017,Nicholl2017,Chornock2017,Drout+17,McCully2017,Tanvir2017,Villar2017}; however, see \citealt{Smartt2017,Tanaka2017}).

Theoretical work has identified several processes that are expected to contribute to mass ejection in NS--NS/NS--BH mergers (e.g.~\citealt{Fernandez&Metzger16}, for a review).  Strong tidal forces between the compact objects just prior to their coalescence eject low-$Y_\el$ matter focused into the equatorial binary plane (e.g.~\citealt{Rosswog+99,Oechslin&Janka06,Hotokezaka2013b,Radice2016,Bovard2017}).  However, the total ejecta mass $\approx 5\times 10^{-2}M_{\odot}$ inferred for GW170817 exceeds the dynamical ejecta obtained by any general-relativistic (GR) NS--NS merger simulation to date (e.g.~\citealt{Shibata2017}); the velocity $v \approx 0.1\,c$ of the red KN is, furthermore, several times lower than that found by the numerical simulations.

An NS--NS merger, or an NS--BH merger resulting in tidal disruption of the NS outside of the innermost stable circular orbit, also produces a massive rotating torus surrounding the central compact remnant.  This accretion torus provides a promising central engine for powering the collimated relativistic jet needed to create a short gamma-ray burst \citep{Narayan+92,Aloy+05,Rezzolla2010,Ruiz+16}. 
Outflows from the same torus over longer timescales of up to seconds provides another contribution to the $r$-process and KN emission, in addition to the dynamical ejecta \citep{Metzger+08c,Metzger2009c,Fernandez2013,Perego+14,Just2015a,Fernandez2015a}.  The torus mass found from numerical simulations can be as high as $\approx 0.1-0.2M_{\odot}$ in an NS--NS merger if the merger remnant goes through a hypermassive neutron star (HMNS) phase\footnote{The formation of an HMNS in GW170817 is supported indirectly by the high and sustained level of neutrino irradiation needed to explain the luminous blue KN (indicative of a large quantity of high-$Y_\el$ polar ejecta).} prior to forming a BH (e.g.~\citealt{Shibata&Taniguchi06,Hotokezaka2013d}).  In this case, the red KN emission from GW170817 could be explained if disk winds carry away $\approx 20-40\%$ of the total initial torus mass.  

The enormous accretion rates achieved after the merger, up to $\gtrsim 1M_{\odot}$ s$^{-1}$, occur under conditions that are highly optically thick to photons. However, the disk can still be cooled by thermal neutrino emission \citep{Popham+99,Narayan+01,Kohri&Mineshige02,DiMatteo+02,Beloborodov2003,Kohri+05,Kawanaka&Mineshige07,Chen2007}, a process that affects the lepton number of the disk in addition to its thermodynamics. The high densities and temperatures achieved in the disk midplane enable weak interactions, particularly the capture of electrons and positrons on free nuclei, to alter $Y_\el$ from the initial value of the merger debris. The precise equilibrium value to which $Y_\el$ is driven depends on the degree of electron/positron degeneracy through the Pauli blocking factors \citep{Beloborodov2003}.

Magnetohydrodynamic (MHD) turbulence, as fed by the magnetorotational instability (MRI; \citealt{Balbus&Hawley92}), is expected to drive accretion in a wide variety of astrophysical environments \citep{Balbus&Hawley98}, including in NS--NS and NS--BH mergers. However, nearly all previous numerical studies of the post-merger accretion flow have been performed under the assumption of hydrodynamics, adopting an effective hydrodynamical $\alpha-$viscosity \citep{Shakura&Sunyaev73} in place of self-consistent MHD turbulence \citep{Fernandez2013,Metzger&Fernandez14,Just2015a,Fernandez2015a,Fujibayashi2017}.\footnote{With the exception of the two-dimensional simulations of \citet{Shibata2007a}; however, the antidynamo theorem \citep{Cowling1933} prevents saturated steady-state MHD turbulence in axisymmetry.} These calculations also generally assume axisymmetry and a pseudo-Newtonian potential to mimic the effects of the GR spacetime.

A properly calibrated $\alpha$-disk model can capture the evolution of the disk surface density and bulk angular momentum reasonably well.  However, in detail, the nature of the hydrodynamical turbulence (convection versus the MRI-driven turbulence) is fundamentally different from that of the MHD case \citep{Balbus&Hawley02,Hawley&Balbus02}. Furthermore, while in $\alpha$-disks the thermal energy generated by viscosity is locally dissipated in proportion to the gas density, numerical simulations of MHD disks show that a disproportionally large fraction of their ``heating" occurs nonlocally through reconnection in low-density coronal regions \citep{Hirose+06,Jiang+14}.  This novel feature of MHD disks may be important in the context of hyperaccretion flows because the energy released in the disk corona as free nuclei recombine into $\alpha$-particles plays a significant role in unbinding mass and driving a mass-loaded outflow \citep{MacFadyen+01}.  

This paper presents the first three-dimensional, general-relativistic magnetohydrodynamic (GRMHD) simulations of the neutrino-cooled BH accretion disks created following NS--NS and NS--BH mergers.  We begin by describing the methodology of the numerical simulations and our implementation of the microphysics (Sect.~\ref{sec:setup}) before discussing the setup of the initial data (Sect.~\ref{sec:initial_data}). We then provide a detailed description of the disk evolution (Sect.~\ref{sec:evo}), including the generation of MHD turbulence; the evolution and self-regulation of the midplane electron fraction; the generation of unbound outflows; and the properties of the disk neutrino emission.  Finally, we describe our calculation of the $r$-process abundance yields of the disk outflows (Sect.~\ref{sec:r-process}). Our results and their immediate implications for the $r$-process in compact object mergers were also summarized in a companion Letter \citep{Siegel2017a}.\footnote{During the preparation of the present manuscript, \citet{Nouri2017} presented evolution of a magnetized, neutrino-cooled accretion disk from a BH--NS merger over $\approx\!60\,\mathrm{ms}$.}

\section{Analytical and numerical setup}
\label{sec:setup}

Our simulations of post-merger accretion disks are performed in ideal
GRMHD using the open-source
\texttt{EinsteinToolkit}\footnote{\url{http://einsteintoolkit.org}}
\citep{Loeffler2012} with the GRMHD code
\texttt{GRHydro} \citep{Moesta2014a}. Although we employ a fixed
background spacetime for computational efficiency in the present
simulations, our code can also handle
dynamical spacetimes. We use a
finite-volume scheme with piecewise parabolic reconstruction
\citep{Colella1984}, the HLLE Riemann solver
\citep{Harten1983,Einfeldt1988}, and constrained transport \citep{Toth2000}
to maintain a divergenceless magnetic field. In this section, we
focus exclusively on changes to GRHydro and features that we have newly
implemented for the current simulations. These include weak
interactions and approximate neutrino transport via a leakage scheme
(Secs.~\ref{sec:GRMHDnu} and \ref{sec:leakage}), a new framework and
methods for the recovery
of primitive variables that support composition-dependent equations of state (EOS; Sec.~\ref{sec:con2prim}), and the Helmholtz EOS as a microphysical EOS
also valid at comparatively low densities and temperatures to accurately
describe the properties of disk outflows (Sec.~\ref{sec:HelmEOS}).

\subsection{GRMHD with weak interactions}
\label{sec:GRMHDnu}

The equations of ideal GRMHD with weak interactions include energy and
momentum conservation, baryon number conservation, lepton number
conservation, and Maxwell's equations,
\begin{eqnarray}
\nabla_\mu T^{\mu\nu} &=& Q u^\nu, \label{eq:ev1}\\
\nabla_\mu (\nb u^\mu) &=& 0, \label{eq:ev2}\\
\nabla_\mu (\nel u^\mu) &=& R, \label{eq:ev3}\\
\nabla_\nu F^{*\mu\nu} &=& 0, \label{eq:ev4}
\end{eqnarray}
where 
\begin{equation}
  T^{\mu\nu}= \left(\rho h + b^2\right) u^\mu u^\nu + \left(p + \frac{b^2}{2}\right) g^{\mu\nu} - b^\mu b^\nu, \label{eq:Tmunu}
\end{equation}
is the energy-momentum tensor, $u^\mu$ is the
four-velocity, $\nb$ is the baryon number density, $\nel$ is the electron
number density, and $F^{*\mu\nu}$ is the dual of the Faraday electromagnetic
tensor. Furthermore, $p$ is the pressure; $h=1+\epsilon + p/\rho$
denotes the specific enthalpy, with $\epsilon$ being the specific internal
energy; $b^\mu\equiv (4\pi)^{-1/2}F^{*\mu\nu}u_\nu$ is the
magnetic field vector in the frame comoving with the fluid;
$b^2\equiv b^\mu b_\mu$; and $g_{\mu\nu}$ is the space-time
metric.\footnote{In this paper, Greek indices take space-time values 0--3, whereas
Roman indices represent the spatial components 1--3 only. Repeated
indices are summed over.} We assume that the thermodynamic properties
of matter can be described by a finite-temperature,
composition-dependent (three-parameter) EOS formulated as a function of
density $\rho = \nb m_\mathrm{b}$,
where $m_\mathrm{b}$ denotes the baryon mass; temperature $T$; and
electron fraction $Y_\el=\nel \nb^{-1}$. The evolution of $Y_\el$ is
described by Eq.~\eqref{eq:ev3}. The source terms $Qu^\nu$ and $R$ on the
right-hand side of
Eqs.~\eqref{eq:ev1} and \eqref{eq:ev3} account for the evolution of $Y_\el$
due to weak interactions, which create neutrinos and antineutrinos
that carry away energy and momentum from the system.

For numerical evolution, Eqs.~\eqref{eq:ev1}--\eqref{eq:ev4} can
essentially be
transformed into a set of conservation equations in flat space by
adopting a 3+1 split of spacetime into nonintersecting
space-like hypersurfaces of constant coordinate time $t$
\citep{Lichnerowicz44,Arnowitt2008}, in which case, the line element can
be written as
\begin{equation}
  \mathrm{d}s^2 = -\alpha^2 \mathrm{d}t^2 + \gamma_{ij}(\mathrm{d}x^i
  + \beta^i\mathrm{d}t) (\mathrm{d}x^j
  + \beta^j\mathrm{d}t),
\end{equation}
where $\alpha$ denotes the lapse function, $\beta^i$ is the shift vector,
and $\gamma_{ij}$ is the metric induced on every spatial
hypersurface. The hypersurfaces are characterized by the time-like unit normal
$n^\mu = (\alpha^{-1},-\alpha^{-1}\beta^i)$ ($n_\mu =
(-\alpha,0,0,0)$), which also defines the Eulerian observer, i.e., the
observer moving through spacetime with four-velocity $n^\mu$
perpendicular to the
hypersurfaces. Equations~\eqref{eq:ev1}--\eqref{eq:ev4} can then be
written as
\begin{equation}
  \partial_t(\sqrt{\gamma}\mathbf{q})
  + \partial_i[\alpha\sqrt{\gamma}\mathbf{f}^{(i)}(\mathbf{p},\mathbf{q})]
  = \alpha\sqrt{\gamma} \mathbf{s}(\mathbf{p}), \label{eq:GRMHDeqns}
\end{equation}
where $\gamma$ is the determinant of the spatial metric $\gamma_{ij}$ and
\begin{equation}
  \mathbf{q} \equiv [D,S_i,\tau,B^i,DY_\el] \label{eq:q}
\end{equation}
denotes the vector of conserved variables. The latter is
composed of the conserved density, the conserved momenta, and the
conserved energy, defined as
\begin{eqnarray}
  D &\equiv& \rho W, \label{eq:D}\\
  S_i &\equiv& -n_\mu T^{\mu}_{\phantom{\mu}i} = \alpha
               T^{0}_{\phantom{0}i}= (\rho h + b^2) W^2 v_i - \alpha b^0 b_i, \label{eq:Si}\\
 \tau &\equiv& n_\mu n_\nu T^{\mu\nu} - D \\
  &=& (\rho h + b^2)W^2 - \left(p +\frac{b^2}{2}\right) -
         \alpha^2(b^0)^2 - D, \label{eq:tau}
\end{eqnarray}
respectively, the three-vector components of the magnetic field $B^\mu
\equiv (4\pi)^{-1/2}F^{*\mu\nu}n_\nu$ as measured by the
Eulerian observer, as well as the conserved electron fraction
$DY_\el$. The Eulerian three-velocity is defined by
\begin{equation}
  v^i \equiv \frac{\gamma^i_{\phantom{i}\mu}u^\mu}{-u^\mu n_\mu} =
  \frac{u^i}{W} + \frac{\beta^i}{\alpha}, \mskip20mu  v_i = \frac{\gamma_{i\mu}u^\mu}{-u^\mu n_\mu} =
  \frac{u_i}{W}, \label{eq:vi}
\end{equation}
where
\begin{equation}
  W \equiv -u^\mu n_\mu = \alpha u^0 = \frac{1}{\sqrt{1-v^2}} \label{eq:W}
\end{equation}
denotes the relative Lorentz factor between $u^\mu$ and $n^\mu$, with
$v^2\equiv\gamma_{ij}v^iv^j$. For completeness, the comoving and Eulerian
magnetic field components are related by
\begin{equation}
  b^i = \frac{B^i}{W} + b^0 (\alpha v^{i} - \beta^i), 
  \mskip20mu  b_i =
  \frac{B_i}{W} + \alpha b^0 v_{i} \label{eq:bi}
\end{equation}
and
\begin{equation}
  b^0 = \frac{W}{\alpha} B^iv_i, \mskip20mu  b^2 = b^\mu b_\mu = \frac{B^2 + (\alpha b^0)^2}{W^2}, \label{eq:b2}
\end{equation}
where $B^2\equiv B^iB_i$. Furthermore,
\begin{equation}
  \mathbf{p} \equiv [\rho,v^i,\epsilon,B^i,Y_\el] \label{eq:p}
\end{equation}
summarizes the set of primitive variables. The fluxes are given by
\begin{equation}
  \mathbf{f}^{(i)}(\mathbf{p,q}) \equiv \left[ \begin{array}{c}
   D \tilde{v}^i \\
   S_j \tilde{v}^i + \left(p +\frac{b^2}{2}\right) \delta^i_j -
                                               \frac{B^i}{W}b_j\\
  \tau \tilde{v}^i + \left(p +\frac{b^2}{2}\right) v^i - \alpha b^0
                                               \frac{B^i}{W}\\
  \tilde{v}^i B^k - \tilde{v}^k B^i\\
  D Y_\el \tilde{v}^i
 \end{array}
 \right]
\end{equation}
and the sources by
\begin{equation}
  \mathbf{s}(\mathbf{p}) \equiv \left[ \begin{array}{c}
   0 \\
   T^{\mu\nu}(\partial_\mu g_{j\nu}-\Gamma^\delta_{\nu\mu}g_{\delta j})+WQv_j\\
  \alpha(T^{0\mu}\partial_\mu\ln\alpha - T^{\mu\nu}\Gamma^0_{\mu\nu})+WQ\\
  0^k\\
  R m_\text{b}
 \end{array}
 \right], \label{eq:sources}
\end{equation}
where $\tilde{v}^i\equiv v^i - \beta^{i} \alpha^{-1}$, and
$\Gamma^\alpha_{\beta\gamma}$ are the Christoffel symbols constructed
from $g_{\mu\nu}$.

\subsection{Neutrino leakage scheme}
\label{sec:leakage}

Weak interactions and neutrino transport determine the source terms on
the right-hand side of
Eqs.~\eqref{eq:ev1} and \eqref{eq:ev3}, and the terms apart from the
geometrical source terms in Eq.~\eqref{eq:GRMHDeqns}
(cf. Eq.~\eqref{eq:sources}). For the present simulations, we employ
an energy-averaged (gray) leakage scheme, which we have newly implemented into
\texttt{GRHydro}. Such leakage schemes are widely used in both
core-collapse supernova and
compact-binary merger simulations (e.g.,
\citealt{vanRiper1981,Ruffert1996b,Rosswog2003b,Sekiguchi2011,Ott2013,Radice2016,Perego2016}). Our
implementation closely follows the one by \citet{Radice2016}, which is
based on \citet{Galeazzi2013}, which, in turn, builds on
\citet{Ruffert1996b} and \citet{Bruenn1985}. We follow the
procedure discussed in \citet{Neilsen2014} to compute optical depths, which is
well suited for aspherical and complex geometries (such as that of an
accretion disk). In the following, we briefly outline some aspects of
our leakage scheme.

We specify the net neutrino
heating/cooling rate per unit volume in the rest frame of the fluid,
$Q$, and the net lepton emission/absorption rate per unit volume in the
rest frame of the fluid, $R$,
(cf. Eqs.~\eqref{eq:ev1},\eqref{eq:ev3}, and \eqref{eq:sources})
as a local balance of absorption and emission of free-streaming neutrinos,
\begin{equation}
  R = \sum_{\nui}\kappa_\nui n_\nui
  - (R_\nue^\mathrm{eff} - R_\nua^\mathrm{eff})
\end{equation}
and
\begin{equation}
  Q = \sum_\nui \kappa_\nui n_\nui E_\nui
  - \sum_\nui Q_\nui^\mathrm{eff}. \label{eq:Q}
\end{equation}
Here $\nui = \{\nue,\nua,\nux\}$, where $\nue$ denotes electron
neutrinos, $\nua$ denotes electron
antineutrinos, and the heavy-lepton neutrinos $\nu_\mu$ and $\nu_\tau$
are collectively labeled as $\nux$. Furthermore, $\kappa_\nui$, $n_\nui$, and
$E_\nui$, denote the corresponding absorption
opacities, number densities, and mean energies of the free-streaming
neutrinos in the rest frame of the fluid, respectively. Finally,
$R_\nue^\mathrm{eff}$, $R_\nua^\mathrm{eff}$, and
$Q_\nui^\mathrm{eff}$, denote the corresponding effective number and
energy emissivities in
the rest frame of the fluid. For the present simulations, we neglect neutrino
absorption, as the accretion disk simulated here remains optically thin to all
neutrino species at all times (cf. \citealt{Siegel2017a}). Neutrino
absorption is only expected to appreciably
change the outflow and disk dynamics for significantly more massive accretion
disks \citep{Fernandez2013}.

The effective emission/cooling rates $R^\mathrm{eff}_{\nu_i}$ and
$Q^\mathrm{eff}_{\nu_i}$ take effects of finite
optical depth into account and
are computed from the intrinsic (free) emission rates
$R_\nui$ and $Q_\nui$ by
(cf. Eq.~(B22) and (B23) of \citealt{Ruffert1996b})
\begin{equation}
  R^\mathrm{eff}_\nui = \frac{R_\nui}{1 +
    \frac{t_{\mathrm{diff},\nui}}{t^{\mathrm{em,R}}_\nui}}, \mskip10mu Q^\mathrm{eff}_\nui = \frac{Q_\nui}{1 +
    \frac{t_{\mathrm{diff},\nui}}{t^{\mathrm{em,Q}}_\nui}}.
\end{equation}
Here
\begin{equation}
  t_{\mathrm{diff},\nui} =
  D_\mathrm{diff}\kappa_\nui^{-1} \tau_\nui^2
\end{equation}
denote the local diffusion timescales, where $\tau_\nui$ are the corresponding
optical depths (see below), and $D_\mathrm{diff}$ is a diffusion
normalization factor, which we set to $D_\mathrm{diff}=6$
\citep{OConnor2010}. Furthermore,
\begin{equation}
   t^{\mathrm{em},R}_\nui = \frac{R_\nui}{n_\nui}, \mskip20mu
   t^{\mathrm{em},Q}_\nui = \frac{Q_\nui}{e_\nui} \label{eq:tem}
\end{equation}
are the local neutrino number and energy emission timescales, where $e_\nui$ refers
to the neutrino energy densities and
\begin{eqnarray}
  R_\nui &=& \delta_{\nui,\nue} R^\beta_\nue + \delta_{\nui,\nua}
  R^\beta_\nua + R^{\mathrm{ee}}_\nui + R^\gamma_\nui, \\
  Q_\nui &=& \delta_{\nui,\nue} Q^\beta_{\nue} + \delta_{\nui,\nua}
  Q^\beta_\nua + Q^{\mathrm{ee}}_\nui + Q^\gamma_\nui
\end{eqnarray}
(cf. Eqs.~(B18)--(B21) of \citealt{Ruffert1996b}). The emission rates $R^\beta_\nui$ and
$Q^\beta_\nui$, $R^\mathrm{ee}_\nui$ and $Q^\mathrm{ee}_\nui$, and $R^\gamma_\nui$ and
$Q^\gamma_\nui$ are computed as in \citet{Galeazzi2013} and reflect the contributing neutrino emission mechanisms
we consider. These are, respectively,
\begin{itemize}
\item[(i)] charged current $\beta$-processes,
\begin{eqnarray}
  e^- + p \rightarrow n + \nue, \label{eq:ep_nnue}\\
  e^+ + n \rightarrow p + \nua, \label{eq:en_pnua}
\end{eqnarray}
the strongest neutrino emission mechanism in hot and dense nuclear matter;
\item[(ii)] electron--positron pair annihilation,
\begin{eqnarray}
  e^- + e^+ \rightarrow \nue + \nua, \label{eq:ee_nuenua}\\
  e^- + e^+ \rightarrow \nux + \nuax, \label{eq:ee_nuxnuax}
\end{eqnarray}
which is most relevant in nondegenerate nuclear matter at low
densities and high temperatures; and
\item[(iii)] plasmon decay,
\begin{eqnarray}
  \gamma \rightarrow \nue + \nua, \\
  \gamma \rightarrow \nux + \nuax,
\end{eqnarray}
which is efficient at intermediate densities and high temperatures.
\end{itemize}

\subsubsection{Calculation of opacities}

The neutrino opacities $\kappa_\nui$ introduced above may be
subdivided into contributions from absorption and scattering,
\begin{equation}
  \kappa_\nui = \kappa_{\nui,\mathrm{abs}} + \kappa_{\nui,\mathrm{scat}},
\end{equation}
where
\begin{itemize}
\item[(i)] $\kappa_{\nui,\mathrm{abs}}$ refers to absorption of
  electron and anti-electron neutrinos only,
 \begin{eqnarray}
    \nue + n \rightarrow p + e^-, \label{eq:nuen_pe}\\
    \nua + p \rightarrow n + e^+; \label{eq:nuap_ne}
  \end{eqnarray}
  and
\item[(ii)] $\kappa_{\nui,\mathrm{scat}}$ refers to coherent
  scattering on heavy nuclei $A$ and scattering on free nucleons,
  \begin{eqnarray}
     \nui + A &\rightarrow& \nui + A, \\
    \nuia + A &\rightarrow& \nuia + A,\\
    \nui + [n,p] &\rightarrow& \nui +[n,p], \\
    \nuia + [n,p] &\rightarrow& \nuia + [n,p].
  \end{eqnarray}
\end{itemize}
The absorption and scattering opacities for these processes are
computed as in \citet{Galeazzi2013}.

\subsubsection{Calculation of optical depths}

We reduce the nonlocal computation of optical depths $\tau_\nui$ to an
effective local problem by applying the method described in
\citet{Neilsen2014}, which is well suited for aspherical geometries
such as an accretion disk. Global integrations are avoided by
decomposing the optical depth at a
given grid point into the optical depth to any neighboring
point plus the already computed optical depth
$\tau_{\nui,\mathrm{neigh}}$ at the neighboring
point, which we compute as
\begin{equation}
  \tau_{\nui,\mathrm{neigh}} + \bar{\kappa}_\nui (\bar{\gamma}_{ab}
  \mathrm{d}x^{a}\mathrm{d}x^{b})^{1/2}, \label{eq:tau_nui}
\end{equation}
where $\mathrm{d}x^a$ is the spatial coordinate distance vector between the two points, and
$\bar{\kappa}_\nui$ and $\bar{\gamma}_{ab}$ denote the
opacities and components of the spatial metric averaged between the two
neighboring points. We define the optical depth at a given grid
point as the minimum over all expressions (Equation \eqref{eq:tau_nui}) computed
for all neighboring points.

\subsection{Recovery of primitive variables}
\label{sec:con2prim}

Conservative GRMHD schemes evolve the conserved variables
$\mathbf{q}$ (cf. Eq.~(\ref{eq:GRMHDeqns})). This involves computing
the flux terms
$\mathbf{f}^{(i)}(\mathbf{p},\mathbf{q})$ and source terms
$\mathbf{s}(\mathbf{p})$ for a given $\mathbf{q}$, which requires
us to obtain the primitive
variables $\mathbf{p}$ from the conserved ones. While the conservative
variables as a function of primitive variables, $\mathbf{q}
=\mathbf{q}(\mathbf{p})$, are given in analytic form by
Eqs.~(\ref{eq:D})--(\ref{eq:b2}), the inverse relation, $\mathbf{p}
=\mathbf{p}(\mathbf{q})$, i.e., the recovery of primitive variables
from conservative ones, is not known in closed form; this instead
requires numerical inversion of the aforementioned set of
nonlinear equations.

We have implemented a new framework for the recovery of primitive
variables in \texttt{GRHydro} that provides support for any composition-dependent,
finite-temperature (three-parameter)
EOS, as well as a recovery scheme based
on a three-dimensional Newton--Raphson solver using Eqs.~(21),(22), and (28)
in \citet{Cerda-Duran2008}. We find that this scheme has particularly fast
convergence properties as compared to other schemes, typically involving a
minimum of EOS calls (\citealt{Siegel2018b}; \citealt{Siegel2018con2primcode}\footnote{Codebase: \url{https://doi.org/10.5281/zenodo.1213306}}). The latter fact is of
particular importance for three-parameter EOS, as most such EOSs are
provided in the form of multidimensional tables, and table lookups can
become computationally expensive. Furthermore, its ability to
recover strongly magnetized regions is important for evolving
low-density magnetized disk winds, as in the present simulation.

\subsection{Helmholtz EOS}
\label{sec:HelmEOS}

We base the microphysical description of matter at the relatively low
densities and temperatures of our present simulation on the Helmholtz
EOS \citep{Timmes1999,Timmes2000}, which we have newly implemented into
\texttt{GRHydro}. Nuclear-reaction networks such as \texttt{SkyNet}
\citep{Lippuner2017b}, which we employ for calculating $r$-process abundance yields, also use
the Helmholtz EOS, which is how we minimize thermodynamical
inconsistencies between the simulation and subsequent postprocessing
to obtain nucleosynthesis abundance yields.

The Helmholtz EOS is formulated in terms of a Helmholtz free energy,
which takes into account contributions from nuclei (treated as ideal
gas) with Coulomb corrections, electrons and positrons with an
arbitrary degree of relativity and degeneracy, and photons in local
thermodynamic equilibrium. As nuclei in the present simulation, we
consider free neutrons and protons, as well as $\alpha$-particles. We
have modified the Helmholtz EOS to include the nuclear binding energy
release from $\alpha$-particle formation. We compute the
abundances of nuclei at given ($\rho$, $T$, $Y_\el$) assuming nuclear
statistical equilibrium (NSE), i.e., by numerically solving the Saha equation
supplemented with baryon number and charge conservation,
\begin{eqnarray}
  n_\mathrm{p}^2n_\mathrm{n}^2 &=& 2n_\alpha
                                   \left(\frac{m_\mathrm{b}k_\mathrm{B}T}{2\pi
                                   \hbar^2}\right)^{9/2}
                                   \exp(-Q_\alpha/k_\mathrm{B}T), \label{eq:Saha1}
  \\
 \nb &=& n_\mathrm{n} + n_\mathrm{p} + 4n_\alpha,\\
 \nb Y_\el &=& n_\mathrm{p} + 2n_\alpha. \label{eq:Saha3}
\end{eqnarray}
Here $k_\mathrm{B}$ is the Boltzmann constant, $\hbar$ is the reduced
Planck constant, $Q_\alpha \simeq
28.3\,\mathrm{MeV}$ is the nuclear binding energy of an $\alpha$-particle, and $n_\mathrm{n}$, $n_\mathrm{p}$, and $n_\alpha$ denote the number
densities of neutrons, protons, and $\alpha$-particles,
respectively. We also include additional terms to the thermodynamical
derivatives that arise from compositional changes with respect to ($\rho$, $T$,
$Y_\el$), i.e., from the fact that $\partial n_\mathrm{n}/\partial \rho$, $\partial
n_\mathrm{n}/\partial T$, $\partial n_\mathrm{n}/\partial Y_\el$
etc. from Eqs.~\eqref{eq:Saha1}--\eqref{eq:Saha3} are nonzero. These
additional terms can be important to the evolution code, as, e.g., the
Riemann solver can depend on thermodynamic derivatives through the
sound speed.

\section{Initial data and grid setup}
\label{sec:initial_data}

We start our long-term disk simulation from an axisymmetric
equilibrium torus around a rotating BH of mass $M_\mathrm{BH} =
3\,M_\odot$ with dimensionless spin $\chi_\mathrm{BH}=0.8$, computed in
horizon-penetrating Kerr--Schild coordinates
\citep{Kerr1963}. We assume a constant
specific angular momentum and a small constant specific entropy of
$8\,k_\mathrm{B}$ per baryon. Under these assumptions, the
GR Euler equations reduce to inverting
the specific enthalpy given by \citep{Stergioulas2011,Friedman2013}
\begin{equation}
  h u_0 = \text{const.}, \label{eq:Euler}
\end{equation}
in order to find all other thermodynamic variables, including density
and temperature. Here the right-hand side is an arbitrary integration
constant and $u_0$ is entirely determined by the metric
components of the Kerr--Schild metric. In numerically inverting Eq.~\eqref{eq:Euler}, we assume a
constant initial electron
fraction $Y_\el = 0.1$ and a torus mass of
$M_{t0}=0.03\,M_\odot$, with a location of maximum density at $R_0 =
30\,\mathrm{km}\,[6.7\,M_\mathrm{BH}]$ (see also
Tab.~\ref{tab:BH_torus}); the inner and outer radii of
the torus are located at
$R_{\mathrm{in},0}=18\,\mathrm{km}\,[4\,M_\mathrm{BH}]$ and
$R_{\mathrm{out},0}= 106\,\mathrm{km}\,[24\,M_\mathrm{BH}]$.

\begin{table}[tb]
\caption{Initial data: BH--Torus configuration with (from left to right) BH mass and 
  dimensionless spin, torus mass, radius
  at maximum density, specific entropy, electron fraction, and maximum magnetic field strength.}
\label{tab:BH_torus}
\centering
\begin{tabular}{ccccccc}
\hline\hline
 $M_\mathrm{BH}$ & $\chi_\mathrm{BH}$ & $M_{t0}$& $R_0$ & $s_0$
  &$Y_{\mathrm{e}0}$ 
  & $B_\mathrm{max} $\\
$(M_\odot)$ &  & $(M_\odot)$& $(\mathrm{km})$ &
                                                $(k_\mathrm{B}/\mathrm{b})$
  & &  $(\mathrm{G})$\\
\hline
$3.00$ & $0.8$ & $0.0$3 & $30$ & 8 & $0.1$ 
                                          & $3.3\times 10^{14}$ \\
\hline
\end{tabular}
\end{table}

We endow the equilibrium torus with a weak initial magnetic seed
field, confined to the interior of the torus and defined by the vector
potential with components $A^r=A^\theta = 0$ and $A^\phi =
A_b\, \mathrm{max}\{p-p_\mathrm{cut},0\}$. Here
$p_\mathrm{cut}=1.3\times 10^{-2}p_\mathrm{max}$, where
$p_\mathrm{max}$ is the pressure at maximum density in the torus;
tuning $A_b$, we set the initial field strength such that the
maximum magnetic-to-fluid pressure ratio in the torus is
$p_B/p_\mathrm{f}<5\times 10^{-3}$, where $p_B=b^2/2$; this ratio
corresponds to a maximum initial magnetic field strength of $3.3\times 10^{14}\,\mathrm{G}$.

The initial parameters of the BH and torus correspond to those of a
typical NS merger remnant. The BH spins resulting from NS--NS mergers leading
to prompt BH formation are typically $\chi_\mathrm{BH}\approx 0.8$
\citep{Kiuchi2009,Rezzolla2010,Bernuzzi2014} and cannot be significantly larger
\citep{Kastaun2013}; the case of delayed BH formation is typically
not much smaller, $\chi_\mathrm{BH}\lesssim 0.7$
\citep{Sekiguchi2016}. Furthermore, $\chi_\mathrm{BH}\sim 0.8$ is a reasonable estimate of the spin of the BH in a BH--NS merger in cases when the NS is tidally disrupted and thus able to form a massive torus \citep{Foucart2012}. The
initial torus mass we adopt is also fairly typical of NS mergers (e.g.,
\citealt{Hotokezaka2011,Foucart2017a}). Furthermore, we have chosen the
initial parameters in such a way that (i) the setup is very
similar to previous 2D Newtonian simulations \citep{Fernandez2015a} and (ii) the resulting configuration after relaxation and having reached a
saturated MRI state (see Sec.~\ref{sec:MRI}) closely resembles the
properties of early post-merger accretion disks obtained from
magnetized NS--NS merger simulations such as, e.g., \citet{Ciolfi2017a}.

The initial torus is embedded in a tenuous atmosphere of uniform
density $\rho=37\,\mathrm{g}\,\mathrm{cm}^{−3}$, temperature
$T=10^5\,\mathrm{K}$, and electron fraction $Y_\el=1$. Both the density
and temperature of the atmosphere are sufficiently low to influence neither the dynamics nor the composition of the disk outflows. This density value translates into a total atmosphere mass
on the entire computational domain of $6.7\times 10^{-5}\,M_\odot$ (and
$7.8\times 10^{-8}\,M_\odot$ over the volume with radius $1000\,\mathrm{km}$,
at which we evaluate bound vs. unbound outflow),
which is safely orders of magnitude smaller than the total ejecta mass in
the disk outflows.  Furthermore, at $T=10^5\,\mathrm{K}$, the
material is sufficiently cold that weak interactions are completely frozen out.

The computational domain consists of a Cartesian grid hierarchy with
the BH at the center, embedded in eight refinement levels extending out to
$1.53\times 10^9\,\mathrm{cm}$ in all coordinate directions. The
initial torus is entirely contained by the finest refinement level,
which has a diameter of $240\,\mathrm{km}$ with a resolution of
$\Delta_{xyz}=856\,\mathrm{m}$, which corresponds to
$\Delta_{xyz}/M_\mathrm{BH}\simeq 0.19$. The simulations are performed
in full 3D without symmetries.

\section{Disk evolution}
\label{sec:evo}

A brief description of the disk evolution corresponding to the initial
data described above was already provided in \citet{Siegel2017a}. Here
we present a more detailed analysis of the evolution and address some
general properties of neutrino-cooled accretion disks for the first
time in GRMHD. In particular, we describe the initial transient phase
in which we witness the onset of MHD turbulence and describe how a
steady turbulent state is achieved (Sec.~\ref{sec:MRI}); we
demonstrate the existence of a self-regulation mechanism to mild
electron degeneracy in the inner parts of the disk, which ensures neutron-rich outflows
 and the production of third-peak $r$-process elements
 (Sec.~\ref{sec:disk_regulation}); and we
present direct evidence for a fully operational magnetic dynamo in the
disk in the presence of neutrino cooling and discuss the physical
processes that generate winds in the hot disk corona
(Sec.~\ref{sec:dynamo}). Finally, we discuss the global structure and
long-term evolution of the disk (Sec.~\ref{sec:global_structure})
and the characteristics of its neutrino radiation (Sec.~\ref{sec:neutrino_emission}).

\subsection{Onset of MHD turbulence and its steady state}
\label{sec:MRI}

\begin{figure}[tb]
\centering
\includegraphics[width=0.49\textwidth]{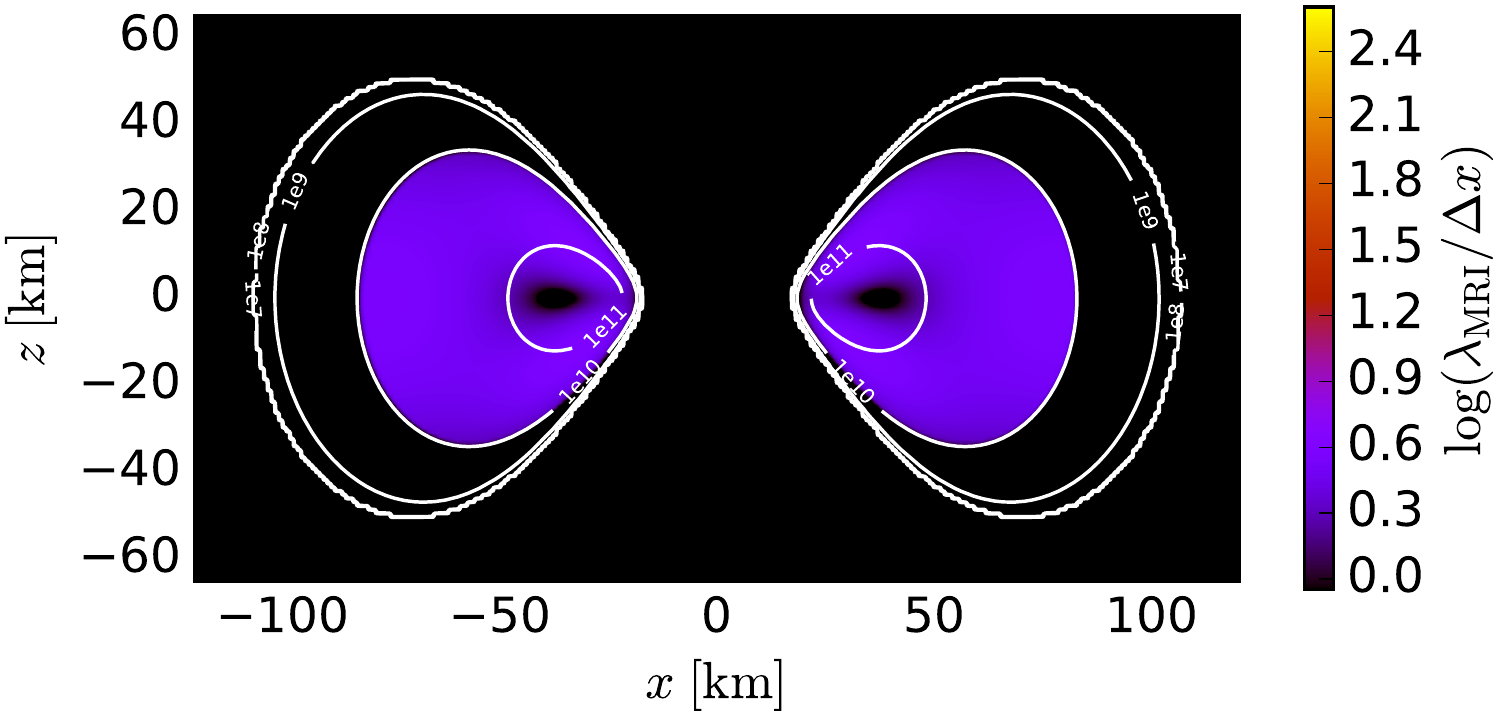}
\includegraphics[width=0.49\textwidth]{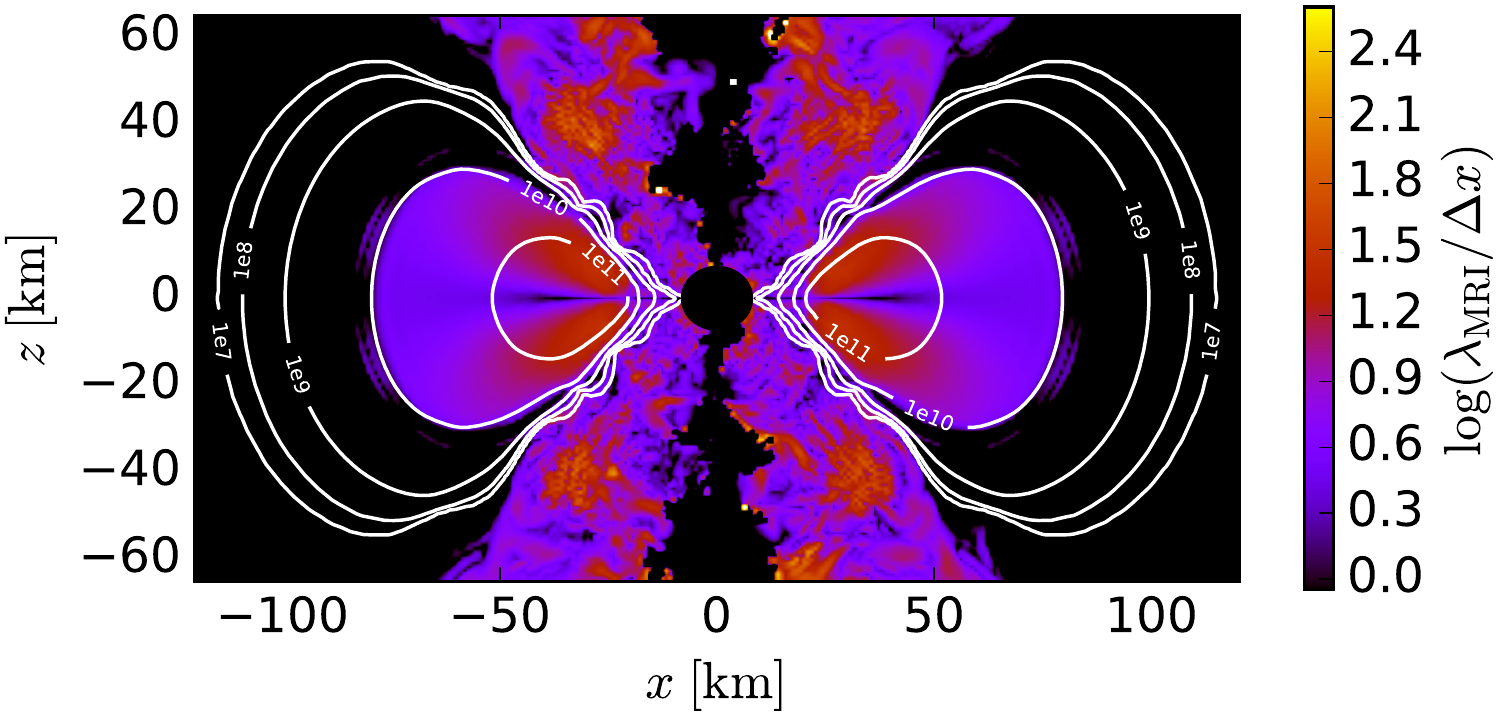}
\includegraphics[width=0.49\textwidth]{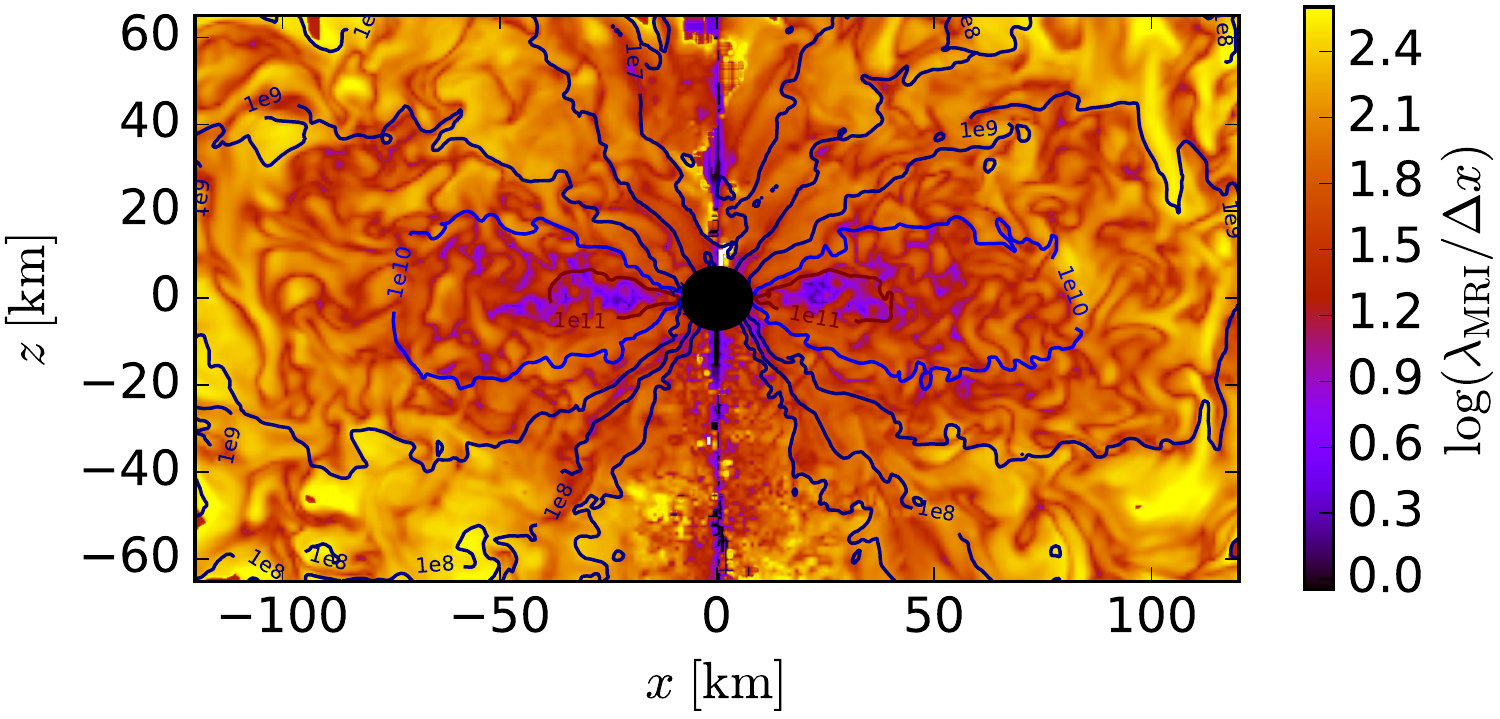}
\caption{Number of grid points per fastest-growing MRI wavelength
  $\lambda_\mathrm{MRI}$ in the meridional plane at $t=0\,\mathrm{ms}$
  (top), at $t=1.1\,\mathrm{ms}$ (center), and at $t=20\,\mathrm{ms}$ (bottom). Also shown are the contours of the
  rest-mass density at $\rho = [10^7, 10^8, 10^9, 10^{10},
  10^{11}]\,\mathrm{g}\,\mathrm{cm}^{-3}$. }
 \label{fig:MRI_start}
\end{figure}

\begin{figure}[tb]
\centering
\includegraphics[width=0.45\textwidth]{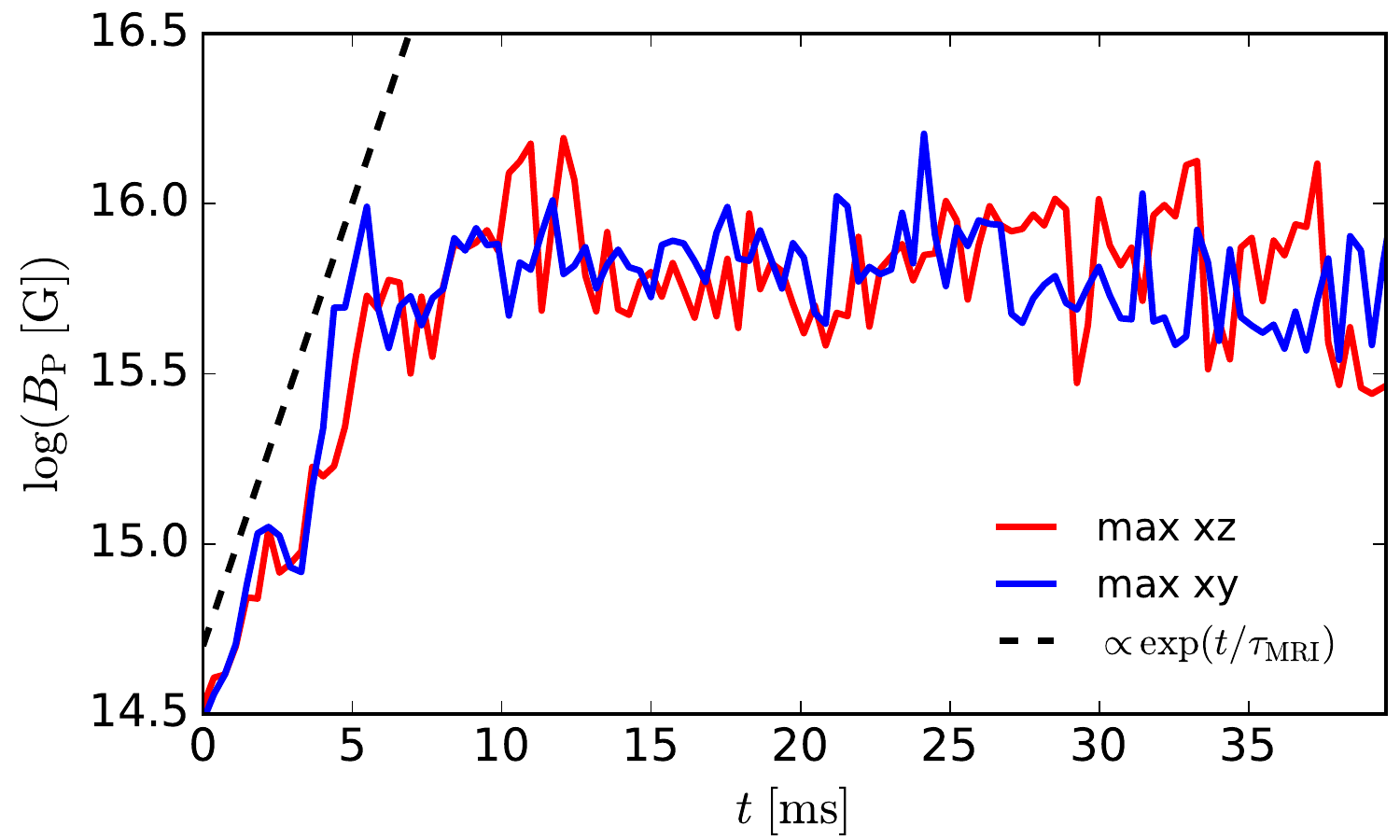}
\includegraphics[width=0.45\textwidth]{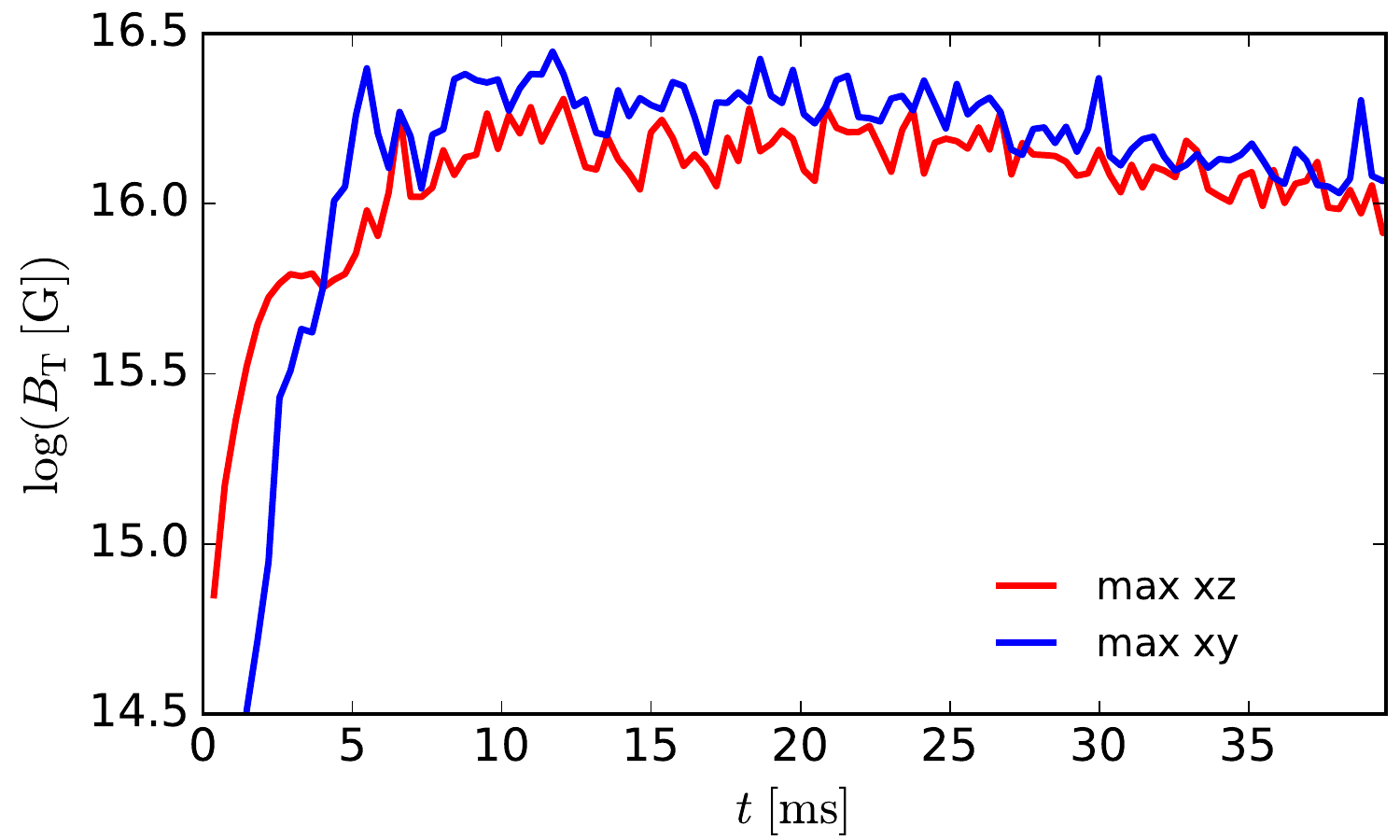}
\includegraphics[width=0.45\textwidth]{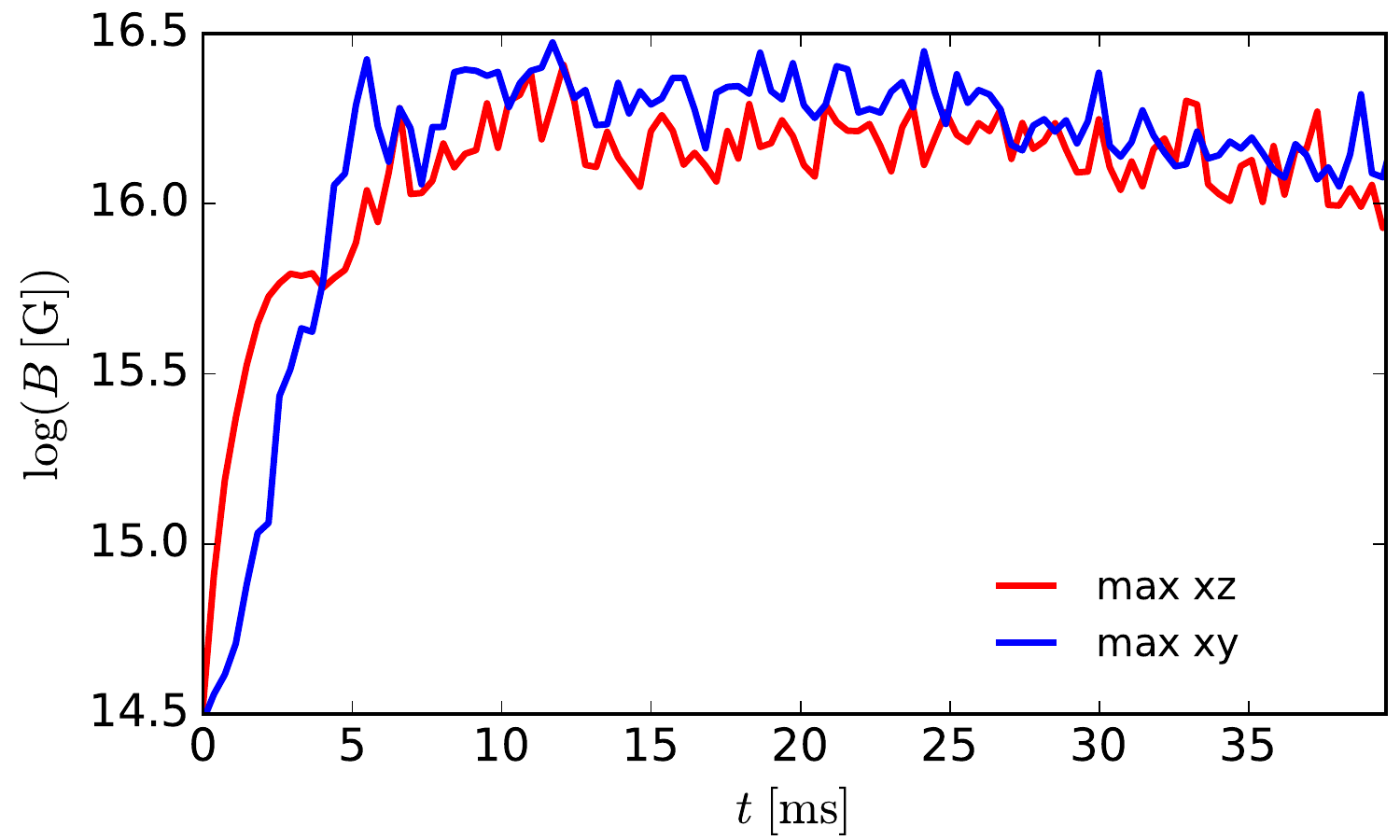}
\caption{Maximum poloidal (top), toroidal (center), and total (bottom) 
  magnetic field strength in the $xy$ and $xz$ planes during the early transient
  phase of the disk evolution. The dashed line indicates the expected
  exponential magnetic field growth due to the MRI for
  typical parameters at maximum density in the disk.}
 \label{fig:MRI_Bnorm}
\end{figure}

\begin{figure}[tb]
\centering
\includegraphics[width=0.49\textwidth]{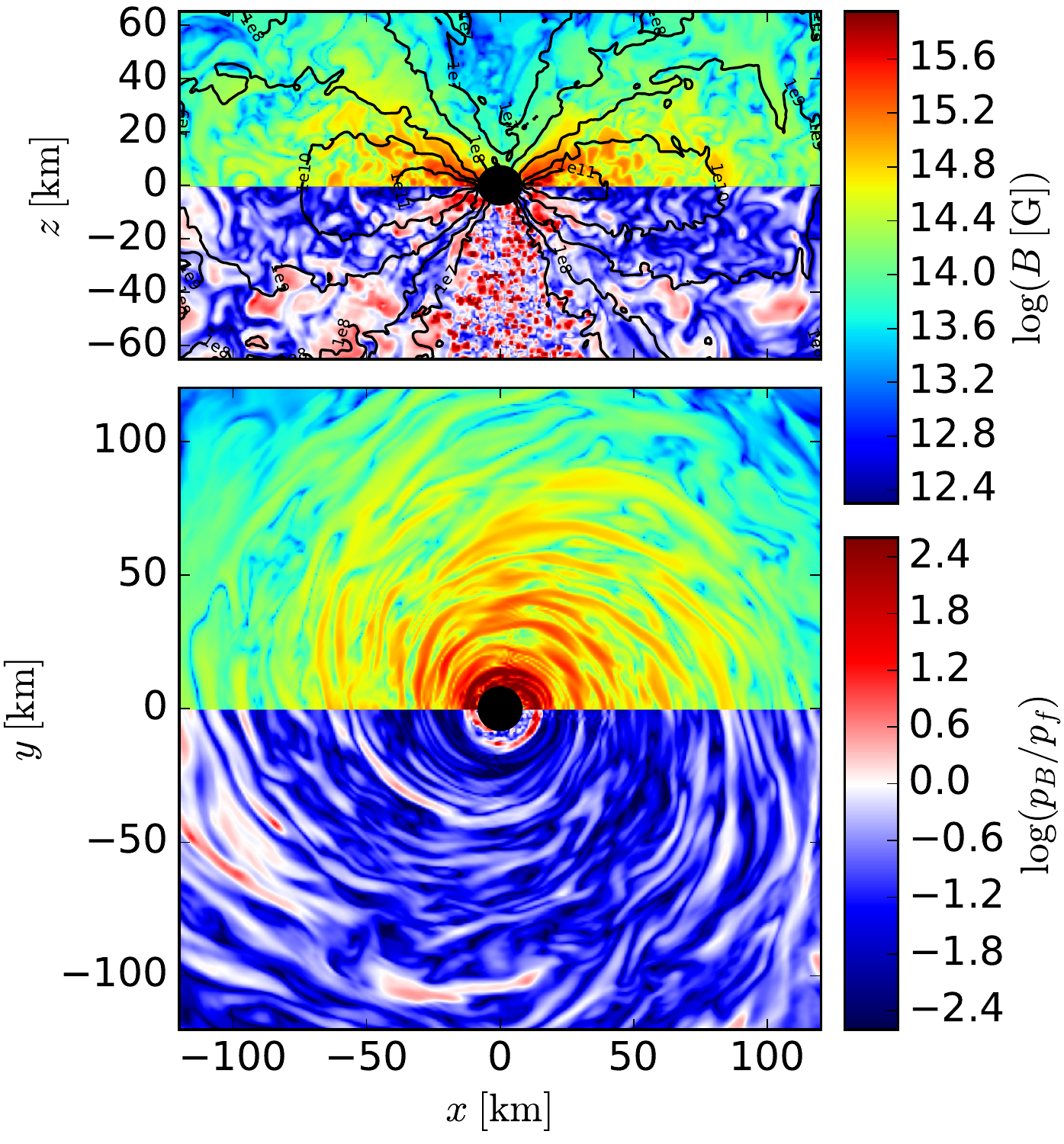}
\caption{Magnetic field strength $B$ and the magnetic-to-fluid pressure
  ratio $p_B/p_f$ in the
  meridional (top) and equatorial (bottom) plane at
  $t=20\,\mathrm{ms}$, when the disk has reached a quasi-stationary
  state. Contours refer to rest-mass density at $\rho = [10^7, 10^8, 10^9, 10^{10},
  10^{11}]\,\mathrm{g}\,\mathrm{cm}^{-3}$. }
 \label{fig:t20snapshot}
\end{figure}

Magnetic stresses generated by turbulence mediate angular
momentum transport and energy dissipation in accretion disks around
compact objects. Turbulence is thought to be generated in this context
by the MRI, which refers to certain exponentially growing modes that can
develop in differentially rotating magnetized fluids
(e.g.,
\citealt{Velikhov1959,Chandrasekhar1960,Balbus1991,Balbus&Hawley98,Balbus2003,Armitage2011}). The
MRI is a local instability, the growth of which is dominated by a
fastest-growing MRI mode; in GRMHD, its wavelength can be estimated by
\citep{Siegel2013,Kiuchi2015a,Kiuchi2017a}
\begin{equation}
  \lambda_\mathrm{MRI}\simeq\frac{2\pi}{\Omega} \frac{b}{\sqrt{4\pi\rho h
  + b^2}},
\end{equation}
where $\Omega=u^\phi/u^0$ is the angular frequency and $b\equiv \sqrt{b^2}$. The MRI is
typically well resolved when $\lambda_\mathrm{MRI}$ is numerically
resolved by at least 10 grid points and partially resolved with more
than $\sim\!5$ grid points (e.g., \citealt{Siegel2013,Kiuchi2015a}).

At $t=0\,\mathrm{ms}$, $\lambda_\mathrm{MRI}$ is only resolved by
$\sim\!5$ grid points in the high-density region of the initial torus
(cf.~Fig.~\ref{fig:MRI_start}, top panel). Within
$\approx\!1\,\mathrm{ms}$, however, by initial
relaxation and magnetic winding, the high-density part
($\sim\!10^{10}-10^{11}\,\mathrm{g}\,\mathrm{cm}^{-3}$) of the torus
rapidly enters a regime in which $\lambda_\mathrm{MRI}$ is resolved by
10 or more grid points (cf.~Fig.~\ref{fig:MRI_start}, center
panel). Indeed, starting at $\approx\!1\,\mathrm{ms}$, we witness the
onset of magnetic field amplification in
the poloidal field at the expected rate for the MRI $\propto
\exp(t/\tau_\mathrm{MRI})$, where \citep{Siegel2013}
\begin{equation}
  \tau_\mathrm{MRI} \simeq \frac{1}{\Omega}
\end{equation}
until saturation (cf.~Fig.~\ref{fig:MRI_Bnorm}, top panel); the onset
of the instability leads to a total amplification by roughly 1.5 orders
of magnitude for the maximum poloidal magnetic field strength.

As we start with a purely poloidal magnetic field configuration, the toroidal
magnetic field component first needs to be amplified by magnetic
winding in order for the grid setup to resolve the MRI in the toroidal field. For the
maximum toroidal magnetic field strength, this initial amplification
process by magnetic winding takes a few ms
(Fig.~\ref{fig:MRI_Bnorm}, center panel) and slightly longer for other parts of
the disk that start with smaller poloidal field strengths. Combined
amplification by winding and the MRI leads to an overall increase of
almost two orders of magnitude in the maximum
total magnetic field strength within the first
$\approx\!5-10\,\mathrm{ms}$ (Fig.~\ref{fig:MRI_Bnorm}, bottom panel).

By $t= 20\,\mathrm{ms}$, the disk has reached a quasi-stationary state,
in which $\lambda_\mathrm{MRI}$ is typically resolved by 10 or more grid points
(Fig.~\ref{fig:MRI_start}, bottom panel). The MRI remains resolved in this way
throughout the torus for the rest of the simulation,
although properly resolving the MRI very close to the BH is a
challenging task with current computational resources; close to the BH,
we do not resolve the MRI with $>10$ grid points at all times and
spatial points. However, we do not expect that this appreciably affects our results for the quantity and composition of the disk outflows, since these are typically generated on larger spatial scales (see, e.g., Sec.~\ref{sec:dynamo}).

The quasi-stationary state reached at $t=20\,\mathrm{ms}$ and depicted
in Fig.~\ref{fig:MRI_start} (bottom panel) and Fig.~\ref{fig:t20snapshot} is very similar to the very early state of accretion
disks obtained in recent NS--NS merger simulations. In particular, the
typical magnetic field strengths of up to $\sim\!10^{15}\,\mathrm{G}$
close to the BH and the disk midplane, as well as the typical
magnetic-to-fluid pressure ratios of $\sim\! 10^{-3}-10^{-1}$
(cf.~Fig.~\ref{fig:t20snapshot}), were also obtained by
\citet{Kiuchi2015a} and \citet{Ciolfi2017a}. This state at $t=20\,\mathrm{ms}$
serves as initial data for the rest of the simulation, and all matter
accreted onto the BH or ejected from the disk during the relaxation
phase $t<20\,\mathrm{ms}$ is discarded from all further analysis.

\subsection{Landau-level quantization}
\label{sec:Landau}

\begin{figure}[tb]
\centering
\includegraphics[width=0.49\textwidth]{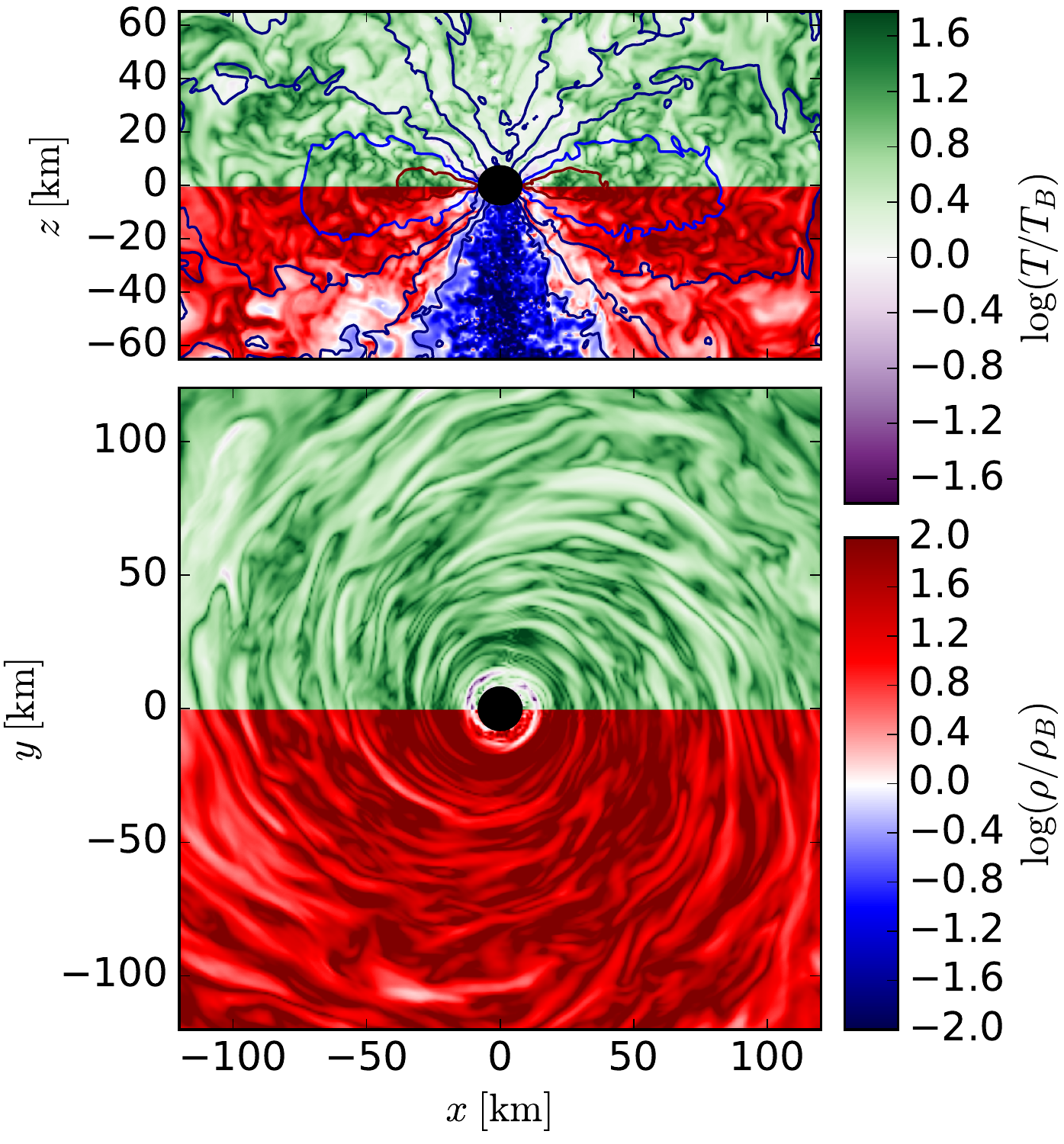}
\caption{Landau-level quantization: temperature in units of the critical temperature $T_B$ and rest-mass density in units of the critical rest-mass density $\rho_B$ (see text) in the
  meridional (top) and equatorial (bottom) plane at
  $t=20\,\mathrm{ms}$, when the disk has reached a quasi-stationary
  state. Also shown are the contours of the
  rest-mass density at $\rho = [10^7, 10^8, 10^9, 10^{10},
  10^{11}]\,\mathrm{g}\,\mathrm{cm}^{-3}$. Since either $\rho\gg \rho_B$ or $T\gtrsim T_B$, the effects of Landau-level quantization are not important.}
 \label{fig:Landau}
\end{figure}

Strong magnetic fields $\sim\!10^{15}-10^{16}\,\mathrm{G}$
(cf.~Fig.~\ref{fig:t20snapshot}) can potentially
modify the EOS and the neutrino emission and
absorption rates (Eqs.~\eqref{eq:ep_nnue}--\eqref{eq:ee_nuxnuax}
and \eqref{eq:nuen_pe}--\eqref{eq:nuap_ne}) through the
quantization of energy levels for electrons and positrons and their
motion perpendicular to the magnetic field
\citep{Lai1998,Duan2004,Duan2005}. Such effects of Landau-level
quantization may become relevant for densities below a critical density \citep{Haensel2007,Harding2006,Kiuchi2015b}
\begin{equation}
	\rho_B = 2.23\times 10^{9} \left(\frac{Y_e}{0.1}\right)^{-1}\left(\frac{B}{10^{15}\mathrm{G}}\right)^{3/2}\,\mathrm{g}\,\mathrm{cm}^{-3}
\end{equation}
and/or below a critical temperature $T_B$ \citep{Harding2006}
\begin{equation}
	T_B = \left\{\begin{array}{cc}
	\frac{m_\el c^2}{k_\mathrm{B}}\left(\sqrt{\frac{2B}{B_Q} + 1}-1\right), & \rho \le \rho_B \\
	\frac{\hbar\omega_\mathrm{c}}{k_\mathrm{B}}(1+x_F^2)^{-1/2}, & \rho \gg \rho_B
	\end{array}\right. .
\end{equation}
Here $m_\el$ is the electron mass, $c$ is the speed of light, $\omega_\mathrm{c}=eB/m_\mathrm{e}c$ is the cyclotron frequency, $x_F=\hbar (3\pi^2 Y_\el \rho)^{1/3}$ is the normalized relativistic Fermi momentum, and $B_Q = 4.414\times 10^{13}\,\mathrm{G}$ is the critical QED magnetic field strength. 

Figure~\ref{fig:Landau}
shows that, typically, $\rho \gg \rho_B$ and $T\gtrsim T_B$ in the
disk. Consequently, many Landau levels are populated, and their thermal widths are larger than the level spacing, such that the magnetic field is nonquantizing. In the polar funnel, $\rho \ll \rho_B$, but still $T\gtrsim T_B$, such that, again, the magnetic field has a nonquantizing effect. Since the disk remains in this state throughout the entire simulation, we conclude that the effects of Landau-level quantization are not important for the disk evolution.



\subsection{Disk self-regulation}
\label{sec:disk_regulation}

\begin{figure*}[tb]
\centering
\includegraphics[width=0.302\textwidth]{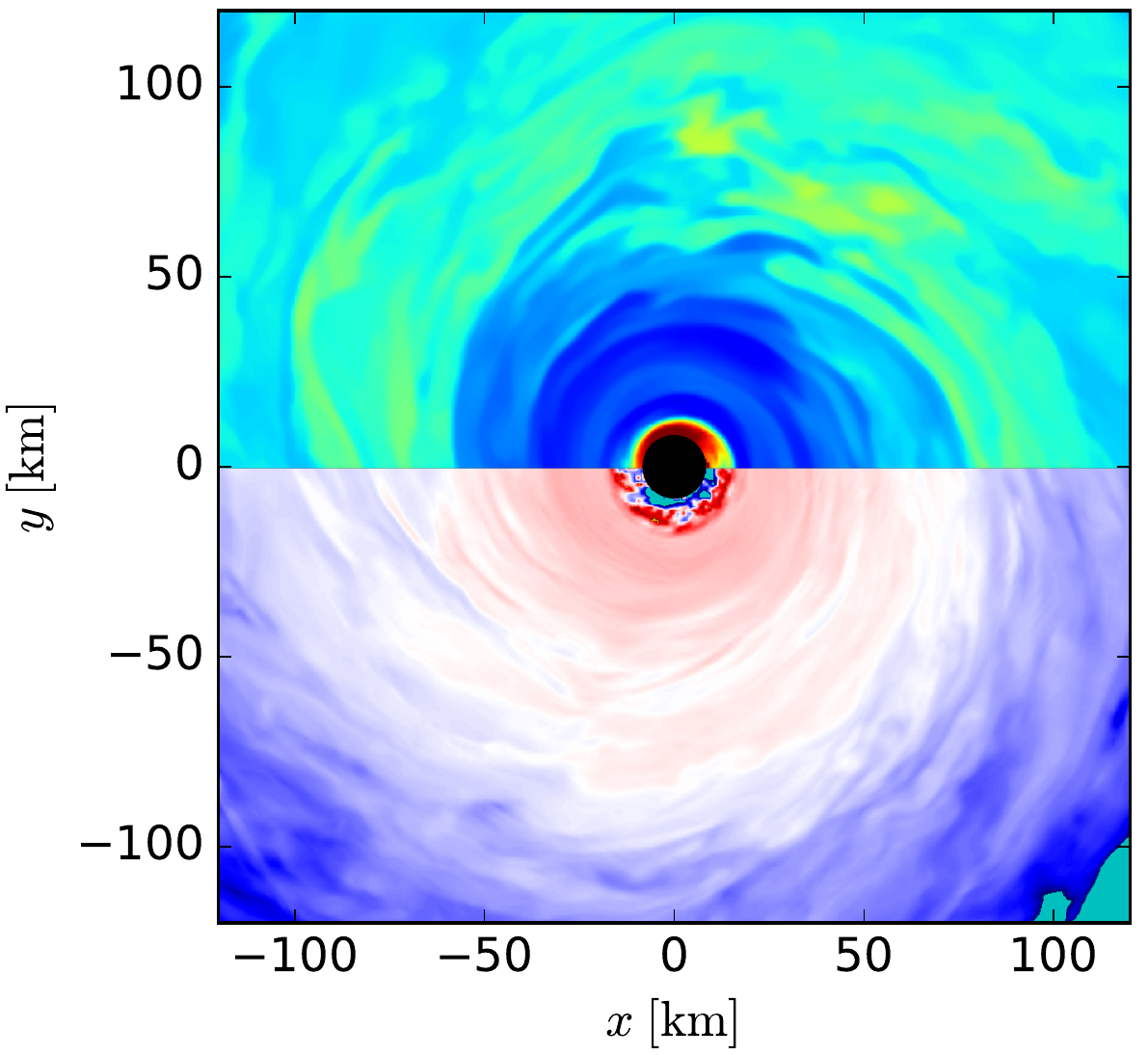}
\includegraphics[width=0.302\textwidth]{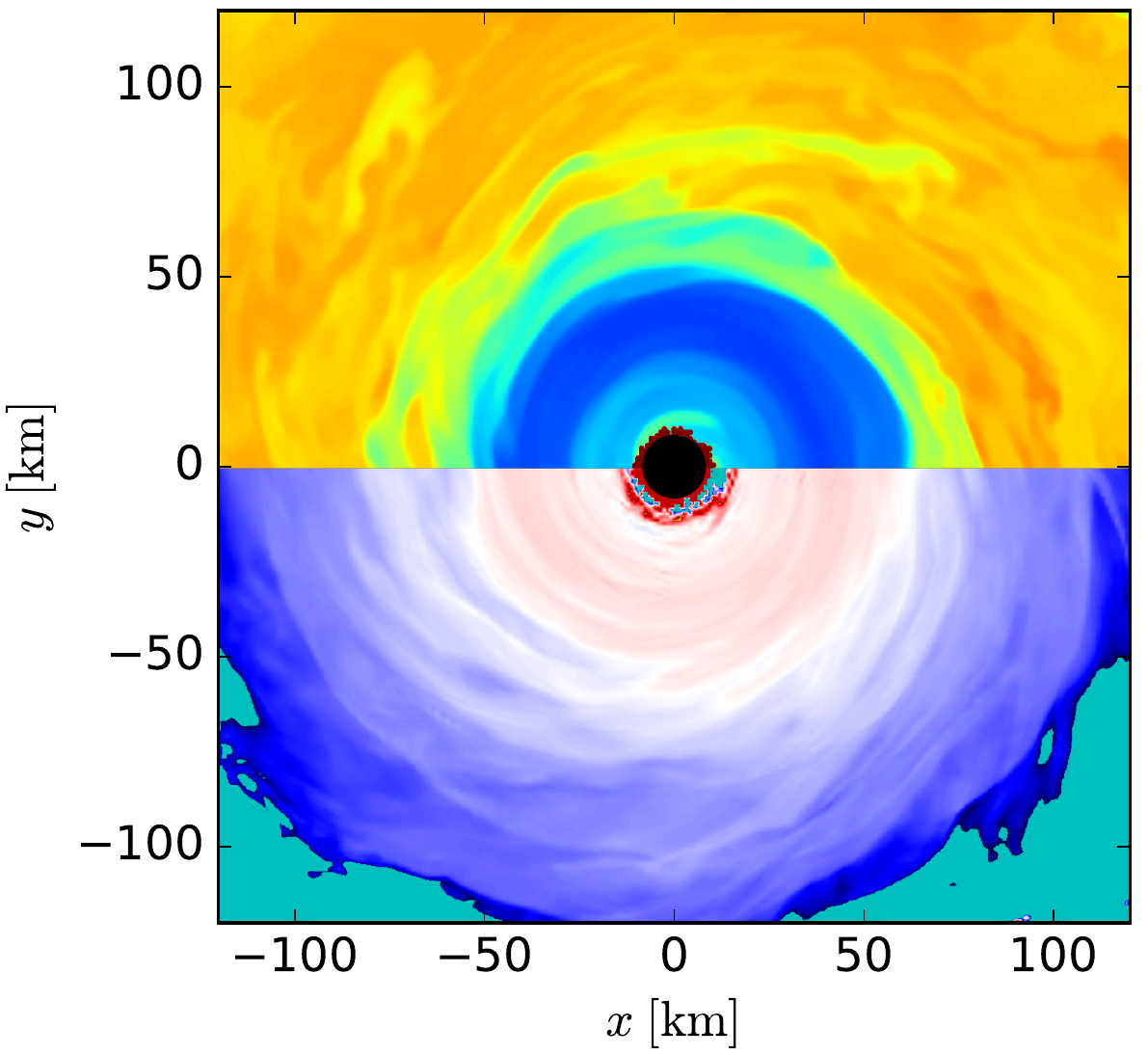}
\includegraphics[width=0.38\textwidth]{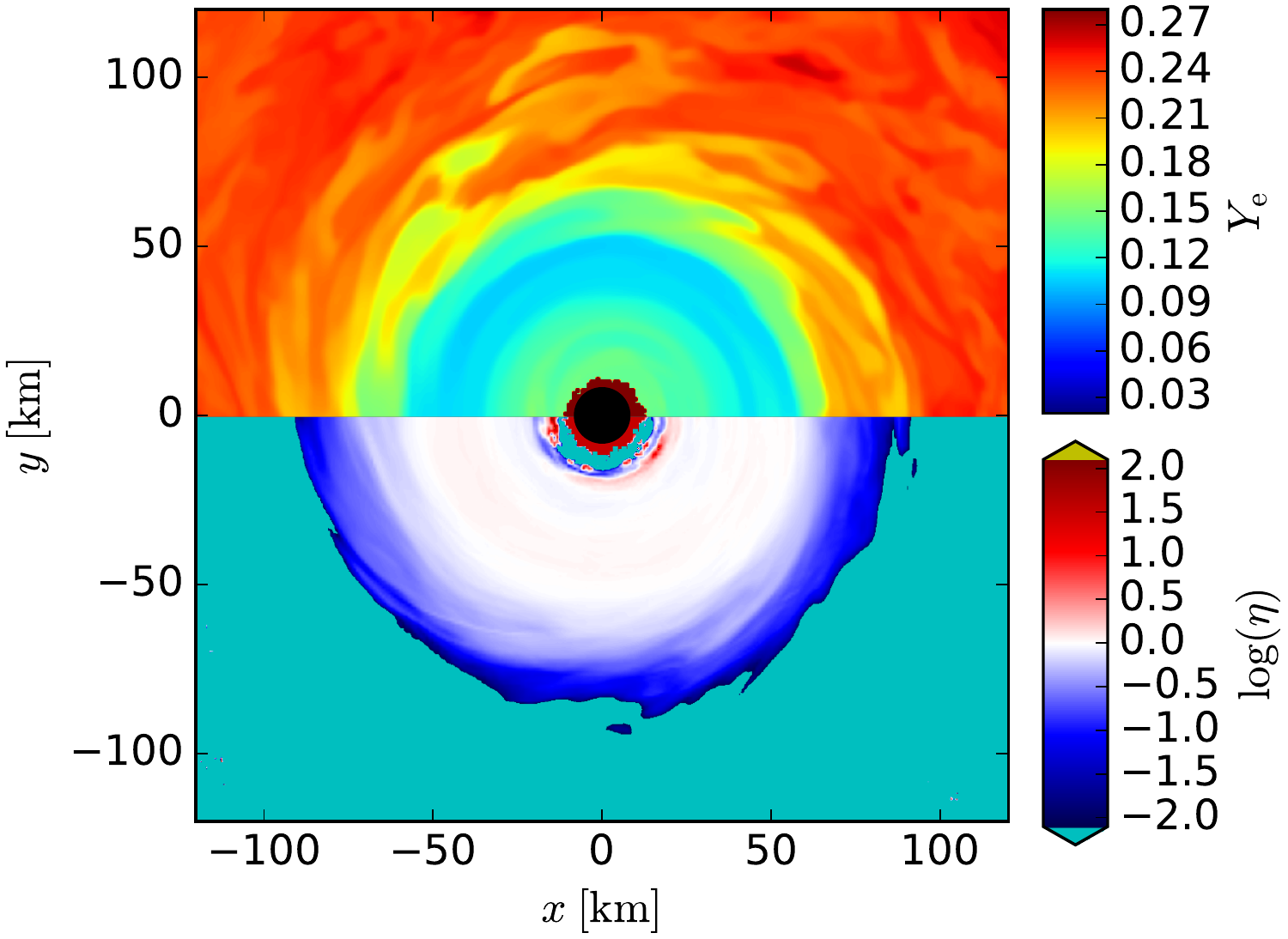}
\caption{Electron fraction $Y_\el$ and normalized electron chemical potential
  $\eta=\mu/\Theta$ at $t=43\,\mathrm{ms}$ (left), $t=130\,\mathrm{ms}$ (center), and $t=250\,\mathrm{ms}$ (right),
  showing a mildly degenerate state and low $Y_\el$ in the inner parts
  of the disk as a result of self-regulation (Sec.~\ref{sec:disk_regulation}).}
 \label{fig:degeneracy}
\end{figure*}

In the neutron-rich environment of the post-merger accretion disk, one might naively expect positron
captures onto neutrons, $e^+ + n \rightarrow p + \bar{\nu}_e$
(Eq.~\eqref{eq:en_pnua}), to be favored over electron captures
(Eq.~\eqref{eq:ep_nnue}), such that the disk matter would protonize over viscous timescales of hundreds of ms, raising the proton/electron fraction $Y_\el$ (e.g.~\citealt{Metzger2009c}). This
effect is indeed evident from Fig.~\ref{fig:degeneracy} in some portions of the disk.  However, a monotonic rise of $Y_\el$ in the disk midplane raises the question of how outflows from the disk can remain sufficiently neutron-rich to synthesize heavy $r$-process elements, even at late times in the disk evolution.  As we now describe, the reason is the existence of a self-regulation mechanism in the inner parts of the disk, which keeps a reservoir of neutron-rich material that is continuously fed into the outflows.

Once the disk has reached a quasi-stationary state (cf.~Sec.~\ref{sec:MRI} and \ref{sec:dynamo}), it regulates itself to mild electron degeneracy, which, in the presence of optically thin neutrino cooling,
results in a low $Y_\mathrm{e}$ state ($Y_\mathrm{e}\sim 0.1$).\footnote{For more massive tori than those we consider here, neutrinos can be ``trapped" in the flow (such that the neutrino diffusion timescale out of the torus exceeds the accretion timescale), and this can result in a somewhat higher midplane electron fraction than that for disks in which neutrinos are free to escape (e.g.~\citealt{DiMatteo+02,Beloborodov2003}).}  This
mechanism has been noted in the context of 1D models
of neutrino-cooled accretion disks on analytical and semi-analytical
grounds \citep{Kawanaka&Mineshige07,Chen2007,Metzger2009c}, and the
first evidence of self-regulation in a full 3D GRMHD simulation has
been presented in \citet{Siegel2017a}. Here we elaborate on these
results and discuss the mechanism in somewhat more detail; the
existence of this mechanism is important for the generation of
neutron-rich outflows from the disk (Sec.~\ref{sec:dynamo}), their
$r$-process nucleosynthesis yields (Sec.~\ref{sec:r-process}), and the
resulting thermal emission (KN).  

In the hot and dense accretion disk, the number densities of electrons
and positrons ($e^\pm$) in thermodynamic equilibrium with the baryonic
matter are given by
\begin{equation}
  n_\pm = \frac{(m_\el c)^3}{\pi^2 \hbar^3}\int_1^{\infty}
  f_\pm(E,T,\mu) E\sqrt{E^2-1}\,\mathrm{d}E, \label{eq:npm}
\end{equation}
where $E$ is the relativistic particle energy in units of $m_\el c^2$. Here $f_\pm$ is the Fermi-Dirac
function,
\begin{equation}
  f_\pm(E,T,\mu) = \frac{1}{\exp[(E\pm\mu)/\Theta]+1}, 
\end{equation}
where $\Theta=k_\mathrm{B}T/m_\el c^2$ and $\mu\equiv\mu_- = -\mu_+$
is the electron chemical potential in units of $m_\el c^2$. Charge
neutrality requires that
\begin{equation}
  n_- - n_+ = Y_\el n_\mathrm{b}, \label{eq:charge_neutrality}
\end{equation}
which, together with Eq.~\eqref{eq:npm}, determines $\mu$ and $n_\pm$
at a given thermodynamic state $(\rho,T,Y_\el)$. For degenerate
relativistic matter ($\mu/\Theta\gg 1$), using the Sommerfeld
expansion of Eq.~\eqref{eq:npm} in terms of $\mu/\Theta$, one can show that the
temperature dependence of $\mu$ is
approximately given by (see Appendix~\ref{app:temp_dependence})
\begin{equation}
  \sqrt{\mu^2-1} = \sqrt{E_\mathrm{F}^2-1} \left(1 - \frac{\pi^2}{6}
    \frac{\Theta^2}{E_\mathrm{F}^2-1} \right),  \label{eq:mu_T}
\end{equation}
where $E_\mathrm{F}\equiv \mu(T=0)$ is the Fermi energy. Furthermore,
for degenerate matter, free $e^\pm$ pairs can only be obtained from around the Fermi edge $E\simeq \mu$ with width $\Delta E\simeq 4\
\Theta$, which is very narrow ($\Delta E/
E \simeq 4\Theta/\mu \ll 1$); from Eq.~\eqref{eq:npm}, one finds that for $\mu/\Theta\gg
1, E\simeq \mu$,
\begin{equation}
\frac{n_+}{n_-} \propto \exp(-2\mu/\Theta), \label{eq:frac_npm}
\end{equation}
i.e., $e^\pm$ creation is heavily suppressed. Higher electron
degeneracy $\eta\equiv \mu/\Theta$ results in less electrons and
positrons (cf.~Eqs.~\eqref{eq:npm} and \eqref{eq:frac_npm}). This
decreases the neutrino emission via charged-current interactions and
pair annihilation
(cf.~Eqs.~\eqref{eq:ep_nnue}--\eqref{eq:ee_nuxnuax}); i.e., it results
in a lower cooling rate and higher temperatures. Higher temperatures,
in turn, decrease $\mu$ (cf.~Eq.~\eqref{eq:mu_T}) and thus increase the
degeneracy, i.e., $\eta$. Because of this negative feedback loop, whenever
the disk enters the (strongly) degenerate regime, it will tend to self-regulate
its degeneracy and maintain a state of mild electron degeneracy $\eta\sim
1$. Indeed, as shown by Fig.~\ref{fig:degeneracy}, soon after reaching
the quasi-stationary state, the disk has regulated itself to mild
degeneracy $\eta\sim
1$ in the inner parts of the disk in which neutrino cooling is
energetically important ($r\lesssim60\,\mathrm{km}$ or $r\lesssim 14$
gravitational radii) and qualitatively remains in this state throughout the
remainder of the simulation.

In the hot and dense matter of the inner parts of the disk, electron and
positron capture (cf.~Eqs. \eqref{eq:ep_nnue} and \eqref{eq:en_pnua}) are
the dominant cooling
reactions. The equilibrium $Y_\el$ that results from conditions of mild degeneracy
in this neutrino-transparent matter is then determined by equal rates of
$e^\pm$ capture,
\begin{equation}
  \dot{n}_{e^-p} = \dot{n}_{e^+n}; \label{eq:enep}
\end{equation}
i.e., Eqs.~\eqref{eq:npm}, \eqref{eq:charge_neutrality}, and
\eqref{eq:enep} determine $Y_\el$ for a given $\rho$ and $T$. For mild
degeneracy $\eta \gtrsim 1$, one can show that from Eq.~\eqref{eq:enep},
the equilibrium $Y_\el$ is approximately given by
\citep{Beloborodov2003}
\begin{eqnarray}
  Y_\el &=& 0.5 + \frac{7\pi^4}{1350\zeta(5)} \left(\frac{Q}{2\Theta} -
    \eta\right) \\
  &=& 0.5 + 0.487\left(\frac{1.2655}{\Theta} - \eta \right),
\end{eqnarray}
where $\zeta$ is the Riemann $\zeta$-function and $Q = (m_\mathrm{n} -
m_\mathrm{p})/m_\el=2.531$ is the neutron--proton mass difference in units
of the electron mass. A very mild electron degeneracy $\eta \gtrapprox 1$ in
hot matter $\Theta \approx 1$ is therefore sufficient to generate
conditions of neutron richness $Y_\el < 0.5$. For the hot
$\Theta\gtrsim 1$ and mildly degenerate conditions $\eta\gtrsim 1$ of
the inner parts of the disk, the resulting neutron richness adjusts to
an equilibrium value of typically $Y_\el \sim 0.1$ or lower (see
Fig.~\ref{fig:degeneracy}). 

The presence of this self-regulation
mechanism to mild electron degeneracy, which implies a low $Y_\el\sim 0.1$,
is important to allow for the generation of neutron-rich outflows that
can undergo $r$-process nucleosynthesis (Secs.~\ref{sec:dynamo} and
\ref{sec:r-process}). It forces the disk to keep a reservoir of
neutron-rich material despite the ongoing protonization process in the
rest of the disk---neutron-rich material that is continuously fed into the outflows to
keep the overall mean electron fraction $\bar{Y}_\el$ of the outflow rather low over the
lifetime of the disk ($\bar{Y}_\el\sim 0.2$, see
Tab.~II of \citealt{Siegel2017a} and Sec.~\ref{sec:neutrino_absorption}). This results in
the possibility of generating a robust second-to-third-peak $r$-process
(cf.~Sec.~\ref{sec:r-process}) and thus the production of a
significant amount of lanthanide material in the outflow. Due to its
high opacity, this material can then produce a red KN, as
observed in the recent GW170817 event.

\subsection{Magnetic dynamo, disk corona, and generation of outflows}
\label{sec:dynamo}

\begin{figure}[tb]
\centering
\includegraphics[width=0.48\textwidth]{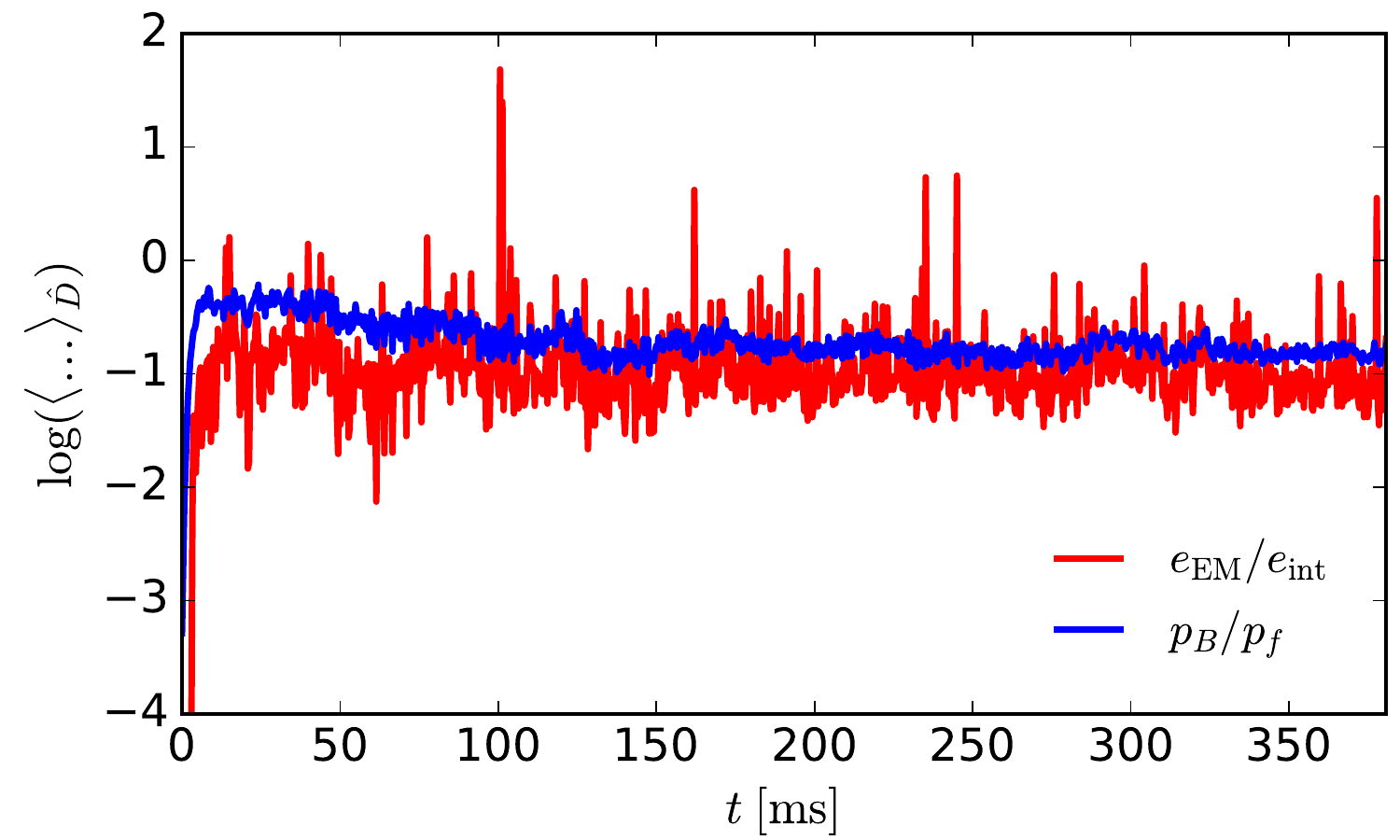}
\caption{Evolution of the density-averaged ratio of the electromagnetic to internal
  energy (red) and of the magnetic-to-fluid pressure ratio (blue), indicating a steady turbulent state of the disk.}
 \label{fig:turbulent_state}
\end{figure}

\begin{figure*}[!tb]
\centering
\includegraphics[width=0.995\textwidth]{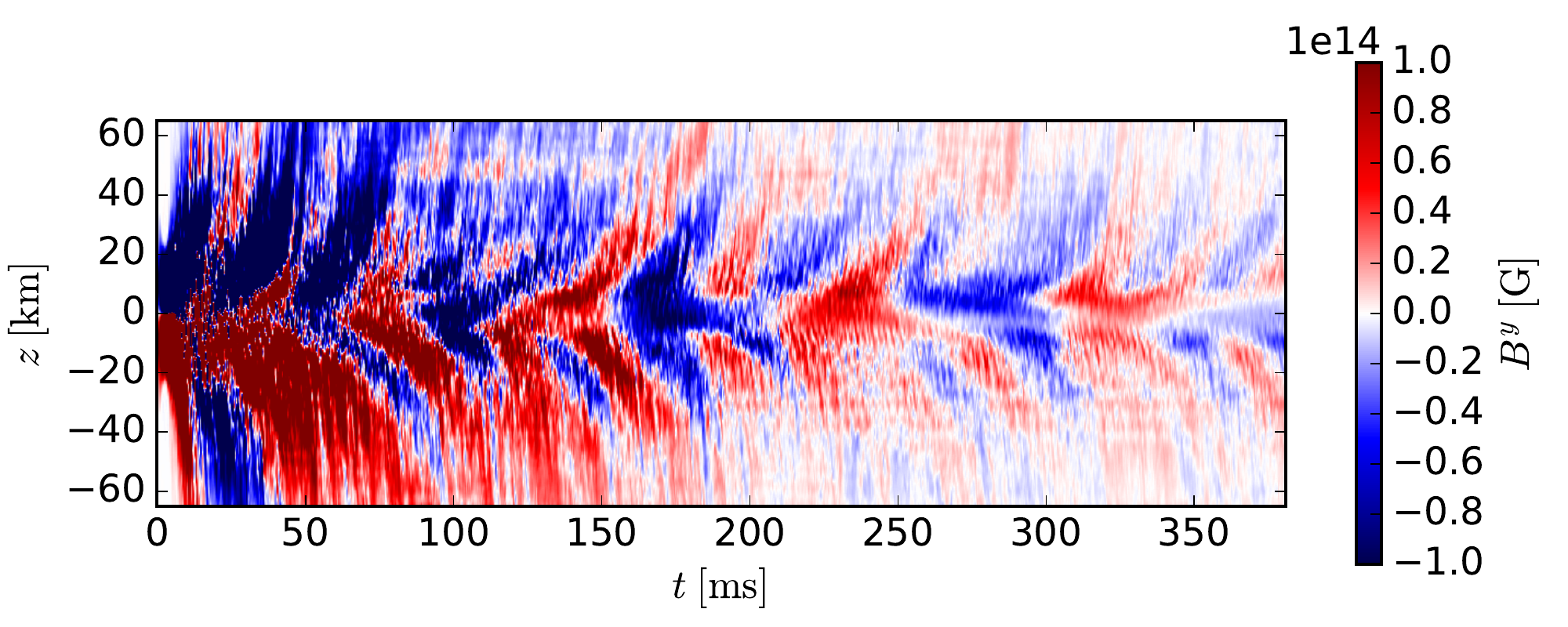}\\
\includegraphics[width=0.995\textwidth]{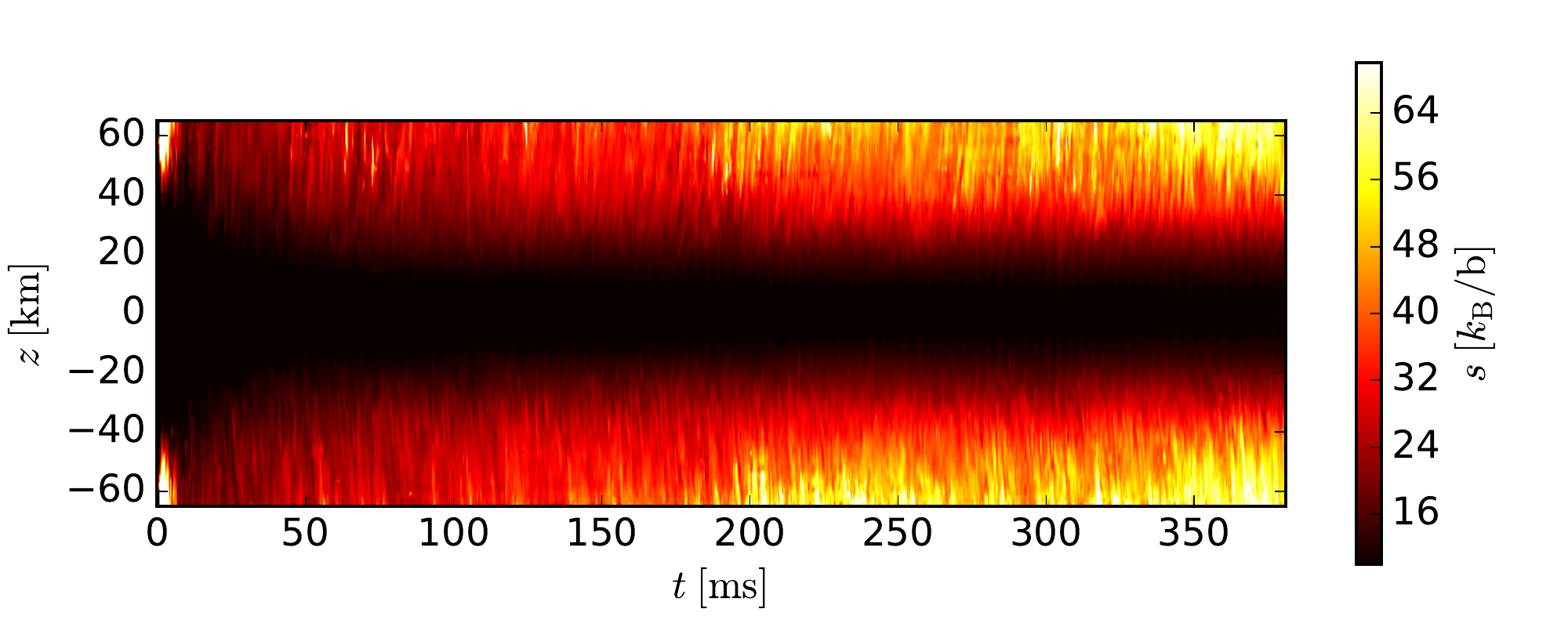}\\
\includegraphics[width=0.995\textwidth]{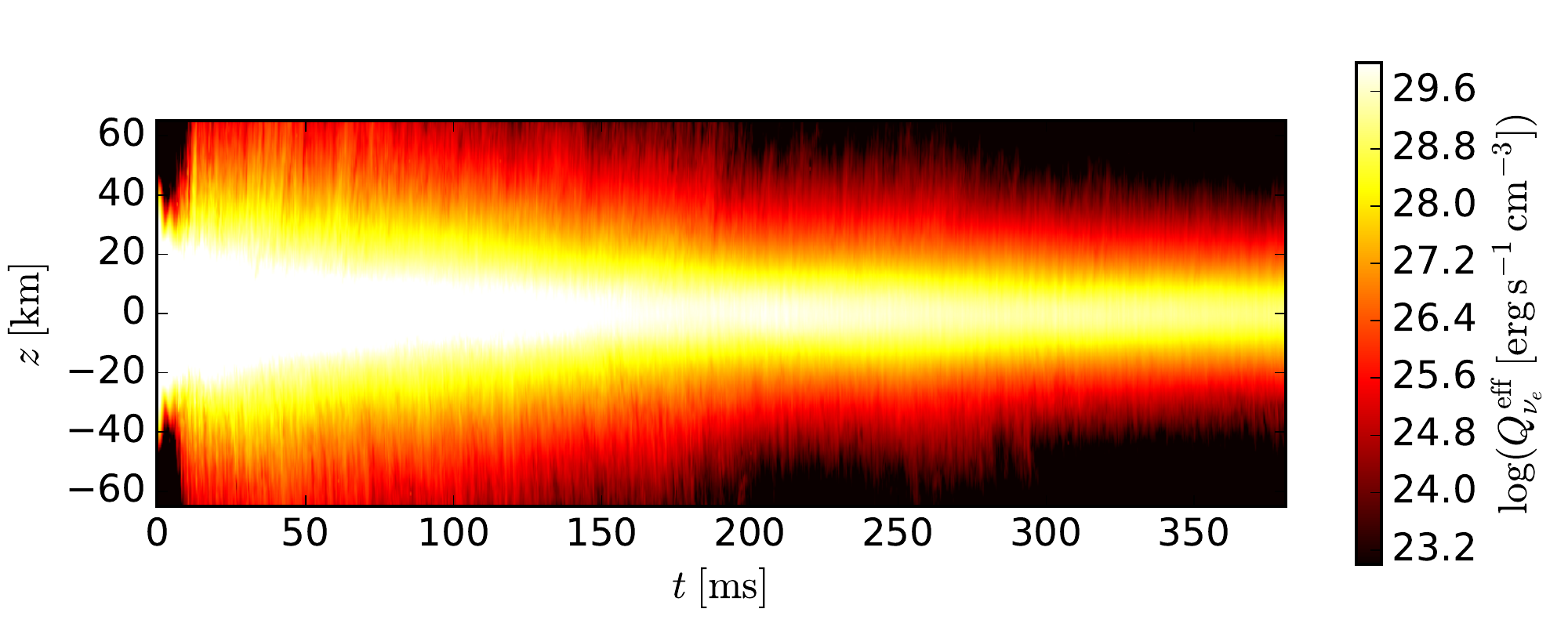}
\caption{Spacetime diagrams of the $y$-component of the magnetic field
  (top), the specific entropy (center), and the effective electron neutrino
  energy emission rate per volume (bottom; representative of neutrino
  cooling), radially averaged between
  30 and $70\,\mathrm{km}$ from the rotation axis in
  the $x$--$z$ (meridional) plane as a
  function of height $z$ relative to the equatorial plane.}
 \label{fig:spacetime}
\end{figure*}

\begin{figure}[tb]
\centering
\includegraphics[width=0.48\textwidth]{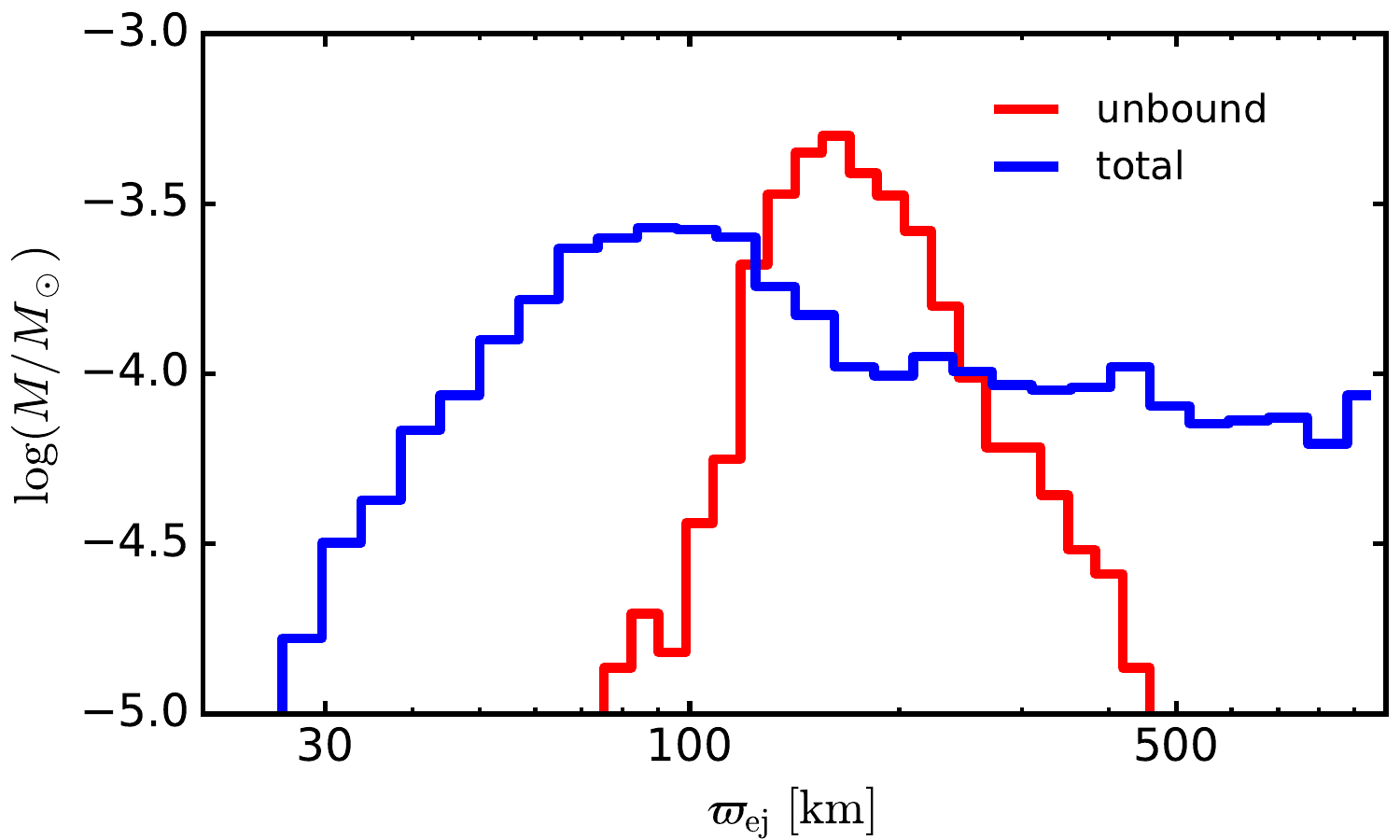}
\includegraphics[width=0.48\textwidth]{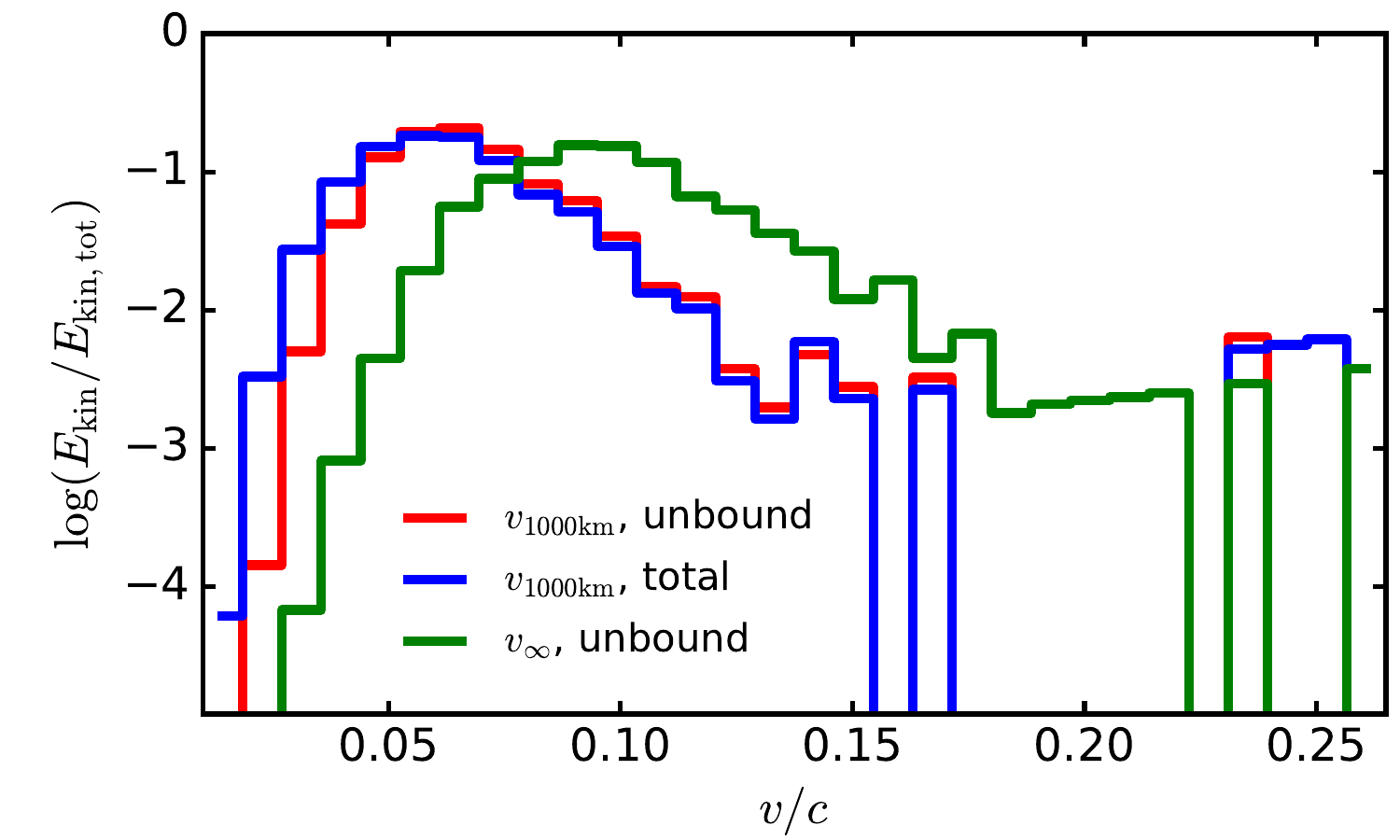}
\caption{Top: mass distributions of the unbound and total disk outflow as
  measured by tracer particles in terms of their cylindrical radius
  $\varpi_\mathrm{ej}$ at the time of ejection from the disk (corona). Bottom:
  distribution of kinetic energy (in units of the respective total kinetic energy) of the unbound and total disk outflow in terms of the outflow velocity $v_{1000\mathrm{km}}$ measured at $r=10^3\,\mathrm{km}$ from the BH and of the unbound outflow in terms of the corresponding asymptotic escape velocity $v_\infty$ (see text).}
 \label{fig:outflow_radii}
\end{figure}

\begin{figure}[tb]
\centering
\includegraphics[width=0.48\textwidth]{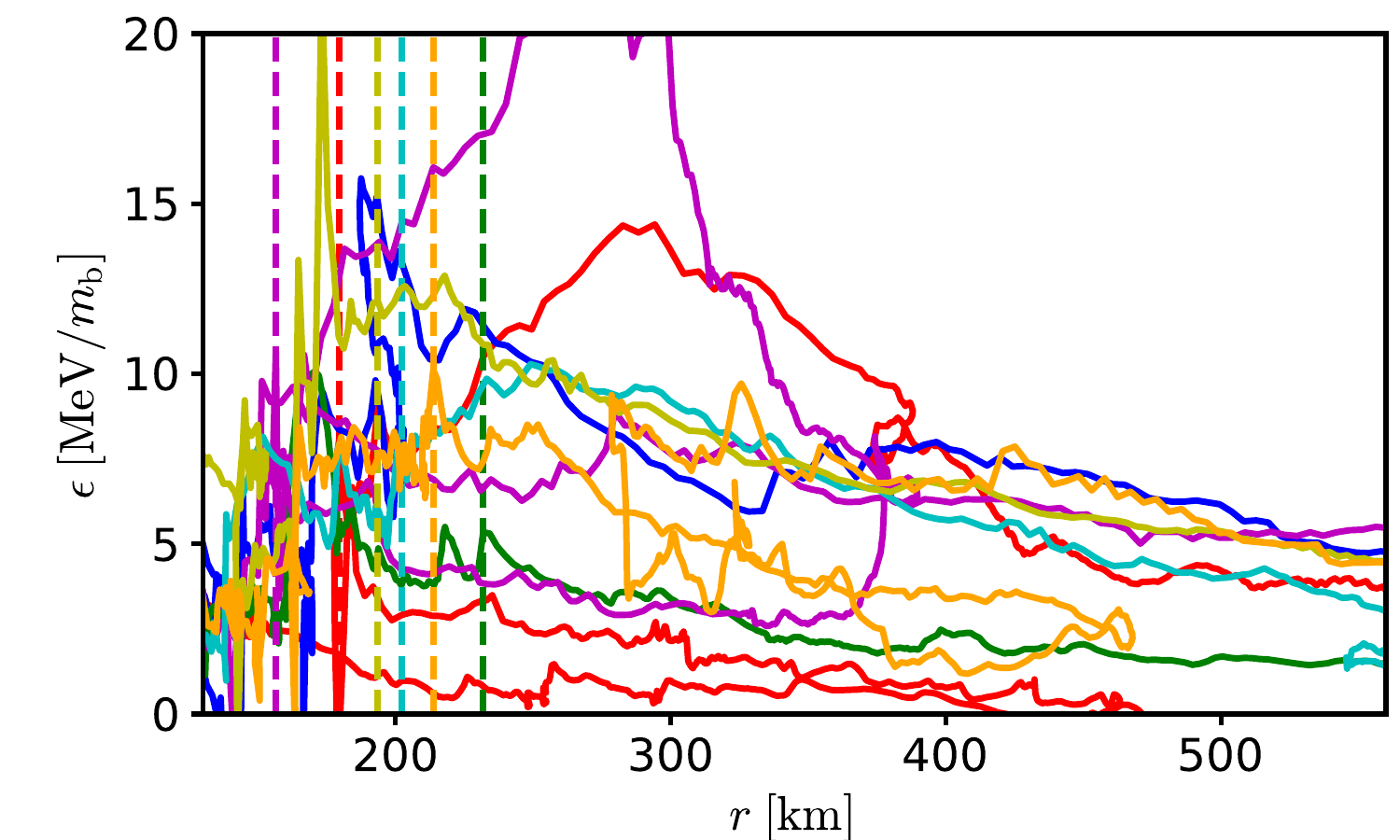}
\includegraphics[width=0.48\textwidth]{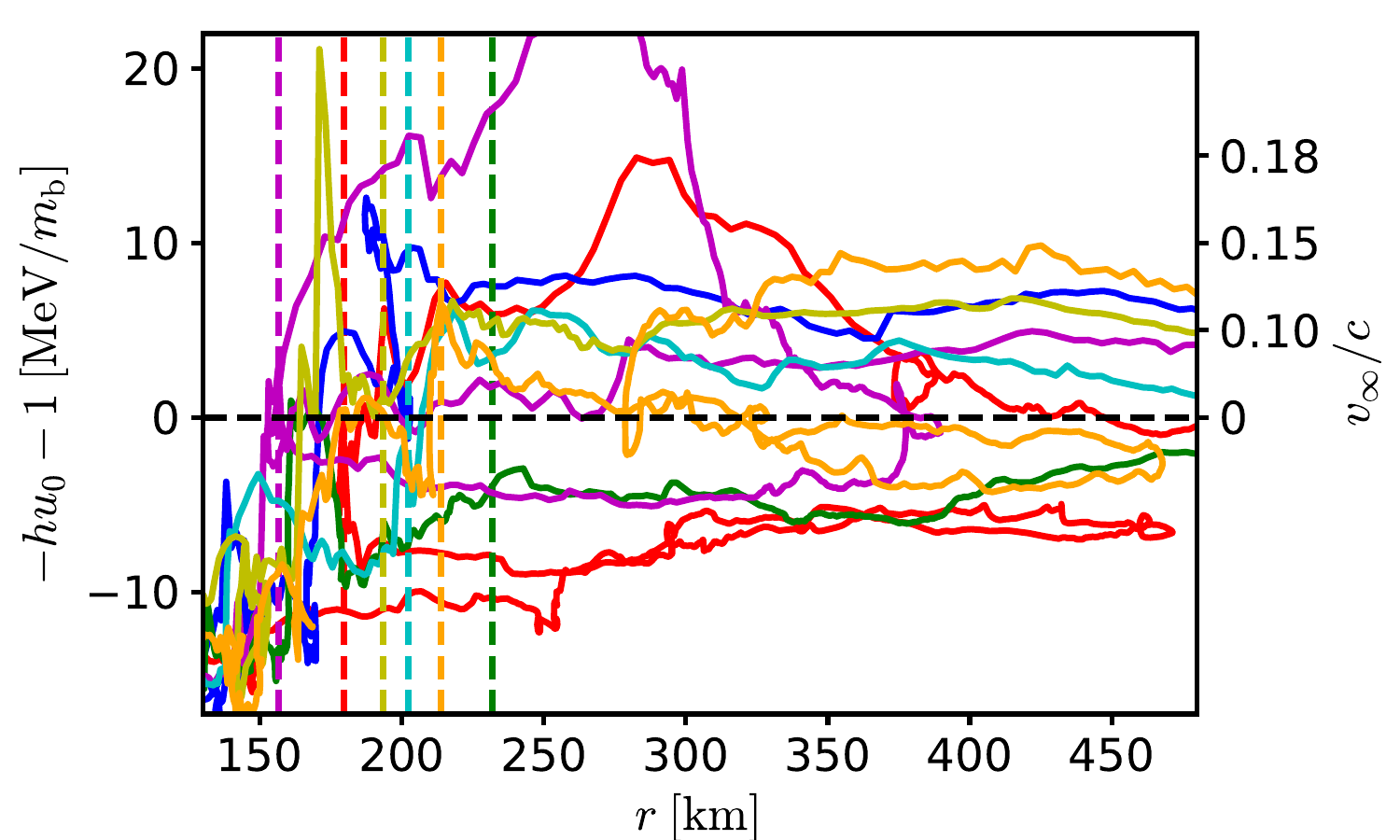}
\caption{Representative tracer particles: specific internal energy (top) and Bernoulli criterion for unboundedness and corresponding asymptotic escape velocity (bottom) as a function of radial coordinate distance from the BH. Vertical dashed lines mark the corresponding radii at which $50\%$ of the total $\alpha$-particle production along the trajectory has been accomplished, i.e., the last time where the $\alpha$-particle mass fraction $X_\alpha=0.5X_{\alpha,\mathrm{max}}$, where $X_{\alpha,\mathrm{max}}=2Y_{\el,\mathrm{max}}$, with $Y_{\el,\mathrm{max}}$ being the maximum electron fraction along the particle trajectory.}
 \label{fig:bernoulli}
\end{figure}

Magnetic stresses generated by MHD turbulence via the MRI mediate
angular momentum transport and thus energy dissipation in the
disk. Turbulence also dissipates magnetic energy, which, however, is
regenerated through a dynamo (e.g.,
\citealt{Parker1955,Brandenburg1995}). The balance of the two processes
results in a saturated steady turbulent, quasi-equilibrium
state, which is characterized by a roughly constant ratio of magnetic
to internal energy in the disk. 

Figure~\ref{fig:turbulent_state} shows the
temporal evolution of the density-averaged ratio of electromagnetic to
internal energy $\langle e_\mathrm{EM}/e_\mathrm{int}\rangle_{\hat D}$
and of the magnetic-to-fluid pressure ratio $\langle
p_B/p_\mathrm{f}\rangle_{\hat D}$, which are
indeed indicative of a disk in a steady turbulent state. We define
the rest-mass density average of a quantity $\chi$ by
\begin{equation}
  \langle \chi \rangle_{\hat D} \equiv \frac{\int \chi \hat
    D\mathrm{d}^3x}{\int \hat D\mathrm{d}^3x}, \label{eq:density_average}
\end{equation}
where $\hat D = \sqrt{\gamma}\rho W$ is the conserved rest-mass
density (cf.~Eqs.~\eqref{eq:GRMHDeqns}--\eqref{eq:D}).\footnote{Here
  and in the following, spatial integrals refer to the entire
  simulation domain, excluding the interior of the BH horizon.} Following
\citet{Duez2006b}, we define the total internal energy
\begin{equation}
  E_\mathrm{int} \equiv \int \epsilon \rho W \sqrt{\gamma}\mathrm{d}^3x
\end{equation}
and the total electromagnetic energy
\begin{equation}
  E_\mathrm{EM} \equiv \int n_\mu n_\nu T^{\mu\nu}_\mathrm{EM} \sqrt{\gamma}\mathrm{d}^3x,
\end{equation}
where $T^{\mu\nu}_\mathrm{EM}$ is the electromagnetic part of the
energy--momentum tensor. We thus define the local ratio of
electromagnetic to internal energy by
\begin{equation}
  \frac{e_\mathrm{EM}}{e_\mathrm{int}} \equiv \frac{n_\mu n_\nu T^{\mu\nu}_\mathrm{EM}}{\epsilon \rho W}.
\end{equation}
Figure~\ref{fig:turbulent_state} shows that for $t>20\,\mathrm{ms}$, this ratio
remains roughly constant in a time-averaged sense and thus indicates
that a steady turbulent state of the disk is indeed achieved and
maintained. Furthermore, Fig.~\ref{fig:turbulent_state} shows that
\begin{equation}
  \left\langle\frac{p_B}{p_\mathrm{f}}\right\rangle_{\hat D} \simeq
  0.1,
\end{equation}
which is also characteristic of such a steady turbulent state
(e.g., \citealt{Jiang2014a,Sadowski2015}). This ratio in the nonlinear saturated state is much larger than the initial value of
$p_B/p_\mathrm{f}< 5\times 10^{-3}$
(cf.~Sec.~\ref{sec:initial_data} and Tab.~\ref{tab:BH_torus}).

The 3D nature of our disk simulation is crucial for generating a
steady turbulent state. Due to the antidynamo theorem
\citep{Cowling1933}, magnetic fields cannot be regenerated by dynamo
action in axisymmetry, and a steady turbulent state cannot thus be
maintained. 

Direct evidence for dynamo action in our disk simulation
is depicted in the top panel of Fig.~\ref{fig:spacetime}, which shows
a spacetime diagram of the radially averaged $y$-component of the
magnetic field in the $x$-$z$ plane. This ``butterfly'' diagram clearly
indicates the presence of magnetic cycles with a period of roughly
$\sim\!20\,\mathrm{ms}$ throughout the entire simulation time
domain. In the disk midplane, magnetic fields of temporally alternating polarity
are generated by MHD turbulence. These fields slowly migrate off the
midplane by magnetic pressure gradients and buoyancy, where they are
gradually dissipated into heat. This migration and dissipation of
magnetic energy contributes to establishing a ``hot'' corona above and below the
midplane, as indicated by the middle panel of
Fig.~\ref{fig:spacetime}. This spacetime diagram of the
specific entropy shows strongly increasing specific entropies off the midplane
where magnetic field strengths decrease. We note that the temperature, however, decreases as a function of height off the midplane. Therefore, the production of high-energy nonthermal neutrinos in the corona by upscattering of thermal neutrinos emitted from the midplane (cf.~bottom panel of Fig.~\ref{fig:spacetime}) is not expected.\footnote{Furthermore, the production of high-energy nonthermal neutrinos by electron--positron pair annihilation in the corona is also not expected, as thermalization processes (e.g., Coulomb scattering) are extremely rapid, which would suppress any nonthermal electron tail above the mean temperature.}

In the hot corona, powerful outflows are generated. In these regions of
lower density, viscous heating from MHD turbulence and dissipation of
magnetic energy exceeds cooling by neutrino emission, which is strongest in the disk midplane
(cf.~Fig.~\ref{fig:spacetime}, bottom panel). This heating-cooling
imbalance results in launching neutron-rich winds from the disk. Above
and below the midplane, the neutrino emissivities decrease as functions of ``height'' $|z|$, and the weak
interactions (and thus $Y_\el$) essentially ``freeze out''; however,
further mixing in the (initially turbulent) outflows can still change
$Y_\el$.

The outflows are tracked by $10^4$ passive tracer particles that are
advected with the plasma. These tracer particles are of equal mass, placed within the initial torus at $t=0\,\mathrm{ms}$ with a probability
proportional to the conserved rest-mass density $\hat
D=\sqrt{\gamma}\rho W$. We distinguish between total outflow, defined
as the entity of all tracer particles that have reached a radial coordinate
distance of $10^3\,\mathrm{km}$ from the center of the BH by the end
of the simulation, and unbound outflow, or ejecta, defined as the
entity of tracer particles that are additionally unbound according to
the Bernoulli criterion $-h u_0 > 1$ (nonvanishing escape velocity at infinity).

Outflows are generated over a wide range of radii. This is illustrated
by the top panel of Fig.~\ref{fig:outflow_radii}, which shows mass histograms of the
outflow tracer particles in terms of their cylindrical coordinate radii
$\varpi=\sqrt{x^2+y^2}$ at the time of ejection from the disk,
$\varpi_\mathrm{ej}\equiv \varpi(t=t_\mathrm{ej})$. We define the
time of ejection from the disk or corona $t=t_\mathrm{ej}$ as the time after
which the radial coordinate position of a tracer particle $r =
\sqrt{x^2+y^2+z^2}$ only increases monotonically with time. The total
outflow shows a broad distribution with significant mass being ejected
between $\varpi_\mathrm{ej}\approx 20\,\mathrm{km}$ and
$\varpi_\mathrm{ej}>600\,\mathrm{km}$ from the BH. However, we find
that mass ejection is most efficient in a narrower range of
ejection radii, as indicated by the histogram of unbound matter, the
latter being ejected essentially in the range $\varpi_\mathrm{ej}\approx
100-400\,\mathrm{km}$ from the BH.

Matter is typically unbound by recombination into $\alpha$-particles. The imbalance of heating and cooling in the hot corona, as mentioned above, lifts material in the BH potential but typically only leads to marginally bound or marginally unbound outflows. Subsequent nuclear binding energy release from recombination of free nucleons into $\alpha$-particles rapidly generates specific enthalpy as matter approaches the recombination temperature and immediately ``unbinds'' the material; this is shown in Fig.~\ref{fig:bernoulli} for a few representative tracer particles. A spike in the specific enthalpy $h$ is created by internal energy that becomes available during the recombination process ($7\,\mathrm{MeV}$ per baryon per $\alpha$-particle produced) plus the resulting pressure increase in a low-density environment. For a stationary relativistic fluid flow (isentropic, constant specific angular momentum), $h u_0$ is constant along a fluid world line (Eq.~\eqref{eq:Euler}). As the material moves away from the disk, the outflows cool ($h\rightarrow 1$) and specific enthalpy is converted into kinetic energy keeping $h u_0$ constant, which sets the asymptotic escape velocity.

The bottom panel of Fig.~\ref{fig:outflow_radii} shows the distribution
of kinetic energy of the unbound and total outflows in terms of their
outflow velocities. We characterize the outflow by two velocities: $v_{1000\mathrm{km}}$, the velocity at a coordinate distance $r=10^3\,\mathrm{km}$ from the BH, and $v_\infty$, the corresponding asymptotic escape velocity when the conversion of internal energy to kinetic energy has been completed. Here $v_\infty$ is computed from the corresponding asymptotic Lorentz factor $W_\infty\equiv -hu_0$, where $hu_0$ is evaluated either when the tracer particle leaves the computational domain or at the final time of the simulation if it stays inside the computational domain for the entire simulation time. Unbound and total outflows have similar velocity distributions in the range
$v_{1000\mathrm{km}}\approx (0.03-0.15)c$. The kinetic energy-weighted mean outflow velocities $\bar
v_{1000\mathrm{km}}\equiv \sqrt{2 E_\mathrm{kin,tot}/M_\mathrm{ej}}$ are
$0.063c$ and $0.058$ for unbound and total outflow,
respectively. Here $E_\mathrm{kin,tot}$ denotes the total kinetic
energy in the outflow type, and $M_\mathrm{ej}$ is the total mass of the
outflow type. The asymptotic kinetic energy distribution of the unbound outflow, however, shows $v_\infty\approx (0.04-0.25)c$, with a higher kinetic energy-weighted mean of $\bar v_\infty = 0.094c\approx 0.1c$. 
 
Though not included in our simulations, the outflows will receive additional nuclear heating from the $r$-process on larger radial scales of $\approx 2-3$ MeV per nucleon \citep{Metzger+10b}, which will boost its speed by an additional $\approx 10-20\%$.  We note that $\bar v_\infty$ of the unbound outflow corresponds to the kinetic energy-averaged value $v_\mathrm{KN}\approx 0.1c$, similar to that required to explain the red KN component observed in the recent GW170817 event (e.g.~\citealt{Chornock2017,Villar2017}).

The total unbound mass from the disk at the end of the simulation amounts to $\approx\!20\%$ of its initial value.  However, the true total ejecta mass, including late times after the simulation has terminated, is likely to be roughly twice as great, as estimated in greater detail in the following subsection.  Additional properties of the outflow are summarized in \citet{Siegel2017a}.

\subsection{Global disk structure and long-term evolution}
\label{sec:global_structure}

\begin{figure}[tb]
\centering
\includegraphics[width=0.49\textwidth]{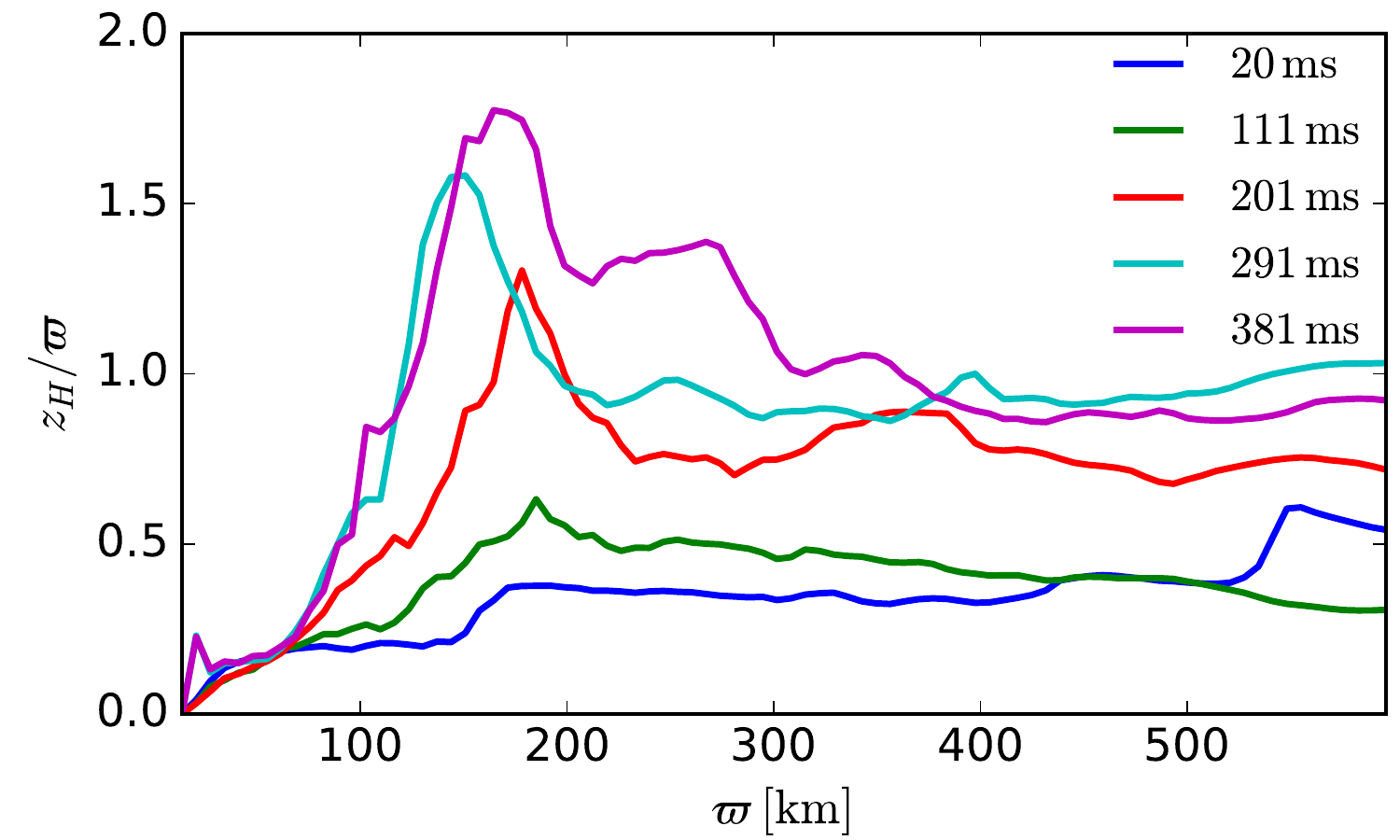}
\caption{Density scale height of the disk at
  different times during the evolution.}
 \label{fig:scale_height}
\end{figure}

\begin{figure}[tb]
\centering
\includegraphics[width=0.49\textwidth]{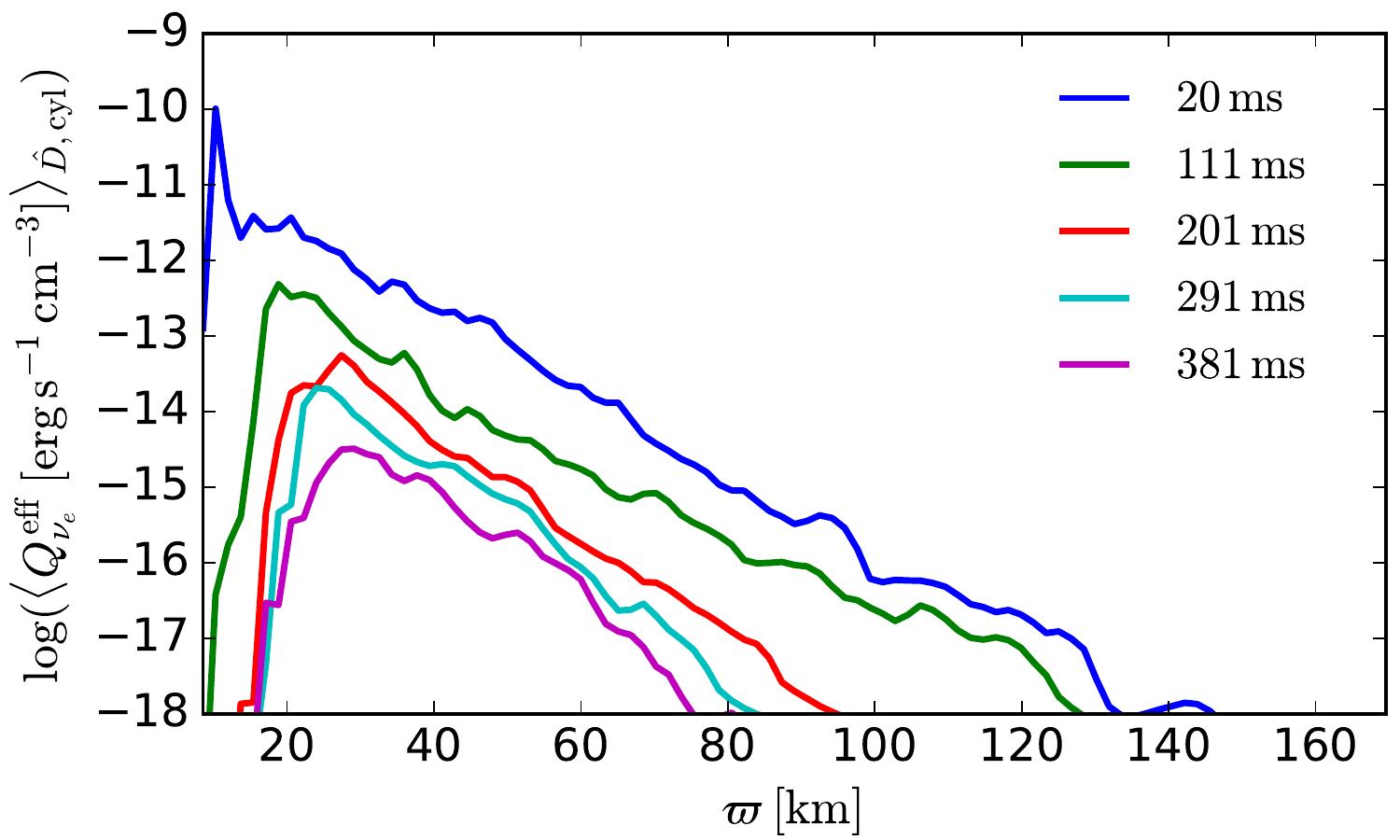}
\includegraphics[width=0.49\textwidth]{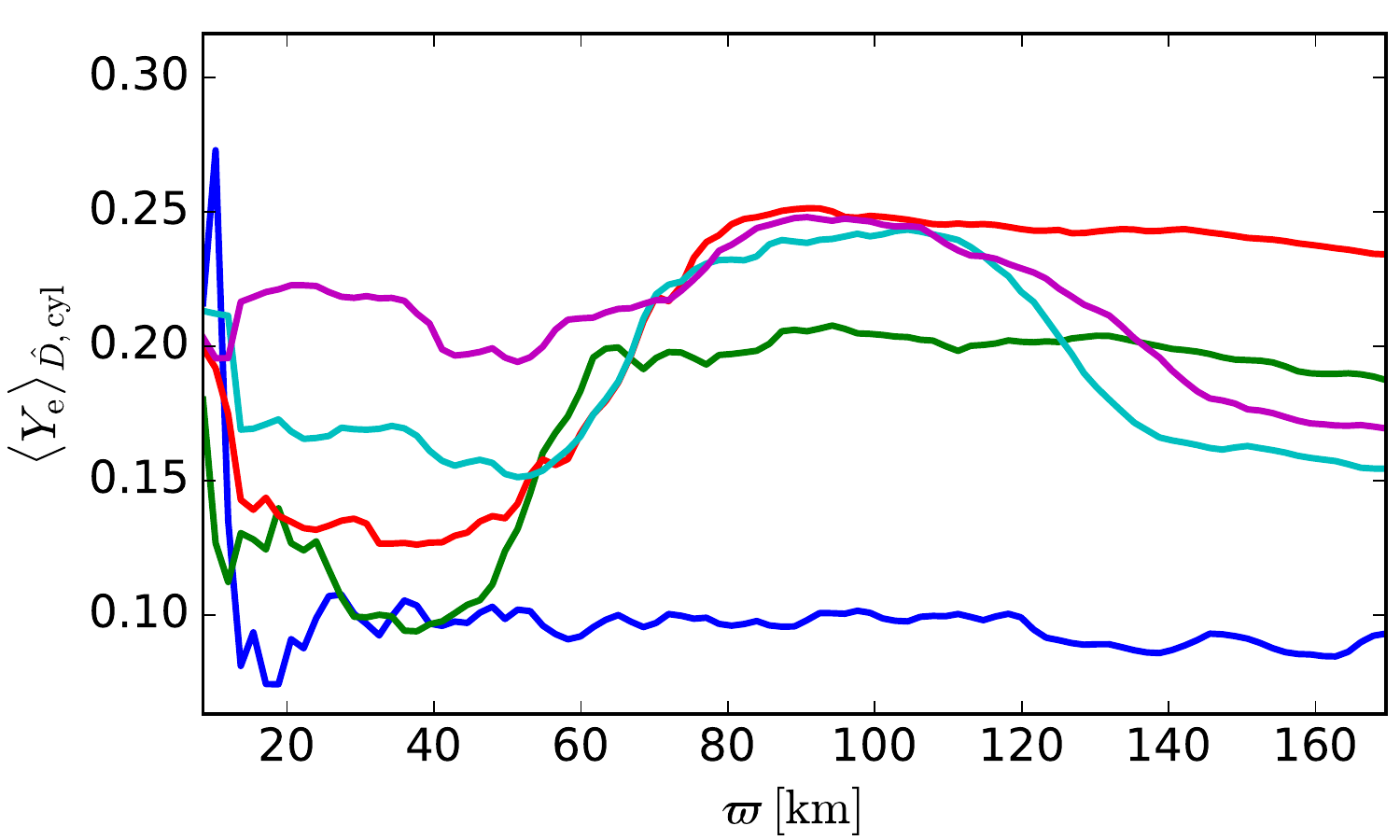}
\includegraphics[width=0.49\textwidth]{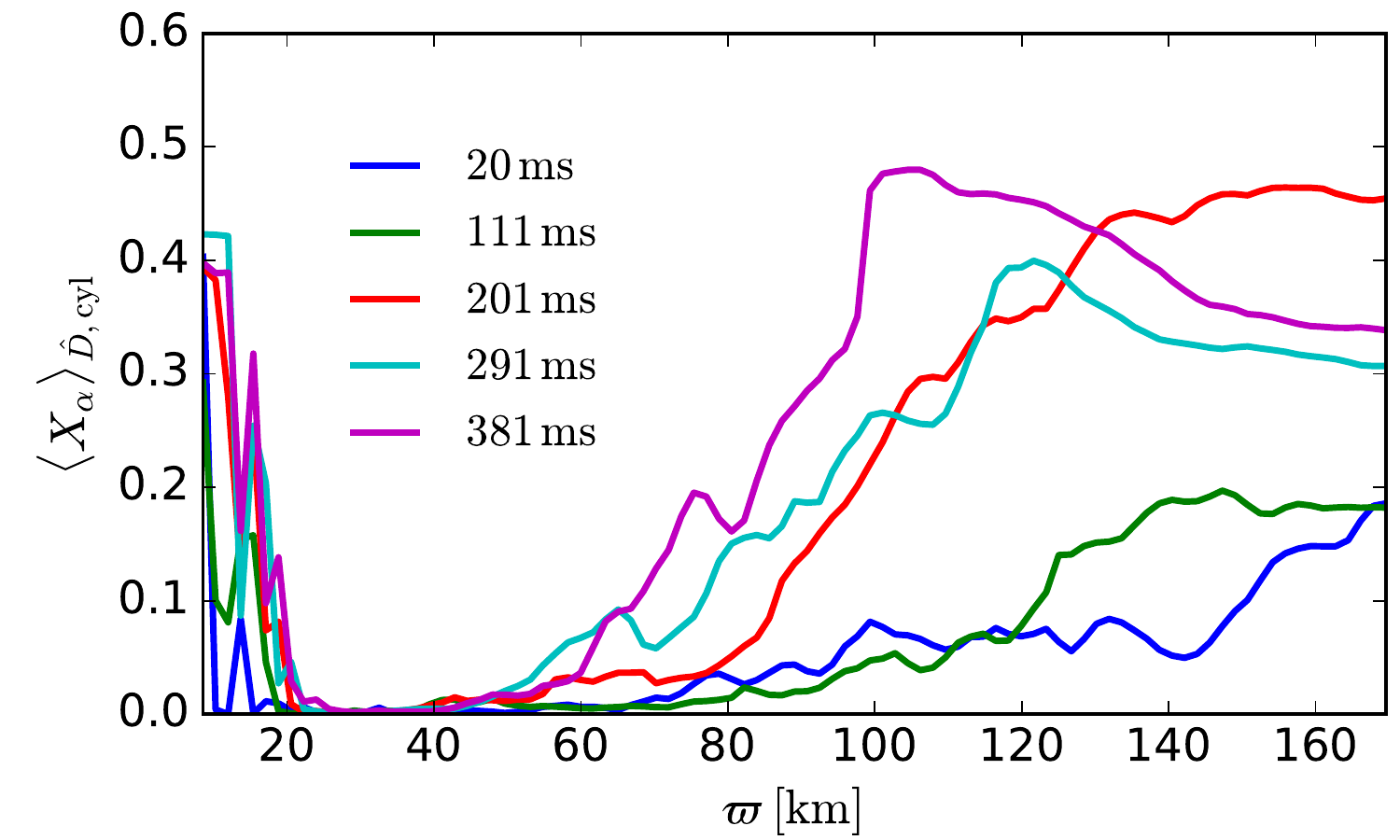}
\caption{Density-averaged radial profiles of (top to bottom) the electron
  neutrino emissivity, electron fraction, and $\alpha$-particle mass
  fraction at different times during the evolution.}
 \label{fig:disk_global_profiles}
\end{figure}

\begin{figure}[tb]
\centering
\includegraphics[width=0.49\textwidth]{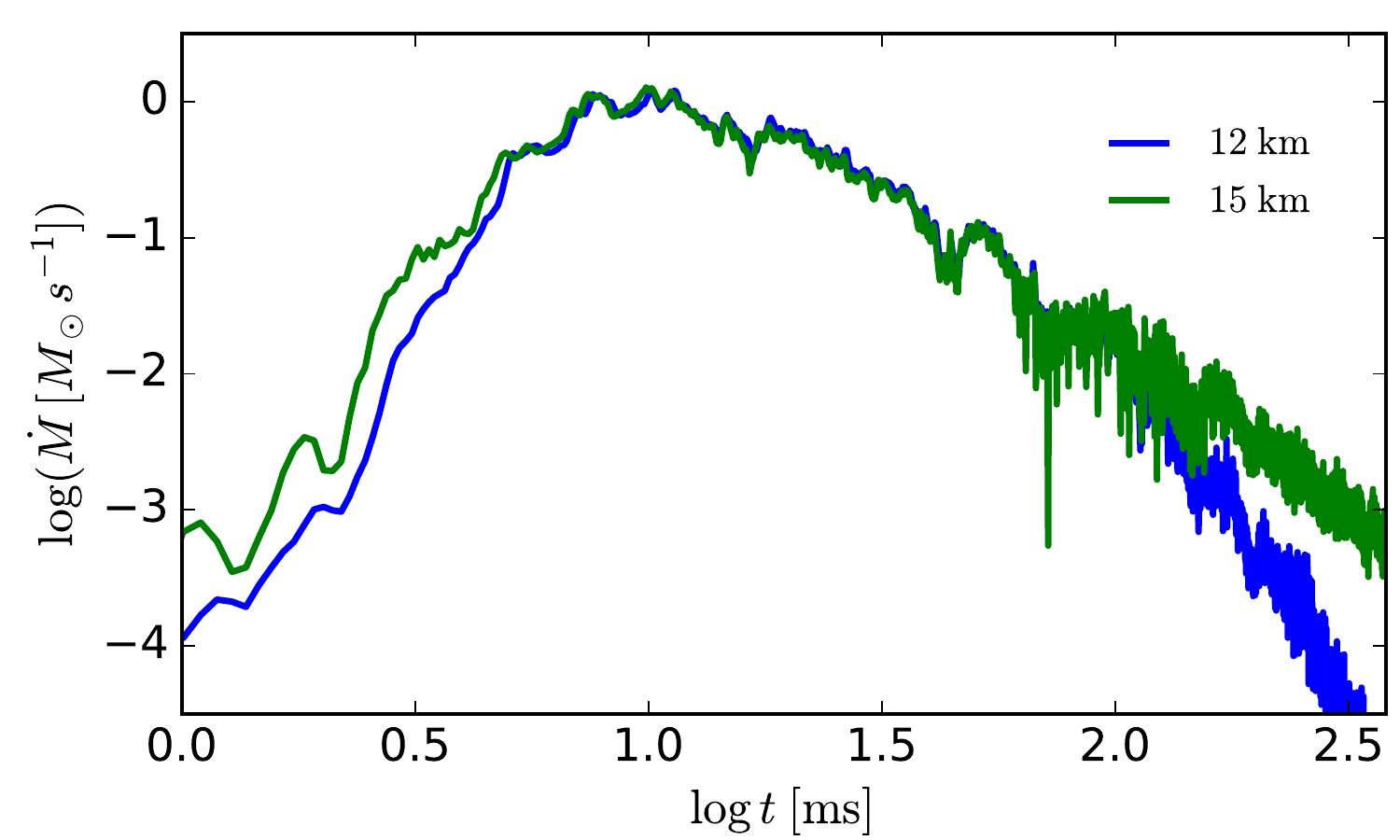}
\includegraphics[width=0.49\textwidth]{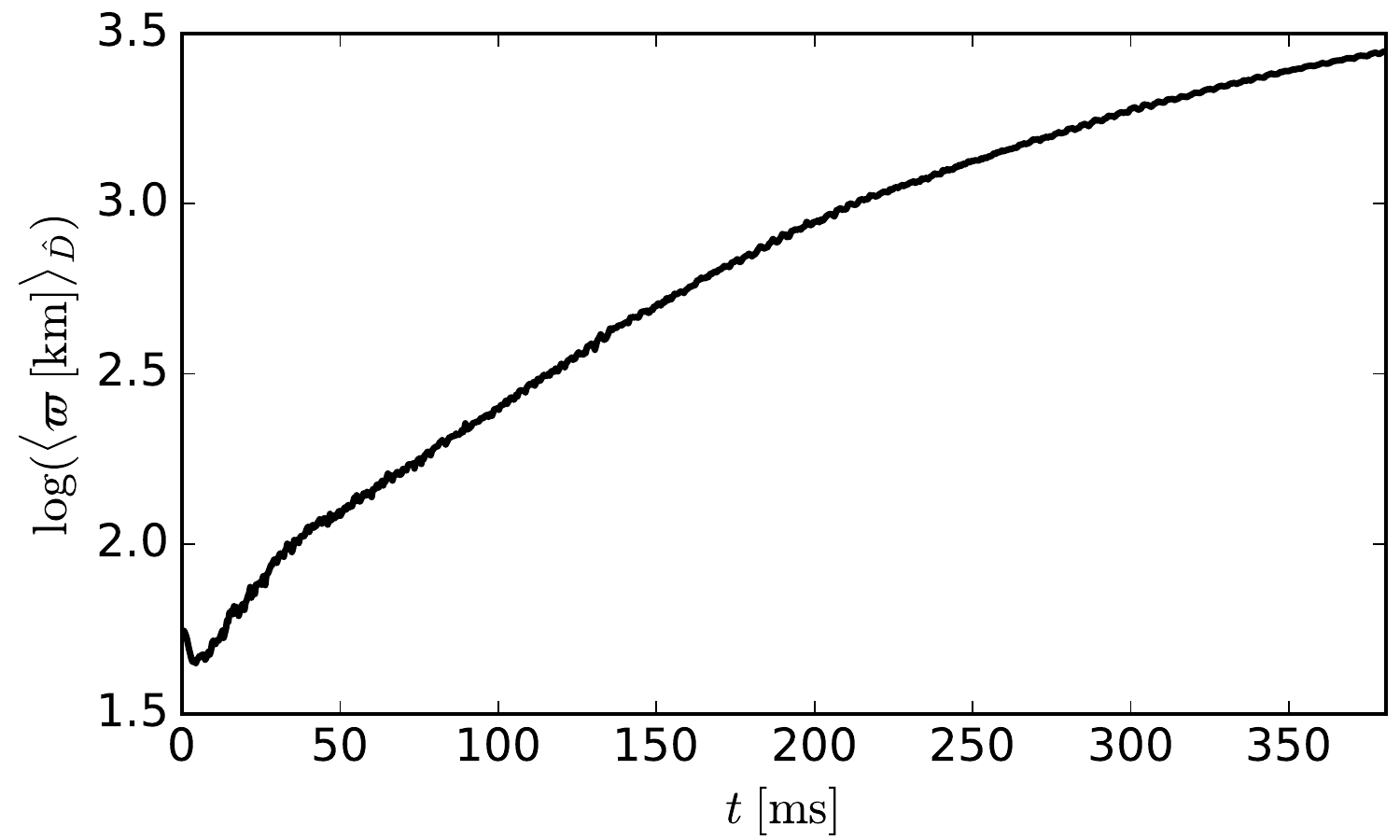}
\caption{Top: accretion rate onto the BH as measured by the mass flux
  through spherical coordinate surfaces with radii 12
  and $15\,\mathrm{km}$. Bottom: evolution of the
  density-averaged cylindrical radius $\varpi$
  of the baryonic matter (cf.~Eq.~\eqref{eq:density_average}), indicating viscous spreading
  of the disk.}
 \label{fig:disk_global}
\end{figure}

The global disk structure as characterized by the radial profile of the vertical density scale height is
shown in Fig.~\ref{fig:scale_height}.  We define the scale height according to 
\begin{equation}
  z_H(\varpi) \equiv\langle |z| \rangle_{\hat D, \mathrm{cyl}}, \label{eq:zH}
\end{equation}
where
\begin{equation}
  \langle \chi \rangle_{\hat D, \mathrm{cyl}} \equiv \frac{\int\int_{0}^{2\pi} \chi \hat D \varpi\mathrm{d}\phi\mathrm{d}z}{\int\int_{0}^{2\pi} \hat D \varpi\mathrm{d}\phi\mathrm{d}z}
\end{equation}
is the rest-mass density average of a quantity $\chi$ over
azimuthal angle $\phi$ and height $z$ as a
function of the cylindrical coordinate radius $\varpi$. 

At large radii, $\varpi \gtrsim 250\,\mathrm{km}$, the disk remains geometrically thick at all times, with a density scale height of $z_H/\varpi \gtrsim
0.4\!-\!1$. This is because neutrino cooling is always inefficient in these low-density regions, as illustrated by the radial profile of the density-averaged electron neutrino emission
rate $\langle Q^\mathrm{eff}_\nue\rangle_{\hat D, \mathrm{cyl}}$ in
Fig.~\ref{fig:disk_global_profiles}.  At late times, $t>200\,\mathrm{ms}$, the density scale height $z_H/\varpi$ exceeds unity
in the radial region $\varpi\approx 100-300\,\mathrm{km}$, which is due to
the outflows being efficiently generated at these radii (see
Sec.~\ref{sec:dynamo}, Fig.~\ref{fig:outflow_radii}). The thickening of the disk as the accretion drops and the concomitant generation of outflows was predicted by 1D (height-integrated) models \citep{Metzger+08c,Metzger2009c}.

The disk becomes thinner at smaller radii, starting at the characteristic radius $\varpi_\alpha$, where
$\alpha$-particles dissociate into free nucleons. The $\alpha$-dissociation consumes
$7\,\mathrm{MeV}$ per nucleon, which acts to cool the accretion flow and results in a geometrically thinner disk. This radius is initially at $\varpi_\alpha\approx 170\,\mathrm{km}$ and decreases to
$\varpi_\alpha\approx 100\,\mathrm{km}$ by the end of the simulation, as indicated by the
radial profile of the density-averaged $\alpha$-particle mass fraction $\langle X_\alpha \rangle_{\hat D, \mathrm{cyl}}$
(cf.~Fig.~\ref{fig:scale_height} and the bottom panel of
Fig.~\ref{fig:disk_global_profiles}).

At yet smaller radii, the accretion flow becomes geometrically even thinner as the result of neutrino cooling, with the density
scale height $z_H/\varpi\sim 0.1$ close to the BH, $\varpi\lessapprox
70\,\mathrm{km}$ (cf.~Fig.~\ref{fig:scale_height}).  This efficient neutrino cooling begins interior to the so-called ``ignition'' radius $\varpi_\mathrm{ign}<\varpi_\alpha$, which is defined as the location where the neutrino-cooling timescale becomes less than the local accretion timescale \citep{Chen2007}. This radius typically coincides with the location at which the energies of electrons and positrons become comparable to the neutron--proton mass difference $(m_\mathrm{n}-m_\mathrm{p})c^2$, triggering the onset of the efficient Urca cooling reactions (Eqs.~\eqref{eq:ep_nnue} and \eqref{eq:en_pnua}; see Fig.~\ref{fig:disk_global_profiles}, top panel). The same weak interactions typically result in further reduction in the electron fraction $Y_\el$, due to the increased degeneracy of the matter, as discussed in the previous subsection (cf.~Fig.~\ref{fig:disk_global_profiles},
middle panel).

By the end of the simulation, the BH has accreted $\approx\!60\%$ of the initial torus mass. The
BH accretion rate as measured by the mass flux through
spherical coordinate detector surfaces is shown in
Fig.~\ref{fig:disk_global} (top panel). It decreases from $\sim
1\,M_\odot \mathrm{s}^{-1}$ at early times to $\sim
10^{-4}\,M_\odot \mathrm{s}^{-1}$ by the end of the simulation. This leads to an essentially converged total
accreted mass onto the BH of $\approx\!1.20\times 10^{-2}\,M_\odot$ or
$\approx\!0.59\,M_{t,\mathrm{in}}$. Here
$M_{t,\mathrm{in}}=2.02\times 10^{-2}\,M_\odot$ is the initial disk
mass at $t=20\,\mathrm{ms}$, excluding all matter that is accreted onto
the BH or ejected from the disk during the initial relaxation phase
(cf.~Sec.~\ref{sec:MRI}). As the accretion rate continues to decrease
as the disk viscously spreads outward (see below), the total accreted disk mass
is unlikely to increase by a significant amount during the subsequent evolution.

The MHD turbulence mediates angular momentum transport in the disk, which
leads to accretion onto the BH but also to viscous radial spreading of the
disk. Evidence for the latter effect is reported in the bottom panel of
Fig.~\ref{fig:disk_global}, which shows that the density-averaged
cylindrical radius $\langle \varpi \rangle_{\hat D}$ of matter in the
simulation domain is monotonically growing after the initial
relaxation phase. The same result is obtained when the disk corona and
winds are explicitly excluded from the integration, i.e., by only
integrating up to the local density scale height $z_H$ of the disk
(Eq.~\eqref{eq:zH}). However, equatorial winds are not straightforward
to distinguish from the disk itself and thus remain in the analysis either way.

About $\approx\!40\%$ of the initial disk mass is unbound in outflows,
which undergo $r$-process nucleosynthesis (Sec.~\ref{sec:r-process}). By
the end of the simulation, roughly $\approx\!20\%$ of the initial disk
mass has already been ejected from the disk; i.e., it has reached
$>1000\,\mathrm{km}$ and is unbound (cf.~Sec.~\ref{sec:dynamo} and
Tab.~II of \citealt{Siegel2017a}). However, the disk is still producing
steady winds by the end of the simulation, which means the total unbound
mass is likely to become significantly higher. Even as the disk dilutes with
time and neutrino cooling becomes less important, viscous heating will
still continue to drive winds. Furthermore, as the disk viscously
spreads, additional material is lifted out of the BH
potential, also aided by nuclear binding energy release
from the formation of $\alpha$-particles and heavier nuclei as the material
cools. With the total accreted mass having already converged, it is thus
reasonable to assume that the remaining disk mass by the end of the
simulation will eventually be evaporated, leading to an estimated
total ejected mass of $\lesssim\!0.4\,M_{t,\mathrm{in}}$.

\subsection{Neutrino emission}
\label{sec:neutrino_emission}

\begin{figure}[tb]
\centering
\includegraphics[width=0.49\textwidth]{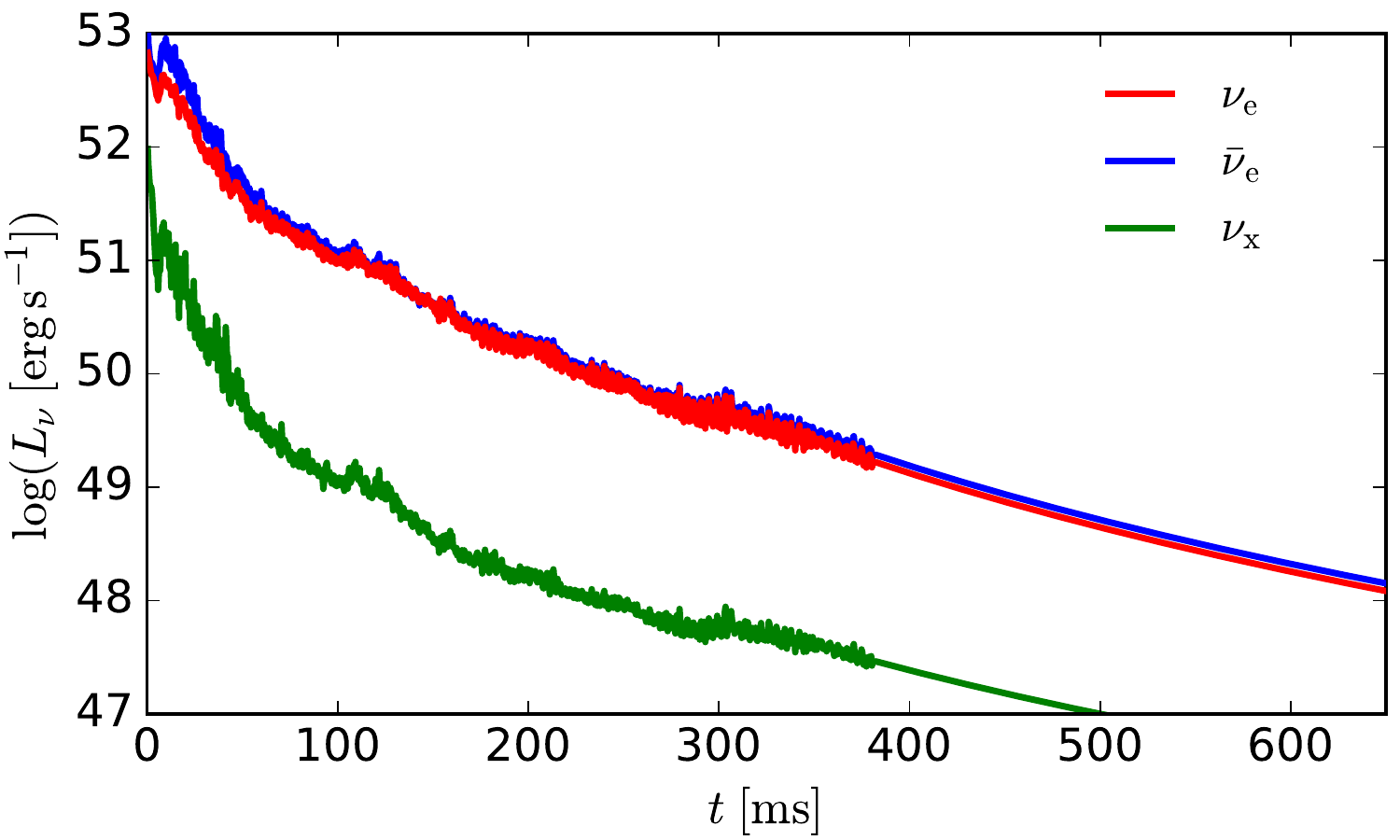}
\includegraphics[width=0.49\textwidth]{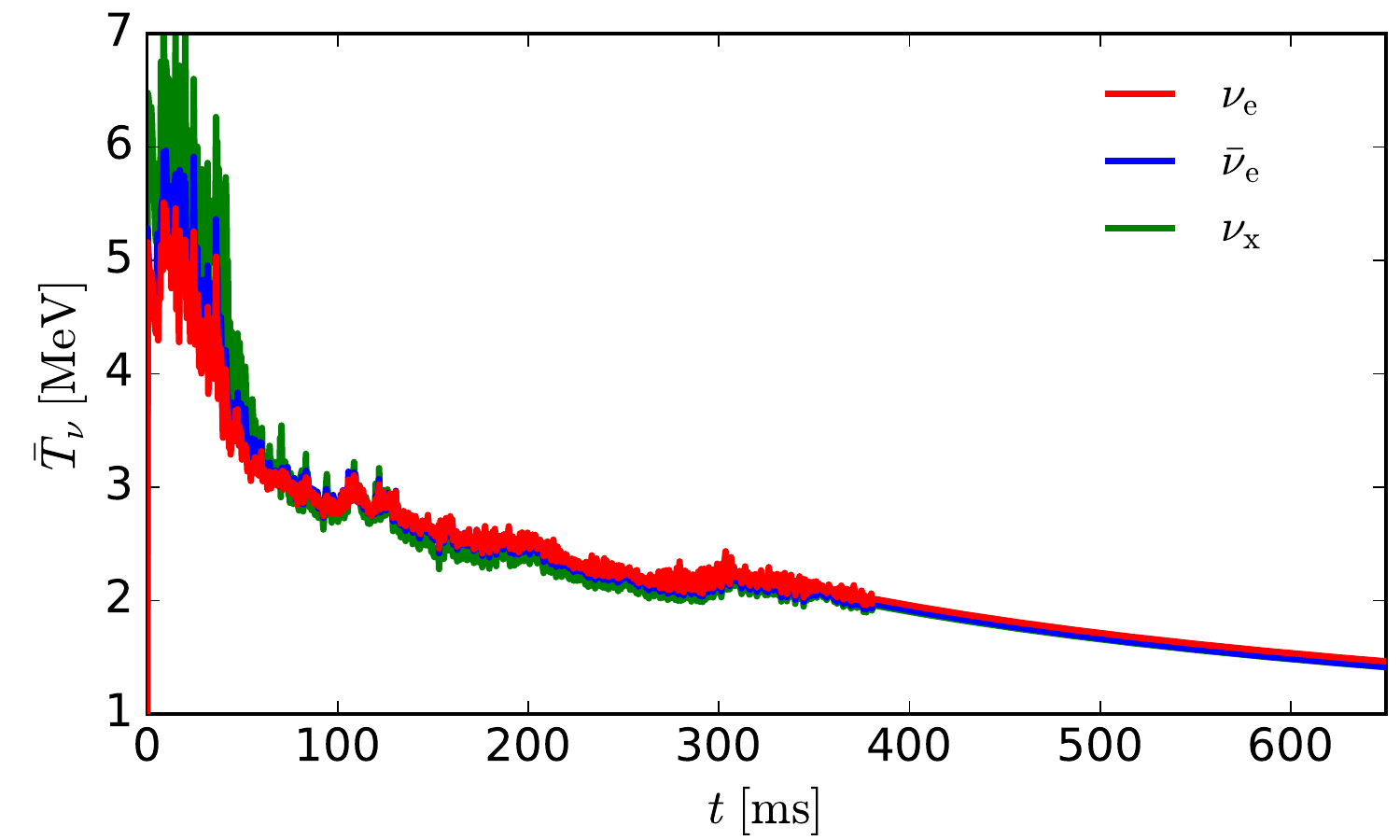}
\includegraphics[width=0.49\textwidth]{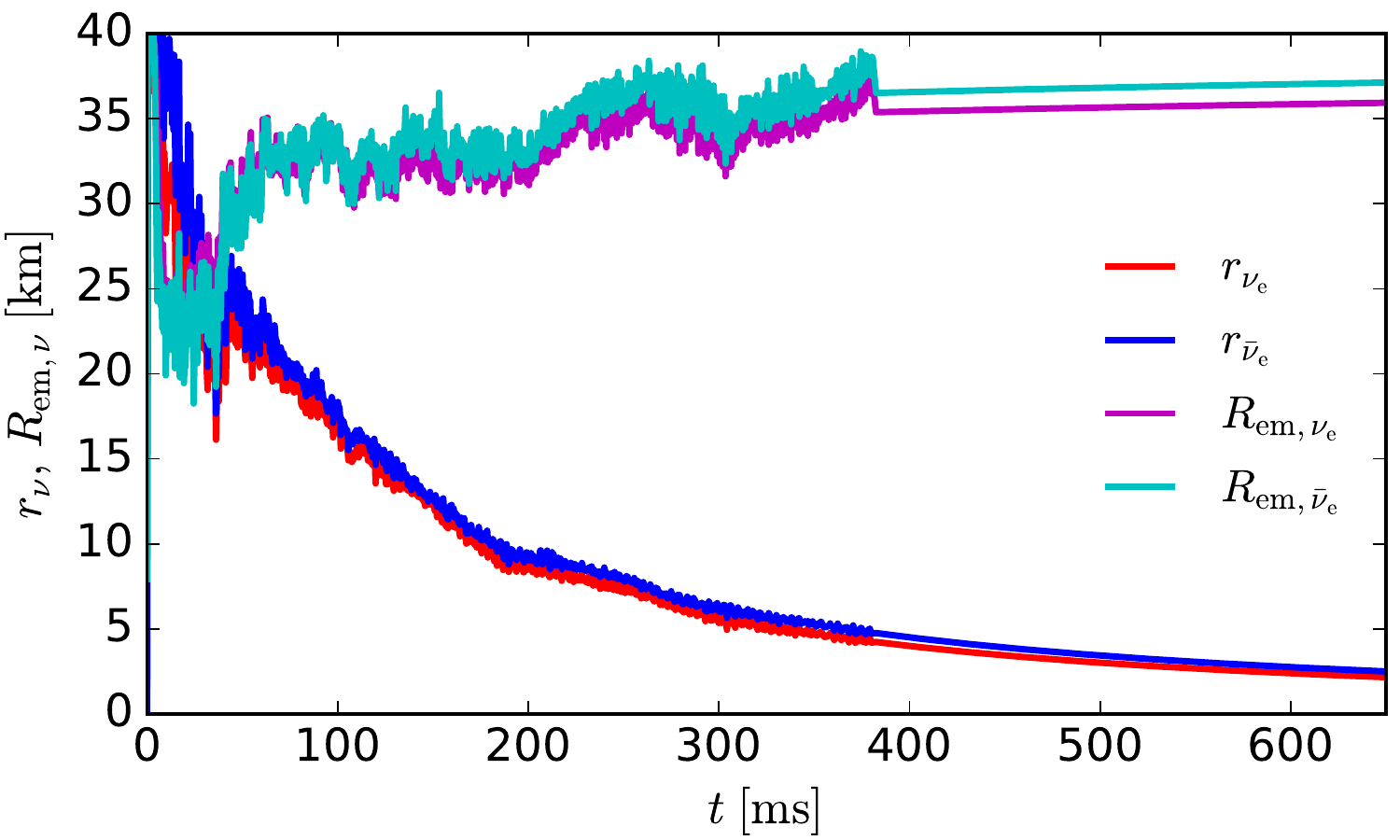}
 \caption{Characteristics of neutrino
   emission from the disk (top to bottom): total neutrino luminosity, mean neutrino
 temperature, and characteristic radii of neutrino emission (see the text). After the end of the simulation
 ($t=381\,\mathrm{ms}$), quantities are extrapolated by power laws
 fit to the late-time simulation data.}
 \label{fig:neutrino_emission}
\end{figure}

The inner parts of the disk are sufficiently hot and dense that
neutrino emission becomes energetically important
(cf.~Fig.~\ref{fig:disk_global_profiles} and
Sec.~\ref{sec:global_structure}). In this section, we discuss the
characteristics of the neutrino radiation from the disk, which will
serve as input to our $r$-process nucleosynthesis calculations presented
in the next section.

We define the total neutrino
luminosity for each neutrino species $\nui \in \{\nue,\nua,\nux\}$ according to
(cf.~Eqs.~\eqref{eq:sources} and \eqref{eq:Q})
\begin{equation}
  L_{\nu_i} = \int \alpha W Q^\mathrm{eff}_{\nu_i} \alpha
  \sqrt{\gamma}\mathrm{d}^3x, \label{eq:Lnui}
\end{equation}
where an additional factor $\alpha$ is included to correct for the gravitational redshift due to the
BH potential. This definition takes into account the effects of finite optical depth; i.e., it is based on the effective
energy emission rates, but it neglects reabsorption of emitted neutrinos
by matter.

Neutrino emission is purely thermal, characterized by the local
emission temperature $T$ (the temperature of matter). We assign mean neutrino emission
temperatures for the
different neutrino species to the disk, defined as
the neutrino energy emission rate averaged quantities
\begin{equation}
  \bar{T}_{\nu_i} \equiv \langle T\rangle_{Q_\nui}. \label{eq:Tnui}
\end{equation}
Here we have defined the neutrino emission rate average of a quantity $\chi$ by
\begin{equation}
  \langle \chi \rangle_{Q_\nui} \equiv \frac{\int \chi
    Q^\mathrm{eff}_\nui W\alpha\sqrt{\gamma}\mathrm{d}^3x}{\int Q^\mathrm{eff}_\nui W\alpha\sqrt{\gamma}\mathrm{d}^3x}.
\end{equation}
Note that $Q^\mathrm{eff}_\nui W\alpha\sqrt{\gamma}$ corresponds to
the energy emitted per unit time and coordinate volume through
neutrinos of species $\nui$
as seen by the Eulerian observer (cf.~Eqs.~\eqref{eq:sources} and
\eqref{eq:Q}). For further reference, we also define a corresponding spherical
blackbody emission radius,
\begin{equation} 
  r_{\nu_i} = \left(\frac{L_{\nu_i}}{4\pi
      \frac{7}{16}\sigma\bar{T}_{\nu_i}^4} \right)^{\frac{1}{2}}, \label{eq:rnui}
\end{equation}
where $\sigma$ is the Stefan--Boltzmann constant and the actual
characteristic neutrino emission radius
\begin{equation}
  R_{\mathrm{em}, \nui} \equiv \langle \varpi \rangle_{Q_\nui}. \label{eq:Rem_nui}
\end{equation}

Figure~\ref{fig:neutrino_emission} shows the total neutrino
luminosities, average neutrino emission temperatures, and blackbody as
well as characteristic emission radii as extracted from our
simulation data. We extrapolate these quantities beyond the end of the
simulation at $t=381\,\mathrm{ms}$ by power laws fitted to the
late-time simulation data.

The neutrino luminosities are initially high, with $L_\nu\sim
10^{52}\,\mathrm{erg}\,\mathrm{s}^{-1}$ for electron and
anti-electron neutrinos and at least an order of magnitude lower for
the heavier neutrino species, but they quickly fade over timescales of
hundreds of ms. We note that these initial neutrino luminosities are very
similar to the values found in the early post-merger accretion systems of
recent hydrodynamic NS--NS and BH--NS merger simulations (e.g.,
\citealt{Sekiguchi2016,Radice2016,Foucart2017a}). The total energy
radiated in neutrinos by the
disk in terms of the various neutrino species is given by $E_\nue, E_\nua,
E_\nux = (4.2, 6.1, 0.083)\times 10^{50}\,\mathrm{erg}$. Despite the
fact that the neutrino luminosities fade rapidly compared to the
evolution timescale of the disk, irradiation by neutrinos during the
early phase of the evolution can still have an appreciable effect on
the composition of the disk outflows and thus on $r$-process
nucleosynthesis. We discuss this effect in the following section.

\section{$r$-process nucleosynthesis}
\label{sec:r-process}

\begin{figure}[tb]
\centering
\includegraphics[width=0.49\textwidth]{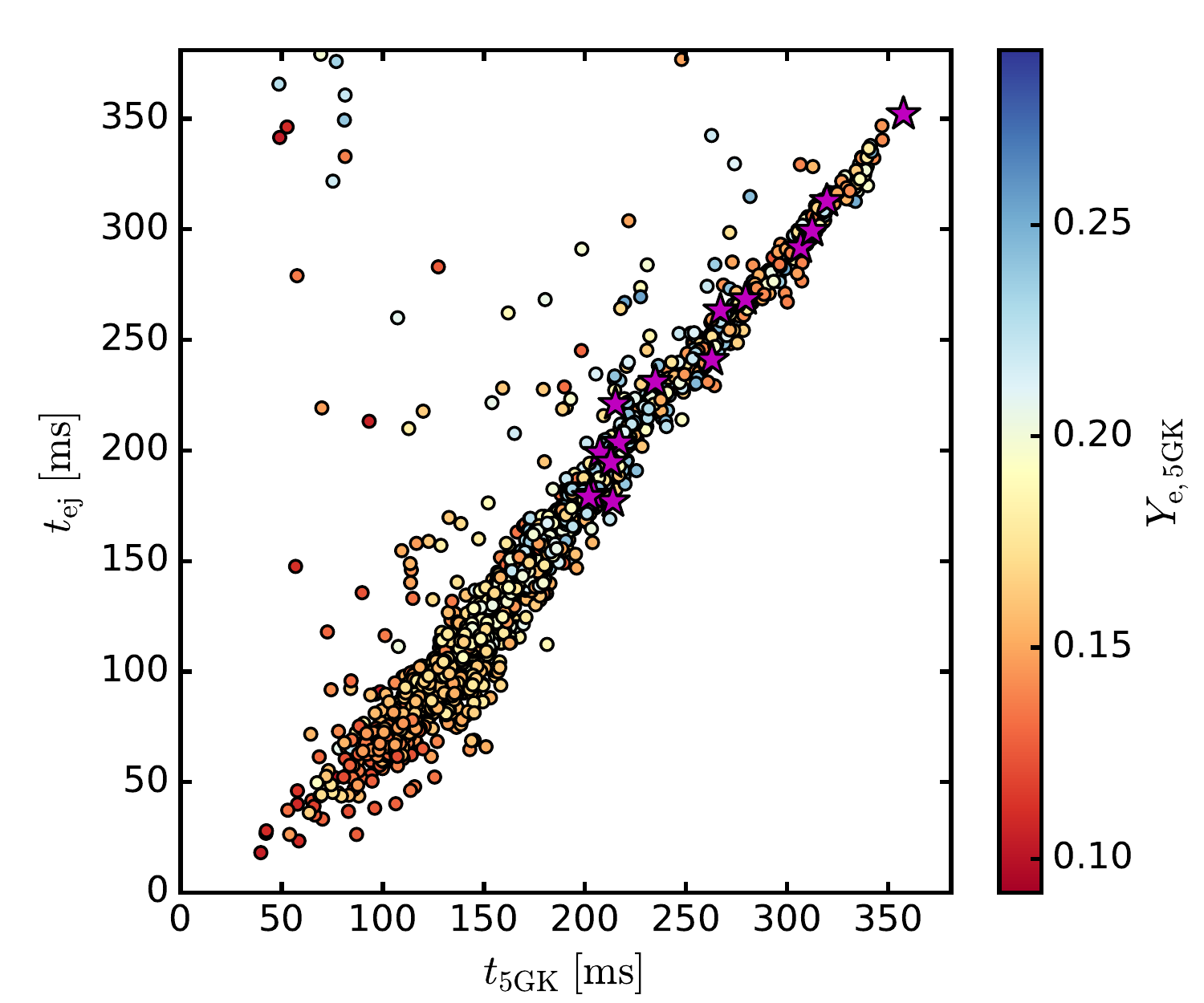}
\caption{Ejection time $t_\mathrm{ej}$ of all unbound tracer particles
  versus the last time $t_{5\mathrm{GK}}$ at
  which the tracer particle reached a temperature of
  $5\,\mathrm{GK}$, color-coded by the electron fraction at
  $t_{5\mathrm{GK}}$. The 15 tracer particles that contribute most to
  the nucleosynthesis anomaly at $A=132$ are marked as magenta stars, which all follow the main correlation between $t_\mathrm{ej}$ and $t_{5\mathrm{GK}}$.}
 \label{fig:rprocess_anomaly}
\end{figure}

\begin{figure}[tb]
\centering
\includegraphics[width=0.49\textwidth]{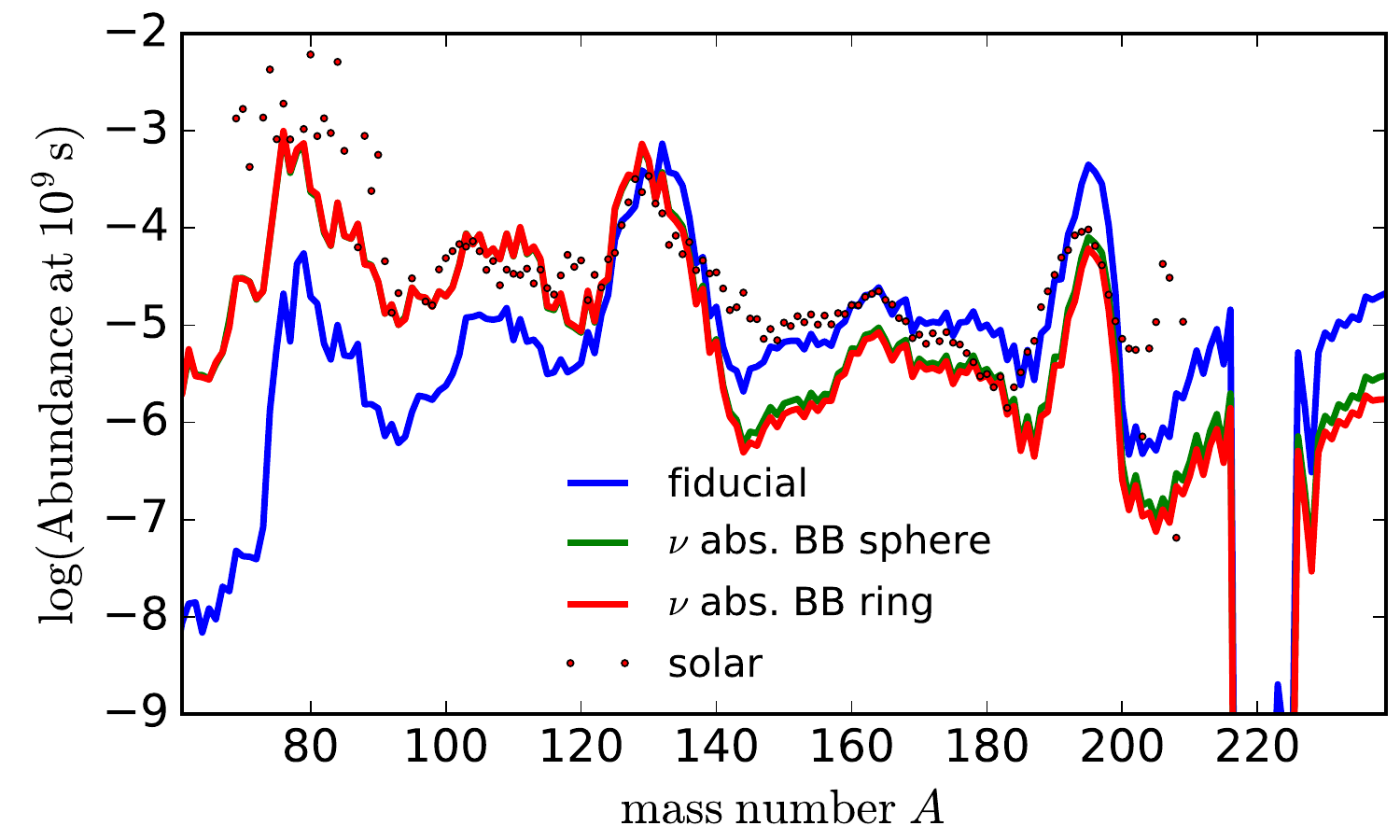}
\includegraphics[width=0.49\textwidth]{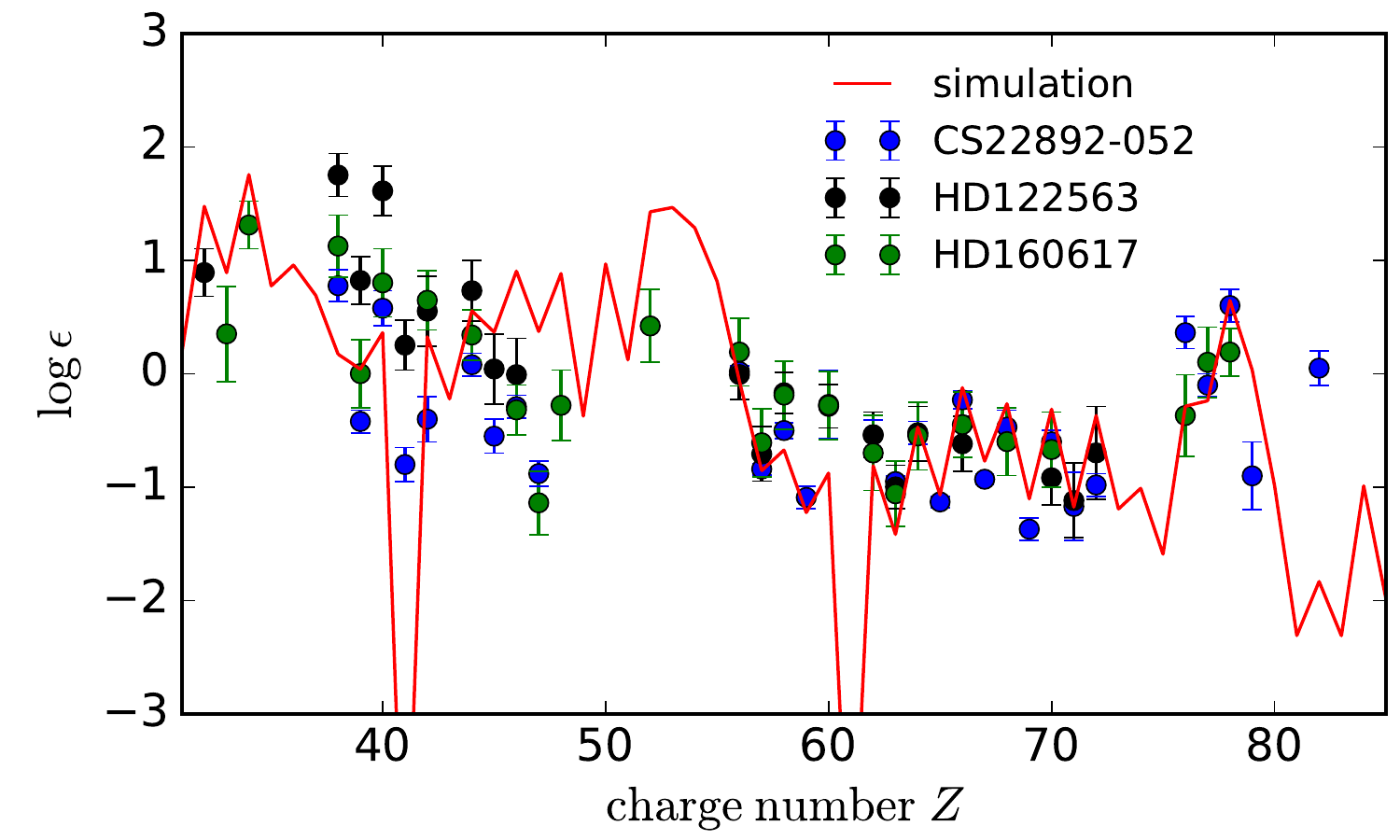}
\caption{Top: final mean elemental abundances for the fiducial case without
  neutrino absorption as in \citet{Siegel2017a} and including
  neutrino absorption according to a spherical blackbody light-bulb
  scheme (see the text; ``$\nu$ abs. BB sphere'') and according to
  ringlike blackbody emission (see the text; ``$\nu$ abs. BB ring'').
  For reference, observed solar system abundances
  from \citet{Arnould2007} are added, scaled to match the fiducial mean
  abundances at $A=130$. Bottom: comparison of abundances including
  neutrino absorption according to the ringlike blackbody emission
  to the observed abundances in metal-poor halo
   stars \citep{Sneden2003,Roederer2012a,Roederer2012b}, showing
   $\log\epsilon = \log Y_Z/Y_1 + 12$, scaled such that $\sum (\log Y_Z/Y_{Z,\mathrm{CS22892-052}})^2$ is minimized in the range $55\le Z \le 75$.}
 \label{fig:rprocess_Lnu_comparison}
\end{figure}

\begin{figure}[tb]
\centering
\includegraphics[width=0.49\textwidth]{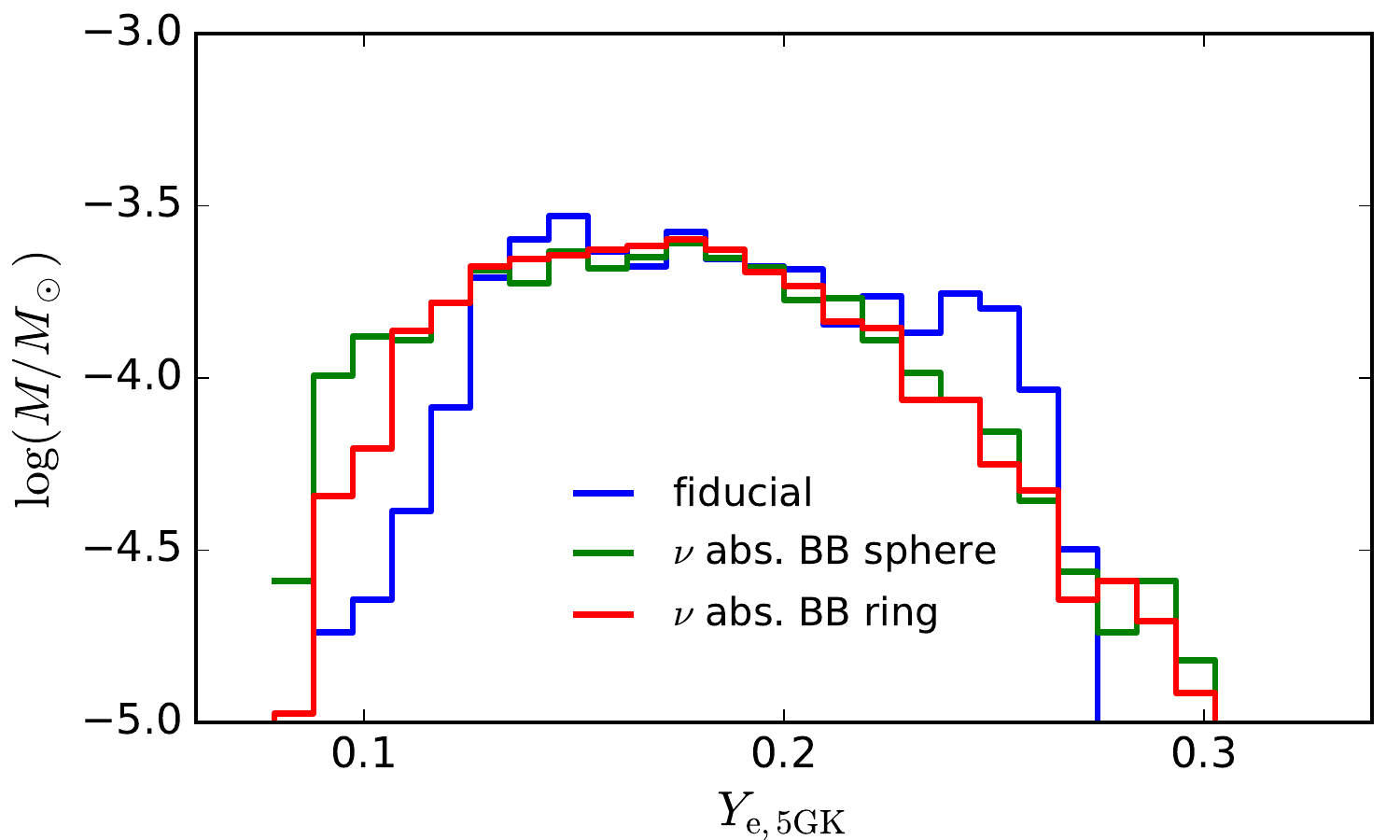}
\caption{Comparison of the mass distributions of unbound tracer particles in terms of their electron fraction at $t=t_{5\mathrm{GK}}$ for the fiducial case without neutrino absorption, as in \citet{Siegel2017a}, and including neutrino absorption according to a spherical blackbody light-bulb scheme (see the text; ``$\nu$ abs. BB sphere'') and according to ringlike blackbody emission (see the text; ``$\nu$ abs. BB ring'').}
\label{fig:rprocess_Ye_comparison}
\end{figure}

Abundance yields from $r$-process nucleosynthesis in the outflows of the
accretion disk were already presented in \citet{Siegel2017a}. Here we
elaborate on these results, discuss the nucleosynthesis anomaly at
$A=132$ (Sec.~\ref{sec:anomaly}), and present results from $r$-process
nucleosynthesis calculations including neutrino absorption, which we perform with
the nuclear-reaction network \texttt{SkyNet} (\citealt{Lippuner2017b}; Sec.~\ref{sec:neutrino_absorption}).

\subsection{The A=132 anomaly}
\label{sec:anomaly}

Previous $r$-process nucleosynthesis analyses of disk outflows from 2D
Newtonian $\alpha$-disk simulations have noted an overproduction of $A=132$
nuclei with respect to the second $r$-process peak ($A=128-130$) when
compared to
observed solar system abundances \citep{Wu2016}. This was ascribed to
late-time, low-temperature convection in the disk outflow, i.e., to
fluid elements, whose ejection time $t_\mathrm{ej}$
(cf.~Sec.~\ref{sec:dynamo}) from the disk is
much greater than $t_{5\mathrm{GK}}$. We define $t_{5\mathrm{GK}}$ as
the last time when the temperature of a fluid element (tracer
particle) decreased below 5\,GK, which is the characteristic
temperature for NSE to break down and the $r$-process to set in.

Although our 3D GRMHD setup is expected to show less large-scale,
low-temperature convection than 2D viscous hydrodynamics (because of
the inverse turbulent cascade in 2D), we still find an overproduction
at $A=132$, which is evident from
Fig.~\ref{fig:rprocess_Lnu_comparison}.

In contrast to \citet{Wu2016}, we find that this anomaly in our 3D
GRMHD setup is not
predominantly due to tracers that undergo late-time low-temperature
convection, i.e., for which $t_\mathrm{ej}\gg t_{5\mathrm{GK}}$. This
is shown in Fig.~\ref{fig:rprocess_anomaly}, which reports
$t_\mathrm{ej}$ vs. $t_{\mathrm{5GK}}$ for all unbound tracer
particles. The dominant contributors to this anomaly all follow the
main correlation between $t_\mathrm{ej}$ and $t_\mathrm{5GK}$, and
tracers with $t_\mathrm{ej}\gg t_{5\mathrm{GK}}$ are not among
those. The origin of this anomaly remains inconclusive at this
point. It may point to a nuclear origin at least for our present
calculations with \texttt{SkyNet}, which requires further
investigation concerning the nuclear physics input.

\subsection{$r$-process nucleosynthesis including neutrino absorption}
\label{sec:neutrino_absorption}

In order to explore the effects of neutrino absorption on $r$-process
nucleosynthesis in the ejecta material, we `light-bulb' irradiate the
ejecta by neutrinos from the disk in a postprocessing step, employing
two different assumptions to bracket the uncertainties in the neutrino
emission geometry.

\textit{Spherical blackbody.} In a first approach, following
\citet{Roberts2017a}, we assume that
neutrinos are emitted with luminosity $L_{\nu_i}$ and
temperature $\bar{T}_{\nu_i}$ from a single spherical
surface centered on the BH of radius $r_\nui$
(cf.~Eqs.~\eqref{eq:Lnui}, \eqref{eq:Tnui}, and \eqref{eq:rnui}) and that they
follow a Fermi-Dirac distribution in energy space,
\begin{equation}
  f_{\mathrm{FD}}(E,\bar{T}_\nui) = \frac{1}{\exp(E/k_\mathrm{B}\bar{T}_{\nu_i}) + 1},
\end{equation}
where $E$ denotes the neutrino energy. The radii of the
neutrinospheres $r_\nui$ are typically on the order of tens of
km and are roughly comparable to or smaller than the actual radii
$R_{\mathrm{em}, \nui}$ of the peak neutrino emission within the disk (see Fig.~\ref{fig:neutrino_emission}, bottom panel). The neutrino distribution
function in energy space as a function of coordinate radius $r$ for
species $\nui$ is then given by
\begin{equation}
  f_{\nu_i}(E,r; \bar{T}_\nui, L_\nui) =
  \frac{1}{2}\left(1-\sqrt{1-\frac{r_\nui^2}{r^2}}\right)
  f_{\mathrm{FD}}(E,\bar{T}_\nui). \label{eq:BB_sphere}
\end{equation}

\textit{Ringlike blackbody.} In a second approach, following the
neutrino emission geometry of \citet{Fernandez2013}, we assume that
neutrinos are emitted with luminosity $L_{\nu_i}$ and temperature
$\bar{T}_{\nu_i}$ from a ring of radius $R_{\mathrm{em}, \nui}$ in the
equatorial plane around the BH (cf.~Eqs.~\eqref{eq:Lnui}, \eqref{eq:Tnui}, and \eqref{eq:Rem_nui}). This geometry more closely
resembles neutrino emission from the disk, as most of the emission is
confined to regions close to the midplane
(cf.~Fig.~\ref{fig:spacetime}, bottom panel) and as the effective emission
rates $Q^\mathrm{eff}_\nui$ are indeed sharply peaked around some
characteristic emission radius $r\simeq
R_{\mathrm{em},\nui}$ (cf.~Fig.~\ref{fig:disk_global_profiles}, top
panel). In analogy to Eq.~\eqref{eq:BB_sphere}, the
neutrino distribution function in this case is given by
\begin{equation}
  f_{\nu_i}(E,r,\theta;\bar{T}_\nui, L_\nui, R_{\mathrm{em},\nui}) =
  \frac{1}{2}N_\nui \mathcal{I}_\nui
  f_{\mathrm{FD}}(E,\bar{T}_\nui), \label{eq:BB_ring}
\end{equation}
where
\begin{equation}
  N_\nui = \frac{L_\nui}{4\pi R_{\mathrm{em},\nui}^2\frac{7}{16}\sigma \bar{T}_\nui^4}
\end{equation}
and
\begin{equation}
  \mathcal{I}_\nui =
  \frac{1}{2\pi}\left(\frac{R_{\mathrm{em},\nui}}{r}\right)^2 \int_0^{2\pi}
    \frac{\mathrm{d}\phi_\mathrm{R}}{2 D(r,\theta,R_{\mathrm{em},\nui},\phi_\mathrm{R})/r^2}.
\end{equation}
Here $r$ and $\theta$ denote the radial coordinate and polar angle,
respectively, and $\phi_\mathrm{R}$ denotes the azimuthal angle that
parameterizes the neutrino emission ring. Furthermore,
\begin{equation}
  D= r\left[1 + \left(\frac{R_{\mathrm{em},\nui}}{r}\right)^2 - 2 \frac{R_{\mathrm{em},\nui}}{r}\sin\theta\cos\phi_\mathrm{R}\right]^{1/2}
\end{equation}
is the distance between a spatial point $(r,\theta)$ and the neutrino
emission ring at position $\phi_\mathrm{R}$ (cf.~Fig.~B2 of \citealt{Fernandez2013}).

Figure~\ref{fig:rprocess_Lnu_comparison} reports detailed abundance
yields, including neutrino absorption, computed with the two methods
outlined above, in
comparison to previous results obtained by neglecting neutrino
absorption \citep{Siegel2017a}. It is reassuring
that these results do not depend on the method by which neutrino
absorption is included; both approaches lead to essentially the same
abundance yields. This is not surprising, given that the source of
neutrino radiation with a diameter of essentially $60-80\,\mathrm{km}$ is
sufficiently compact compared to the spatial
size of the entire disk and outflows
(cf.~Sec.~\ref{sec:global_structure}).

With neutrino absorption included, the production of the entire range
of $r$-process nuclei from the first to the third peak of the $r$-process can be
explained. Including neutrino absorption dramatically improves the
agreement between the abundance
yields of the lighter nuclei from the first to the second $r$-process
peak ($A\sim 80-120$) compared to the observed solar system
abundances. This is due to neutrinos irradiating part of
the outflow and the outer parts of the disk, thereby raising $Y_\el$ in
part of the outflow (see Fig.~\ref{fig:rprocess_Ye_comparison}), which enhances the production of lighter
$r$-process nuclei. However, a strong second-to-third-peak $r$-process is still
maintained. The fact that the outflow well reaches the production of third-peak
elements at the required level to explain solar abundances, even in the presence of strong neutrino irradiation, is at least
in part due to the self-regulation mechanism discussed in
Sec.~\ref{sec:disk_regulation}, which continuously releases very
neutron rich-material into the outflow. The excellent agreement with
observed abundances is also reflected in the bottom panel of
Fig.~\ref{fig:rprocess_Lnu_comparison}, which compares the abundance
yields from our simulation including neutrino absorption with observed
abundances in metal-poor stars in the halo of the Milky Way.

\section{Conclusion}
\label{sec:conclusion}

Below, we summarize our main results and conclusions.

\begin{itemize}

\item[(i)] We witness the onset of MHD turbulence, which quickly results in a steady turbulent state (Sec.~\ref{sec:MRI}) and an effective initial disk configuration that is very similar to results from recent NS--NS or NS--BH merger simulations. The disk remains in this steady turbulent state for the rest of the simulation time (Fig.~\ref{fig:turbulent_state}). The butterfly diagram (Fig.~\ref{fig:spacetime}) indicates a fully operational magnetic dynamo with a secular cycle of roughly $\sim\!20\,\mathrm{ms}$. The dynamo generates magnetic fields of alternating polarities in the disk midplane that slowly migrate to higher latitudes, where they gradually dissipate into heat in a ``hot corona.''

\item[(ii)] We find the emergence of a hot disk corona at higher latitudes. There, viscous heating from MHD turbulence and dissipation of magnetic fields is not balanced by neutrino cooling (which tracks density and thus rapidly falls off with latitude; Fig.~\ref{fig:spacetime}), and powerful thermal outflows are launched.  The energy released by $\alpha$-particle formation also plays a crucial role in unbinding matter from the disk after it is lifted out of the BH gravitational potential by coronal heating.  The asymptotic velocity scale of $v_\infty\approx 0.1c$ of the unbound outflows is largely set by the energy released from $\alpha$-particle recombination (Fig.~\ref{fig:bernoulli}). Our results agree qualitatively with previous work by \citet{Barzilay&Levinson08}, who explored models of steady-state outflows driven from the midplane of neutrino-cooled disks, including those powered by the dissipation of turbulent energy in the disk corona, finding that such outflows can preserve the neutron richness of the disk midplane (see also \citealt{Metzger+08b}).

\item[(iii)] We observe a regulation of the electron fraction in the disk midplane by weak interactions. We identify a self-regulation mechanism based on electron degeneracy in the inner parts of the disk (where viscous heating is roughly balanced by neutrino cooling), which regulates the electron fraction to $Y_\el\sim 0.1$ irrespective of the initial conditions (Sec.~\ref{sec:disk_regulation}). This results in the formation of a reservoir of neutron-rich material, despite the ongoing protonization in the outer parts of the disk over viscous timescales (Fig.~\ref{fig:degeneracy}). This reservoir continuously feeds very neutron-rich material into the outflows, which thus keeps the overall mean electron fraction of the outflows comparatively low ($\bar Y_\el\sim0.2$) over viscous timescales and guarantees the production of third-peak $r$-process nuclei.

\item[(iv)] We demonstrate that the EOS and weak interactions in the disk are not affected by magnetic field effects (Fig.~\ref{fig:Landau}).

\item[(v)] We find that unbound outflows carry away $\lesssim\!40\%$ of the initial disk mass with asymptotic escape velocities centered around $v_\infty\approx 0.1c$, with a roughly spherical geometry (Secs.~\ref{sec:dynamo} and \ref{sec:global_structure}; Fig.~\ref{fig:outflow_radii}). The total ejecta mass is given by
\begin{equation}
	M_\mathrm{ej} \simeq 10^{-2} \left(\frac{f_\mathrm{ej}}{0.35}\right)\left(\frac{M_\mathrm{disk}}{3\times 10^{-2}M_\odot}\right)M_\odot, \label{eq:M_ej}
\end{equation}
where $f_\mathrm{ej}$ denotes the fraction of mass ejected from the original disk of mass $M_\mathrm{disk}$. This is larger than that found by previous 2D Newtonian viscous-hydrodynamic simulations \citep{Fernandez2015a,Just2015a}, which we attribute to additional nonlocal coronal heating that quickly evaporates disk material.  With $M_\mathrm{disk}\simeq\mathrm{few}\times 10^{-2}M_\odot$ being a rather conservative lower limit on disk masses from NS mergers (e.g., \citealt{Hotokezaka2013d,Ciolfi2017a}), we conclude that {\it post-merger disk winds likely represent the dominant mass ejection mechanism in NS--NS mergers}; in BH--NS mergers, tidal ejecta may still dominate, depending on the binary parameters due to the more extreme binary mass ratios expected in this case.

The asymptotic escape velocities and the quantity of wind ejecta, if extrapolated to a moderately higher initial torus mass $\approx 0.1M_{\odot}$, provide a natural explanation for the red KN from the recent GW170817 event (e.g.~\citealt{Chornock2017,Cowperthwaite2017,Villar2017}).

\item[(vi)] The disk radiates thermal neutrinos at characteristic temperatures of $T\sim \mathrm{few}\,\mathrm{MeV}$ with rapidly declining luminosities starting at $L_\nu\sim 10^{52}\,\mathrm{erg}\,\mathrm{s}^{-1}$ and total radiated energies of $E_\nue, E_\nua, E_\nux = (4.2, 6.1, 0.083)\times 10^{50}\,\mathrm{erg}$ (Fig.~\ref{fig:neutrino_emission}).

\item[(vii)] Outflows from the accretion disk are sufficiently neutron-rich to synthesize $r$-process elements extending up to the third peak, a result that we find is insensitive to our treatment of neutrino heating. Neutrino heating can have a moderate impact on $r$-process nucleosynthesis (Fig.~\ref{fig:rprocess_Lnu_comparison}), which is likely to be greater in the case of a more massive torus \citep{Just2015a}.  We find that by including neutrino absorption, the entire range of $r$-process nuclei from the first to the third $r$-process peak can be synthesized in the unbound outflows, in agreement with the findings of previous $\alpha-$disk simulations (e.g.~\citealt{Wu2016}).

\item[(viii)] The production of first-to-third-peak $r$-process elements with relative abundances in good agreement with observed solar abundances and those on metal-poor stars in the halo of our galaxy, together with the inferred total ejecta masses (Eq.~\eqref{eq:M_ej}) and the relatively high rate of NS--NS mergers inferred from the discovery of GW170817 \citep{LIGO+17DISCOVERY}, arguably provide the strongest evidence yet, backed by first-principle simulations, for NS mergers being the prime production site of $r$-process elements in the universe.    

\end{itemize}


\acknowledgments

We thank A.~Beloborodov, R.~Fern\'andez, R.~Haas, W.~Kastaun,
J.~Lippuner, G.~Mart\'inez-Pinedo, P.~Moesta, C.~Ott, Y.~Qian, D.~Radice, L.~Roberts, and M.-R. Wu for valuable discussions. Resources supporting this work were provided by the NASA High-End Computing (HEC) Program through the NASA Advanced Supercomputing (NAS) Division at Ames Research Center. Support for this work was provided by the National Aeronautics and Space Administration through Einstein Postdoctoral Fellowship Award Number PF6-170159 issued by the Chandra X-ray Observatory Center, which is operated by the Smithsonian Astrophysical Observatory for and on behalf of the National Aeronautics and Space Administration under contract NAS8-03060.  BDM and DMS acknowledge support from NASA ATP grant NNX16AB30G and NSF grant AST-1410950.
 

\appendix

\section{Temperature dependence of electron chemical potential}
\label{app:temp_dependence}

In this appendix, we derive the temperature dependence of the chemical
potential $\mu$ of electrons in relativistic degenerate matter
(Eq.~\eqref{eq:mu_T}). We start by writing the electron number density
(Eq.~\eqref{eq:npm}) as
\begin{equation}
  n_- = \frac{(m_\el c)^3}{\pi^2 \hbar^3}\int_{-\infty}^{\infty}
  f_-(E,T,\mu) g(E)\,\mathrm{d}E, \label{eq:nel_app}
\end{equation}
with
\begin{equation}
  g(E)\equiv \left\{ \begin{array}{ccl}
E\sqrt{E^2-1} &, & E \ge 1 \\
0           &, & E < 1.
\end{array} \right.
\end{equation}
Noting that (i) $g(E)$ only diverges as a power of $E$ as
$E\rightarrow \infty$, (ii) $g(E)\rightarrow 0$ as $E\rightarrow
-\infty$, and (iii) $g(E)$ is well behaved at $E\sim \mu >
1$, we can make use of the Sommerfeld expansion and write
\begin{equation}
  n_-= \frac{(m_\el c)^3}{\pi^2 \hbar^3}\left\{\int_{-\infty}^{\mu}
      g(E) \,\mathrm{d}E
  +   2\sum_{n=1}^{\infty} (1 - 2^{1-2n}) \zeta(2n) \Theta^{2n}
      \left[\frac{\mathrm{d}^{2n-1}g(E)}{\mathrm{d}E^{2n-1}}
      \right]_{E=\mu}\right\}, \label{eq:nel_Sommerfeld}
\end{equation}
where $\zeta$ is the Riemann $\zeta$-function. One can easily convince
oneself that, at least for the first few derivatives of $g(E)$,
\begin{equation}
  \left[\frac{\mathrm{d}^{n}g(E)}{\mathrm{d}E^{n}}\right]_{E=\mu} \simeq \frac{g(\mu)}{\mu^n}\mathcal{O}(1),
\end{equation}
where $\mathcal{O}(1)$ refers to terms of order unity. Thus, the ratio
of subsequent terms in the sum of Eq.~\eqref{eq:nel_Sommerfeld} scales
as $\eta^{-2}$, and for degenerate matter $\eta = \mu/\Theta \gg 1$,
the sum converges rapidly. Only retaining the first two terms in
Eq.~\eqref{eq:nel_Sommerfeld} results in
\begin{equation}
  n_-\simeq \frac{(m_\el c)^3}{\pi^2
    \hbar^3}\left\{\frac{1}{3}(\mu^2-1)^{3/2} + \frac{\pi^2}{6} \Theta^2 (\mu^2-1)^{1/2}\right\}. 
\end{equation}
Again to first order, this can be rewritten as
\begin{equation}
  \left(\frac{\mu^2-1}{E_\mathrm{F}^2-1}\right)^{1/2}\simeq \left\{1 - \frac{\pi^2}{6} \frac{
\Theta^2}{E_\mathrm{F}^2-1}\right\}, 
\end{equation}
where $E_\mathrm{F}\equiv \mu(T=0)$ is the relativistic Fermi
energy. This is the relation to be derived.


\bibliographystyle{aasjournal}
\bibliography{references}

\end{document}